\documentclass[12pt]{article}

\usepackage[colorlinks,linkcolor=Cobalt,citecolor=Cobalt,urlcolor=Cobalt,bookmarks,bookmarksnumbered]{hyperref}
\usepackage{XoohmF}

\def\hj{{\hat\jmath}}
\def\hl{{\hat\ell}}
\def\sB{\mathscr{B}}
\def\cB{{\cal B}}
\def\sC{\mathscr{C}}
\def\rD{{\rm D}}
\def\bD{\boldsymbol{D}}
\def\bDb{\hbox to0em{\kern.25em\rule[1.8ex]{.6em}{.15ex}}\boldsymbol{D}}

\def\sD{\mathscr{D}}
\def\vdt{\partial_\tau}
\def\sF{\mathscr{F}}
\def\cF{{\cal F}}
\def\cF{\mathcal{F}}
\setbox0=\hbox{$\IL$}\newdimen\IGaH\IGaH=\ht0
\def\IGa{\raisebox{\IGaH}{\rotatebox{180}{\reflectbox{$\IL$}}}}

\def\cG{\mathcal{G}}
\def\sH{\mathscr{H}}
\def\iL{\IL^{\!\sss-1}}
\def\cM{{\cal M}}
\def\rO{\mathop{\rm O}\limits}
\def\cX{\mathcal{X}}
\def\cY{\mathcal{Y}}
\def\Bm#1{\left[\begin{smallmatrix}#1\end{smallmatrix}\right]}

\def\cb#1#2{\setlength\fboxsep{1pt}\colorbox{#1}{\color{#1}\fbox{\color{black}#2}}}
\def\cB#1{\hbox to0pt{\setlength\fboxsep{0pt}\hss\color{grey3}\fbox{\cb{white}{#1}}\hss}}
\def\bB#1{\hbox to0pt{\setlength\fboxsep{0pt}\hss\color{grey3}\fbox{\cb{black}{\color{white}#1}}\hss}}

\def\Lx#1{\makebox[0pt][l]{#1}}
\def\Rx#1{\makebox[0pt][r]{#1}}

\def\hCS{\ha{\text{tCS}}}
\def\HCS{\text{($\ha{\text{tCS}}$)}}
\def\rot{{\text{rot}}}
\def\exR{{\text{exR}}}

 \font\rOpe=cmsy10                        
 \def\ktl{{\hbox{\rOpe\char'170}}}        
 \def\kbl{{\hbox{\rOpe\char'170}}}        
 \def\kcr{{\reflectbox{\rOpe\char'170}}}        
 \def\ktr{{\reflectbox{\rOpe\char'170}}}        
 \def\kbr{{\reflectbox{\rOpe\char'170}}}        
 \def\Border{\vbox{\hsize0pt
        \setlength{\unitlength}{1mm}
        \newcount\xco
        \newcount\yco
        \xco=-21
        \yco=12
        \begin{picture}(0,0)(-7.5,0)
        \put(\xco,\yco){$\ktl$}
        \advance\yco by-1
        {\loop
        \put(\xco,\yco){$\kcr$}
        \advance\yco by-2
        \ifnum\yco>-240
        \repeat
        \put(\xco,\yco){$\kbl$}}
        \xco=170
        \yco=12
        \put(\xco,\yco){$\ktr$}
        \advance\yco by-1
        {\loop
        \put(\xco,\yco){$\kcr$}
        \advance\yco by-2
        \ifnum\yco>-240
        \repeat
        \put(\xco,\yco){$\kbr$}}
        \put(-19.5,13){\scalebox{.6067}{%
         University of Maryland Center for String and Particle  Theory \&\ Physics Department%
        |University of Maryland Center for String and Particle  Theory \&\ Physics Department}}
        \put(-19.5,-241.5){\scalebox{.58385}{%
         Howard University Department of Physics and Astronomy%
        |Howard University Department of Physics and Astronomy%
        |Howard University Department of Physics and Astronomy}}
        \end{picture}
        \par\vskip-8mm}}
\definecolor{UMred}{rgb}{.9,0,0}
\definecolor{HUblue}{rgb}{.0,.3,.7}
 \def\UMbanner{\vbox{\hsize0pt
        \setlength{\unitlength}{.4mm}
        \thicklines\color{UMred}
        \begin{picture}(0,0)(-30,-10)
        \put(165,16){\line(1,0){4}}
        \put(170,16){\line(1,0){4}}
        \put(180,16){\line(1,0){4}}
        \put(175,0){\line(1,0){4}}
        \put(180,0){\line(1,0){4}}
        \put(185,0){\line(1,0){4}}
        \put(169,0){\line(0,1){16}}
        \put(170,0){\line(0,1){16}}
        \put(179,0){\line(0,1){16}}
        \put(180,0){\line(0,1){16}}
        \put(184,0){\line(0,1){16}}
        \put(185,0){\line(0,1){16}}
        \put(169,16){\oval(8,32)[bl]}
        \put(170,16){\oval(8,32)[br]}
        \put(179,0){\oval(8,32)[tl]}
        \put(185,0){\oval(8,32)[tr]}
        \color{HUblue}
        \put(167.75,-2){\line(1,0){4}}
        \put(172.75,-2){\line(1,0){4}}
        \put(177.75,-2){\line(1,0){4}}
        \put(182.75,-2){\line(1,0){4}}
        \put(167.75,-2){\line(0,-1){16}}
        \put(171.75,-2){\line(0,-1){16}}
        \put(172.75,-2){\line(0,-1){16}}
        \put(176.75,-2){\line(0,-1){16}}
        \put(181.75,-2){\line(0,-1){16}}
        \put(182.75,-2){\line(0,-1){16}}
        \put(181.75,-2){\oval(8,32)[bl]}
        \put(182.75,-2){\oval(8,32)[br]}
        \put(167.75,-18){\line(1,0){4}}
        \put(172.75,-18){\line(1,0){4}}
        \end{picture}
        \par\vskip-6.5mm
        \thicklines}}
\def\Ft#1{\,\footnote{#1}}
\newdimen\parshift\parshift=\parindent
\catcode`@=11
 \long\def\@footnotetext#1{\insert\footins{\reset@font\footnotesize
           \interlinepenalty\interfootnotelinepenalty\splittopskip%
            \footnotesep\splitmaxdepth\dp\strutbox\floatingpenalty\@MM%
             \hsize\columnwidth\addtolength{\hsize}{-4pc}
              \@parboxrestore\protected@edef\@currentlabel%
              {\csname p@footnote\endcsname\@thefnmark}%
                \color@begingroup%
                 \@makefntext{\rule\z@\footnotesep\ignorespaces#1%
                  \@finalstrut\strutbox}%
                \color@endgroup}}
 \long\def\@makefntext#1{\hglue\parshift%
           \vbox{\noindent\baselineskip=11pt plus.5pt minus.5pt%
                  \hb@xt@0em{\hss\@makefnmark\kern1pt}#1}}
\catcode`@=12

 \SfTitles
 \allowdisplaybreaks
 \numberwithin{equation}{section} 

\begin{document}
\setcounter{page}{0}
\thispagestyle{empty}
\vbox{\Border\UMbanner}
\vglue0mm
 \noindent
 \today
  \hfill\smash{\parbox[t]{50mm}{\raggedleft\small
                           UMDEPP-014-007
 }}
 \vspace*{10mm}
 \begin{center}
{\Large\bsf\boldmath
  On Clifford-Algebraic ``Holoraumy,''\\*[2mm]
  Dimensional Extension, and SUSY Holography
 }\\*[7mm]\vfill
{\bsf 
      S.J.\,Gates, Jr.$^*$,
      T.\,H\"{u}bsch$^\dag$
      and
      K.\,Stiffler$^\ddag$
}\\*[5mm]
\parbox[t]{80mm}{\small\centering\it
      $^*$Center for String and Particle Theory\\[-1mm]
Department of Physics, University of Maryland\\[-1mm]
College Park, MD 20742-4111 USA
  \\[-4pt] {\tt\slshape  gatess@wam.umd.edu}}
\\*[4mm]
\parbox[t]{80mm}{\small\centering\it
      $^\dag$Department of Physics \&\ Astronomy,\\[-1mm]
      Howard University, Washington, DC 20059
  \\[-4pt] {\tt\slshape  thubsch@howard.edu}}
\\*[4mm]
\parbox[t]{120mm}{\small\centering\it
      $^\ddag$Department of Chemistry, Physics, and Astronomy\\[-1mm]
      Indiana University Northwest, Gary, Indiana 46408 USA
  \\[-4pt] {\tt\slshape  kmstiffl@iun.edu}}
\\*[7mm]\vfill
{\sf\bfseries ABSTRACT}\\[3mm]
\parbox{141mm}{\addtolength{\baselineskip}{-2pt}\parindent=2pc\noindent
We analyze the group of maximal automorphisms of the $N$-extended world-line supersymmetry algebra, and its action on off-shell supermultiplets. This defines a concept of ``holoraumy'' that extends the notions of holonomy and curvature in a novel way and provides information about the geometry of the supermultiplet field-space. In turn, the ``holoraumy'' transformations of 0-brane dimensionally reduced supermultiplets provide information about Lorentz transformations in the higher-dimensional spacetime from which the 0-brane supermultiplets are descended. Specifically, Spin(3) generators are encoded within 0-brane ``holoraumy'' tensors. World-line supermultiplets are thus able to holographically encrypt information about higher dimensional
spacetime geometry.
}
\end{center}
\vspace*{10mm}\vfill
\noindent
\parbox[t]{60mm}{PACS: 11.30.Pb, 12.60.Jv}\hfill
\parbox[t]{100mm}{\raggedleft\small\baselineskip=12pt\sl
            If the facts are right, then the proofs are a matter of\\*[-1pt]
             playing around with the algebra correctly.\\*
            |\,Richard P.~Feynman}
\vspace*{10mm}\vfill\vfill
\clearpage\thispagestyle{empty}
\ToC{3}{12pt}
\setcounter{page}{1}
\section{Introduction, Results, and Summary}
In standard (bosonic/commutative) differential geometry, covariant derivatives generate parallel translations; their {\em\/commutator\/} defines the torsion $(T_{\m\n}{}^\rho$) and the curvature ($\ha{R}_{\m\n}$) tensors:
\begin{equation}
 [\N_\m,\N_\n]=T_{\m\n}{}^\r\N_\r+\ha{R}_{\m\n},
 \label{e:TdR}
\end{equation}
where $\ha{R}_{\m\n}$ is valued in the Lie algebra of transformations of the objects (vectors, tensors,\ldots) upon which it acts. For example, when acting on covariant vectors, $\ha{R}_{\m\n}$ takes on the familiar form of the Riemann tensor, $[\ha{R}_{\m\n}(V)]_\r\Defl R_{\m\n\r}{}^\s V_\s$. Algebraically, the curvature tensor $\ha{R}_{\m\n}$ is valued in the Lie algebra of linear, homogeneous transformations of the tensors upon which it acts. The geometric interpretation of the commutator $[\N_\m,\N_\n]$ is the comparison of a concatenation of two infinitesimal parallel translations with the oppositely ordered concatenation of those same parallel translations. When acting on tangent vectors on a manifold, the curvature term determines the linear, homogeneous transformation, \ie, {\em\/holonomy\/} associated to the quadrilateral loop formed by the alternating concatenations of infinitesimal parallel transports.

In this paper, we discuss a supersymmetric analogue to bosonic curvature and torsion which we dub ``holoraumy\footnote{We ask the forbearance of linguistic purists who will decry the mixing of the Greek word ``{\em\/holos\/}''~(complete) with the German word ``{\em\/Raum\/}''~(space); the linguistically pure ``holochory'' (from {\em\/holos\/}+{\em\/horos\/}) seems considerably less euphonious, at least to our ears, and much more prone to misunderstanding.}:''  
\begin{subequations}\label{e:holoR}
\begin{alignat}9
	[ {\rm D}_a , {\rm D}_b  ] ( \text{fermions} )= & -2 i \mathscr{F}_{ab}  (\text{fermions} ) , \label{e:holoRF} \\
	[ {\rm D}_a , {\rm D}_b  ] ( \text{bosons} )= & -2 i \mathscr{B}_{ab}  ( \text{bosons} ) .\label{e:holoRB} 
\end{alignat}
\end{subequations}
The structure~\eq{e:holoR} is contained within off-shell representations of spacetime supersymmetry, and in this paper we focus on the case where $D_a$ is the $4D$, $\mathcal{N}=1$ supercovariant derivative. Notice that holoraumy is defined off of the \emph{commutator} of the supercovariant derivative $D_a$ analogous to how holonomy is defined off of the commutator of the bosonic covariant derivative $\nabla_\mu$. As suggested in~\eq{e:holoR}, we will use the convention that $\mathscr{F}_{ab}$ refers to a holoraumy tensor that acts on fermions and $\mathscr{B}_{ab}$ refers to a holoraumy tensor that acts on bosons. 

As the goal of this paper is to show how holoraumy of lower-dimensional supersymmetry encodes information on Lorentz transformations in higher-dimensions,  we define holoraumy off of $1D$, $N=4$ supercovariant derivatives as
\begin{subequations}\label{e:holoR1D}
\begin{alignat}9
	[ {\rm D}_I , {\rm D}_J  ] (\text{fermions} )= & -2 i \mathscr{F}_{IJ}  (\text{fermions} ) , \label{e:holoRF1D} \\
	[ {\rm D}_I , {\rm D}_J  ] ( \text{bosons} )= & -2 i \mathscr{B}_{IJ}  ( \text{bosons} ).\label{e:holoRB1D} 
\end{alignat}
\end{subequations}
Our convention used throughout this paper is that $D_I$ with an upper case Latin index is reserved for $1D$, $N$-extended supersymmetry where as $D_a$ with a lower-case Latin index is reserved for $4D$, $\mathcal{N}=1$ supersymmetry. 

A seldom discussed fact about gamma matrices is that in $4D$, their products separate into two separate $su(2)$ algebras that generate $3D$ spatial rotations ($\spin(3)_\rot$) and extended R-symmetry ($\spin(3)_\exR$):
\begin{align*}
	\spin(3)_\rot: \left( \gamma^{12}~,~\gamma^{23},~\gamma^{31}\right)~~~,~~~	\spin(3)_\exR: \left( \gamma^{0}~,~\gamma^{123}~~,~~\gamma^{0123}\right)~~~.
\end{align*}
The product $\gamma^{12} = \gamma^{1}\gamma^2$ and all other such products in our conventions are shown in Table~\ref{t:16}. 

The main result of this paper is that the symmetries $\spin(3)_\rot$ and/or $\spin(3)_\exR$ are manifest in the holoraumy of $1D$, $N=4$ adinkra ``shadows"  of $4D$, $\mathcal{N}=1$ supersymmetric systems. That is, lower-dimensional supersymmetric systems contain information on higher-dimensional Lorentz transformations. 

We now give a very brief description of adinkras. Further detailed descriptions are found throughout the paper where adinkras are used. For a more complete review of adinkras, see for instance section 2 of Ref.~\cite{SGG}.  Consider as an example the adinkra for the chiral multiplet as shown in~\eq{e:CSadinkra}:

\begin{equation}
 \vC{\begin{picture}(40,35)(0,-2)
   \put(0,0){\includegraphics[height=31mm]{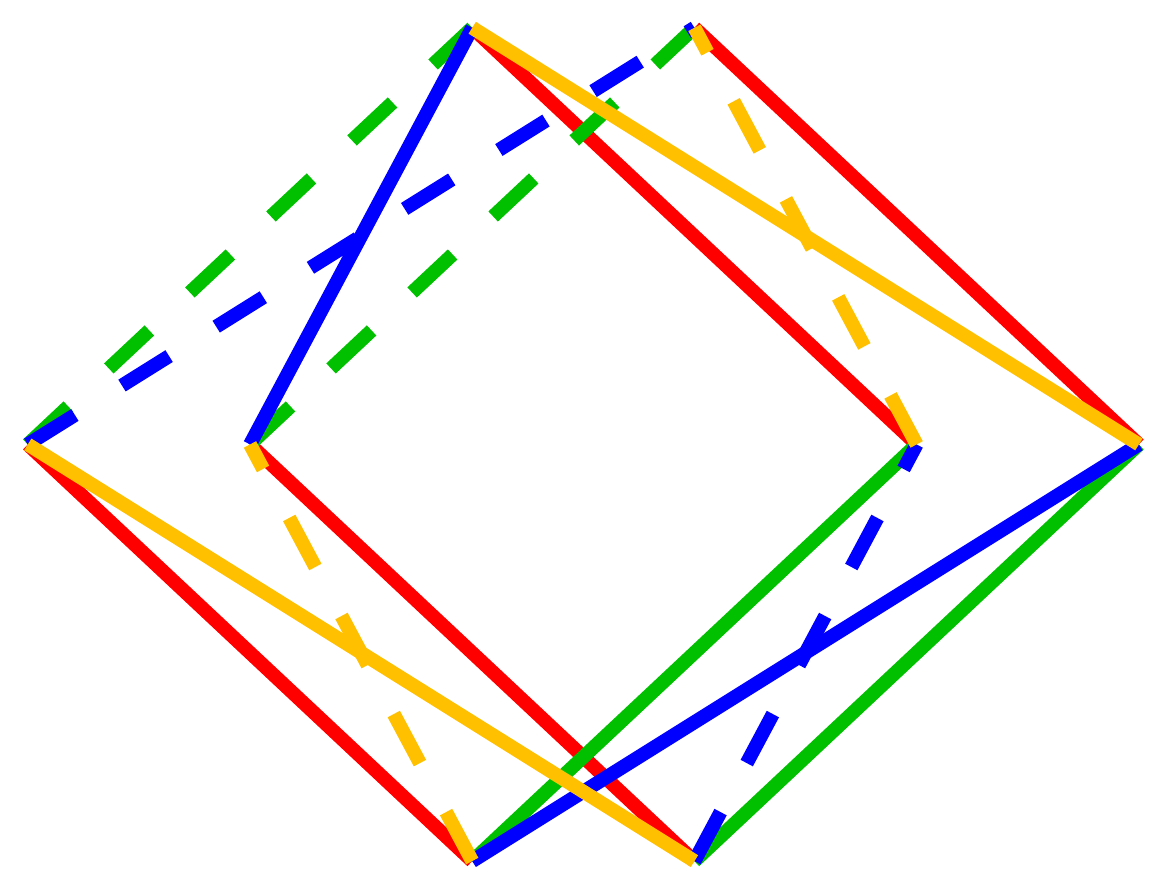}}
    \put(16,0){\cB{$A$}}
    \put(25,0){\cB{$B$}}
    \put(1,14){\bB{$\j_1$}}
    \put(11,14){\bB{$-\j_4$}}
    \put(30,14){\bB{$\j_2$}}
    \put(39,14){\bB{$\j_3$}}
    \put(16,29){\cB{$\cF$}}
    \put(25,29){\cB{$-\cG$}}
 \end{picture}}
 \label{e:CSadinkra}
\end{equation}
The first step in creating this adinkra is to reduce the supersymmetric transformation laws for the component fields of the  $4D$, $\mathcal{N}=1$ chiral multiplet
\begin{subequations}
 \label{e:CS4Dfull}
\begin{alignat}9
  \rD_a A
 &=\j_a,&\qquad
  \rD_a B
 &= i (\g^5)_a{}^b \, \j_b,\mkern100mu \label{e:CS4DfullAB} \\*
  \rD_a \mathcal{F}
  &= (\g^\mu)_a{}^b \, \partial_\mu \j_b,&\qquad
  \rD_a \mathcal{G}
  &= i \,(\g^5\g^\mu){}_a{}^b \partial_\mu \,\j_b, \label{e:CS4DfullFG}\\*[1mm]
  \rD_a \j_b
  &=\makebox[0pt][l]{$
    i (\g^\mu)_{a \,b}\,\partial_\mu A - (\g^5\g^\mu)_{a \,b}\,\partial_\mu B
    -i C_{a\,b}\, \mathcal{F} +(\g^5)_{ a \, b}\, \mathcal{G},$} \label{e:CS4Dfullj}
\end{alignat}
\end{subequations}
to the 0-brane by considering only temporal ($\tau$) dependence of all fields:
 \begin{subequations}
 \label{e:CS1Dfull}
\begin{alignat}9
  \rD_a A
 &=\j_a,&\qquad
  \rD_a B
 &= i (\g^5)_a{}^b \, \j_b,\mkern100mu \label{e:CS1DfullAB} \\*
  \rD_a \mathcal{F}
  &= (\g^0)_a{}^b  \partial_\tau \,\j_b,&\qquad
  \rD_a \mathcal{G}
  &= i \,(\g^5\g^0){}_a{}^b \partial_\tau \,\j_b, \label{e:CS1DfullFG}\\*[1mm]
  \rD_a \j_b
  &=\makebox[0pt][l]{$
    i (\g^0)_{a \,b}\,\vdt A - (\g^5\g^0)_{a \,b}\,\vdt B
    -i C_{a\,b}\,\mathcal{F} +(\g^5)_{ a \, b}\,\mathcal{G}.$} \label{e:CS1Dfullj}
\end{alignat}
\end{subequations}
The 0-brane transformations \eq{e:CS1Dfull} are encoded within the adinkra \eq{e:CSadinkra} with complete fidelity. At first glance, it would appear that an adinkra only ``knows'' about the temporal dimension. Upon further investigation we find that adinkras are indeed ``aware'' of higher dimensions as well.  We specifically show in this paper that the adinkra~\eq{e:CSadinkra} encodes information on $3D$ spatial rotations through the appearance of the $\spin(3)_\rot$ algebra elements within the holoraumy tensors.  Similar results are found for three other $4D$, $\mathcal{N}=1$ supersymmetric multiplets as well: the vector, tensor, and twisted chiral multiplets. This is strong evidence for a suspicion that we have long thought in working with adinkras: that supersymmetry itself keeps track of spatial information even after dimensional reduction.

 The holoraumy structure reveals itself upon following the line of work that began in Refs.\cite{rGR-1,rGR0} and eventually coalesced with the idea of ``SUSY holography''\cite{rGLPR,rGLP}---the proposal that supermultiplets retain enough information in the process of dimensional reduction (even to world-line supersymmetric quantum mechanics!) so as to enable a full reconstruction of the original, higher-dimensional spacetime supersymmetry structure.

In general, these holoraumy transformations {\em\/compose\/} linear homogeneous transformations in the target space with the full Poincar\'e group of transformations or perhaps even the conformal transformations and central charge action in the domain space.
 Herein, we focus on the simplest version (see Eqs.\eq{e:ValEnv} below), exhibited on unitary, finite-dimensional, off-shell representations of world-line $N$-extended supersymmetry. This framework is the common denominator in all physics applications of supersymmetry, and will provide building blocks in all developments of a fully quantum description.
 Also, in the special case of so-called valise adinkras, the simplest off-shell linear representations of world-line $N$-extended supersymmetry, these holoraumy transformations turn out to be a uniform composition of target-space $\SO(n)$ rotations and domain-space $\t$-translations. 
 Fixed points of a similar type of composition of domain- and target-space transformations comprise ``orientifolds,'' which have been studied for well over two decades\cite{rSagn87,rHor89,rDLP89-OrF,rL89-OrF}. It is then a little surprising that this type of transformation, existing in all of field theory, has not been studied more systematically, and seems to have remained nameless.
 
Being a {\em\/composition\/} of target- and domain-space transformations, holoraumy differs significantly from holonomy in the field-space (thought of as the total space of the bundle of vector spaces spanned by all the fields and fibered over the domain-space), which is a linear combination of those two types of transformation. Also, the very definition and computation of holoraumy differ significantly from holonomy, as will be shown herein. It is this inherent novelty that motivates our present study.

In the following, we introduce two holoraumy tensors that naturally appear in the context of so-called valise supermultiplets. The definition of these tensors is reminiscent of the familiar definition of curvature and torsion tensors in differential geometry: On a curved general manifold, the commutator of two covariant derivatives  produces the Riemann curvature and the torsion tensors. In turn, applying {\em\/commutators\/} (rather than anti-commutators!) of the covariant super-derivatives on supermultiplets produces specific transformations among the bosonic and fermionic components separately.
 These transformations generate a group and the objects encoding them we dub the {\em\/quadratic\/} holoraumy tensors.

The first of the two particular holoraumy tensors we will study herein has implicitly appeared in our previous work \cite{rSJG12,rGHS-HoloSuSy}.  Here we formalize its definition in such a way that it can be easily generalized.  This holoraumy tensor seems particularly suited to play a fundamental role in understanding SUSY holography.  Just as the Riemann curvature tensor may be used to define holonomy groups for curved manifolds, the holoraumy tensors likewise provide a similar characteristic for supermultiplets, as well as the supersymmetric field theory models built from such supermultiplets. Just as the holonomy group of a real $n$-dimensional Riemannian manifold must be a subgroup of $\Spin(n)$ and that of a complex $n$-dimensional K\"ahler manifold a subgroup of $\rOp{U}(n)$, the holoraumy groups acting on both bosonic and fermionic component fields of any supermultiplet must be subgroups of\ft{\label{f:Pin}$\Pin(p,q)$ extends $\Spin(p,q)$ by elements of negative determinant, and is the double-cover of the orthogonal group $\rOp{O}(p,q)$ wherein spinors are faithful representations; $\rOp{O}(p,q)$ preserves a metric with $p$ negative and $q$ positive eigenvalues. By definition, $\Pin(n)\Defl\Pin(0,n)$; $\Spin(p,q)\approx\Spin(q,p)$ but $\Pin(p,q)\not\approx\Pin(q,p)$.} $\Pin(N)=\Aut(\SSp^{1|N})$, the maximal group of outer automorphisms of the $N$-extended world-line supersymmetry without central extensions\eq{e:SuSyD}; see below.

Perhaps the most remarkable feature about the holoraumy tensors is that they are inherent characteristics of representations of world-line supersymmetry. Yet, as we show below, they provide ample information about higher-dimensional supermultiplets to which those world-line supermultiplets may extend, as well as obstructions that prevent extension to certain supermultiplets in certain higher-dimensional spacetimes.
In particular, the fermionic holoraumy tensor contains information about Fierz identities that hold within the 3+1-dimensions, ${\cal N}\,{=}\,1$ chiral and vector supermultiplets when they are dimensionally reduced to the world-line 0-brane and upon a type of field redefinitions called ``node-lowering''\cite{r6-1}.

This paper is organized as follows: the basic definitions and notation is provided in the remainder of this introduction, and Section~\ref{s:HoloR} provides the general framework, definitions and results pertaining to the holoraumy tensors. These ideas are then applied to several 3+1-dimensional ${\cal N}\,{=}\,1$ supermultiplets and their world-line dimensional reduction in  Section~\ref{s:3LPs}. Finally, Section~\ref{s:Coda} collects our concluding comments, while the technical details are deferred to the Appendices.

\paragraph{Notation and Definitions:}
We now focus on world-line general $N$-extended supersymmetry without central extensions, for which the covariant super-derivatives and the $\t$-derivative satisfy the algebra\ft{Throughout, and without loss of generality\cite{rHTSSp08}, we use superspace methods and notation\cite{r1001,rWB,rBK}.}
\begin{equation}
  \SSp^{1|N}\,:\quad
  \big\{\,\rD_I\,,\,\rD_J\,\big\} = 2i\,\d_{IJ}\,\vdt,\qquad
  \big[\,\vdt\,,\,\rD_I\,\big] = 0,\qquad
  I,J=1,\cdots,N.
 \label{e:SuSyD}
\end{equation}
Unitary and finite-dimensional representations of this algebra---supermultiplets---are provided by collections of intact\ft{By ``intact superfields,'' we mean un-constrained, un-projected, un-gauged and in no other way restricted Salam-Strathdee superfields\cite{rSSSS1}, but will assume them to be real unless explicitly stated otherwise.} superfields, related by first order super-differential relations.

It has been proven recently\cite{rDHIL13} that every unitary, finite-dimensional and engineerable\ft{A supermultiplet is engineerable if all component fields have a consistent assignment of engineering dimension, \ie, physical units; this is a natural requirement in all physics applications.} off-shell supermultiplet of world-line $N$-extended supersymmetry without central charges\eq{e:SuSyD}:
\begin{enumerate}\itemsep=-3pt\vspace*{-3mm}
 \item becomes, through an iterative sequence of local component field transformations (so-called ``node-raising''\cite{r6-1}, \ie, ``dressing''\cite{rKRT}, and simple linear combinations), a direct sum of minimal ``valise'' supermultiplets\cite{r6-3.2},
and conversely
 \item may be ``synthesized'' from this direct sum of minimal ``valise'' supermultiplets, by reversing the linear combination and node-raising procedure.
\end{enumerate}\vspace*{-2mm}
 For this reason, we focus herein on these valise supermultiplets, wherein the component superfields, $(\F_i|\J_\hj)$, are related by means of the first order super-differential system of equations:
\begin{subequations}
 \label{e:valise}
\begin{alignat}9
  \rD_I\,\F_i   &= i(\IL_I)_i{}^\hj\,\J_\hj,\qquad &\hj&=1,\cdots,d; \label{e:DF=J}\\
  \rD_I\,\J_\hj &=  (\IR_I)_\hj{}^i\,\vdt\F_i,\qquad &i&=1,\cdots,d.  \label{e:DJ=F}
\end{alignat}
\end{subequations}
Within such a valise supermultiplet, the engineering dimensions (physical units) of all bosons $\F_i$ are the same, as are those of the fermions $\J_\hj$, and the two are related by
 $[\J_\hj]=[\F_i]+\frc12$, with $[\vdt]=1=2[\rD_I]$.
For this to be an off-shell representation of the algebra\eq{e:SuSyD}, the second order super-differential identities\eq{e:SuSyD} must continue to hold on each of the component superfields, $(\F_i|\J_\hj)$, without requiring any of these superfields to satisfy any $\t$-differential equation, which could be derived as an equation of motion from some Lagrangian. In turn, this requires the $\IL_I$- and $\IR_I$-matrices to close the so-called ${\cal GR}(d,N)$ algebra\cite{rGR2}:
\begin{subequations}
 \label{e:GRdN}
  \begin{alignat}9
    (\IL_I)_i{}^\hj\,(\IR_J)_\hj{}^k + (\IL_J)_i{}^\hj\,(\IR_I)_\hj{}^k
     &= 2\,\d_{IJ}\,\d_i{}^k, \label{e:GRdNLR}\\
    (\IR_I)_\hj{}^i\,(\IL_J)_i{}^\hl + (\IR_J)_\hj{}^i\,(\IL_I)_i{}^\hl
     &= 2\,\d_{IJ}\,\d_\hj{}^\hl. \label{e:GRdNRL}
  \end{alignat}
\end{subequations}
Following Refs.\cite{r6-3,r6-3.1}, we note that the $I=J$ cases of\eq{e:GRdN} imply the identity $\IR_I=\IL^{\!-1}_I$, and we use this result hereafter.

Using the well-known relationship between the covariant super-derivatives and the supercharges, $Q_I = -i(\rD_I+2i\d_{IJ}\q^J\vdt)$ and the projection to the purely bosonic part of superspace, the system\eq{e:valise} produces the supersymmetry transformations within the supermultiplet $(\F_i|\J_\hj)$. Herein, however, we explore\eq{e:valise} as a system of super-differential equations, which restrict the geometry of the field-space spanned by the component fields $\f_i\Defl\F_i|$ and $\j_\hj\Defl\J_\hj|$.

\section{Clifford-Algebraic Structure of World-Line Supermultiplets}
\label{s:HoloR}
Writing the super-differential system\eq{e:valise} as
\begin{equation}
  (\F_i|\J_\hj)~:\quad
  \rD_I\BM{\bm\F\\\bm\J\\} = \BM{\bm0&\IL_I\\\iL_I&\bm0}\BM{\vdt\bm\F\\i\bm\J\\},
 \label{e:Valise}
\end{equation}
defines:
\begin{equation}
  \IGa_I\Defl\bM{\bm0&\IL_I\\\iL_I&\bm0} \qquad\5{\text{(\ref{e:GRdN})}}\Longrightarrow\qquad
  \big\{\,\IGa_I\,,\,\IGa_J\,\big\} = 2\,\d_{IJ}\,\Ione,
 \label{e:Cl0N}
\end{equation}
and the matrices $\IGa_I$ generate the $\Cl(0,N)$ Clifford algebra. Refs.\cite{r6-3,r6-3.1} also introduce the fermion number operator, $\IGa_0\Defl(-1)^F$, which acts on the supermultiplet $(\F_1|\J_\hj)$ as a diagonal matrix $\IGa_0=\text{diag}(+1,\cdots,+1|{-}1,\cdots,{-}1)$. It is easy to show that $\{\IGa_0,\IGa_I\}=0$ and $(\IGa_0)^2=\Ione$ and solely by virtue of\eq{e:Cl0N}, so that $\IGa_0$ and the $\IGa_I$'s jointly generate the $\Cl(0,N{+}1)$ Clifford algebra.

\subsection{General Facts about World-Line Holoraumy}
\paragraph{The Enveloping System:}
The realization that\eq{e:valise} is equivalent to\eq{e:Valise} where $\IGa_I$ generate the $\Cl(0,N)$ Clifford algebra\eq{e:Cl0N} has a standard but important consequence\cite{rLM,rFRH}:
 The Clifford algebra $\Cl(0,N)\Defl\otimes^*\Span(\IGa_I)/\text{(\ref{e:Cl0N})}$ and the exterior algebra $\wedge^*\IGa$ are isomorphic as vector spaces, and so are both spanned by the matrices familiar to physicists as the Dirac algebra\ft{Throughout this paper, square brackets indicate normalized antisymmetrization of the enclosed indices:
$\IGa_{IJ}\Defl\inv{2!}(\IGa_I\IGa_J{-}\IGa_J\IGa_I)$,
$\IGa_{IJK}\Defl\inv{3!}(\IGa_I\IGa_J\IGa_K{-}\IGa_J\IGa_I\IGa_K{+}\ldots)$, \etc}:
\begin{equation}
  \Ione,\quad
  \IGa_I,\quad
  \IGa_{IJ},\quad
  \IGa_{IJK},\quad
   \ldots\quad
  \IGa_{I_1I_2\cdots I_N},
 \label{e:Dirac}
\end{equation}
which is a basis that is canonically induced from the choice of $\IGa_1,\cdots,\IGa_N$---using only Eq.\eq{e:Cl0N} to reduce un-symmetrized tensor products to the antisymmetric products,
 $\IGa_{I_1}\cdots\IGa_{I_n}\too{\sss\text{(\ref{e:Cl0N})}}\IGa_{[I_1}\cdots\IGa_{I_n]} +\text{lower-order terms}$.
This prompts us to consider the ``enveloping system'' of\eq{e:Valise}:
\begin{subequations}
 \label{e:ValEnv}
\begin{alignat}9
 \rD_{[I}\rD_{J]}
  \BM{\bm\F\\\bm\J\\}
 &=-\BM{\IL^{}_{[I}\iL_{J]} & \bm0\\  \bm0 &\iL_{[I}\IL^{}_{J]}}
     \BM{i\vdt\bm\F\\[2pt]i\vdt\bm\J\\[1pt]},
 \label{SomethingRandom} \\[2mm]
 \rD_{[I}\rD_{J}\rD_{K]}
  \BM{\bm\F\\\bm\J\\}
 &=-\BM{\bm0 &\IL^{}_{[I}\iL_{J}\IL^{}_{K]}\\
        \iL_{[I}\IL^{}_{J}\iL_{K]}  & \bm0}
     \BM{i\vdt^2\bm\F\\[2pt]-\vdt\bm\J\\[1pt]},
 \label{SomethingRandom2} \\[2mm]
 \rD_{[I}\rD_{J}\rD_{K}\rD_{L]}
  \BM{\bm\F\\\bm\J\\}
 &= \BM{\IL^{}_{[I}\iL_{J}\IL^{}_{K}\iL_{L]} & \bm0\\
        \bm0 &\iL_{[I}\IL^{}_{J}\iL_{K}\IL_{L]}}
     \BM{-\vdt^2\bm\F\\[2pt]-\vdt^2\bm\J\\[1pt]},
 \label{SomethingRandom3}
\end{alignat}
\end{subequations}
and so on. The antisymmetrized products of the $\IL$- and $\IL^{\!\sss-1}=\IR$-matrices appearing in\eq{e:ValEnv} have been studied previously in Ref.\cite{rGLP}, without exploring the holoraumy that they generate.

 In precise analogy with\eq{e:Dirac}, un-symmetrized composition of the super-derivatives is reduced to the antisymmetric products using only the defining relations of the supersymmetry algebra,
 $\rD_{I_1}\cdots\rD_{I_n}\too{\sss\text{(\ref{e:SuSyD})}}\rD_{[I_1}\cdots\rD_{I_n]} +\text{lower-order super-derivatives}$. We emphasize that both sides of the super-differential systems\eq{e:ValEnv} have been obtained solely by using the defining equations,\eq{e:SuSyD} and \eq{e:Cl0N} respectively. Therefore, they apply equally to all valise supermultiplets\eq{e:valise} and their geometric content is independent of any concrete choice of the $\IL_I$-matrices.
 The list of such higher-order super-differential relations is straightforwardly generated by iterating the relations from the original super-differential system\eq{e:valise}, and is finite: the progression effectively stops with the $N^\text{th}$ order super-derivative: Every product of more than $N$ super-differential operators $\rD_I$ necessarily reduces to one of an order at most equal to $N$, composed with a suitable power of $\vdt$. Thereby, every super-differential relation\eq{e:ValEnv} of order higher than $N$ simply produces a (multiple) $\vdt$-derivative of a lower-order relation.
 
Given the superspace relation $Q_I = -i(\rD_I+2i\d_{IJ}\q^J\vdt)$, the Taylor expansion of the ``finite'' supersymmetry transformation operator,
 $\exp\{i\e\,{\cdot}Q\} = \exp\{\e^I[\rD_I+2i\d_{IJ}\q^J\vdt]\}$, straightforwardly
reproduces the enveloping system\eq{e:ValEnv}, order-by-order and ignoring the terms explicitly containing $\q^I$ as they vanish upon component evaluation. The enveloping system of super-differential relations\eq{e:ValEnv} in addition to\eq{e:Valise} therefore spans the closed orbit of the supersymmetry transformations. Considering\eq{e:Valise} together with\eq{e:ValEnv} is therefore akin to considering the ``finite'' (super)symmetry transformation instead of just its infinitesimal generator.

Amongst the differential relations\eq{e:ValEnv}---and very much shadowing the situation in the structure of Clifford algebras---the even-order super-derivatives, $\rD_{[I}\rD_{J]}$, $\rD_{[I}\rD_J\rD_K\rD_{L]}$, \etc, are of special interest in that their application on the supermultiplet clearly elicits transformations that are {\em\/uniform compositions\/} of:
\begin{enumerate}\itemsep=-3pt\vspace*{-3mm}
 \item (possibly higher-order) translations in the domain space (here, time $\t$), and
 \item spin/statistics-preserving, linear and homogeneous transformations in the field-space.
\end{enumerate}\vspace*{-2mm}
This general nature of these transformations makes it possible to regard them as a natural generalization of the concept of holonomy as described in the text surrounding Eq.~(\ref{e:TdR}). 

In Eq.\eq{e:SuSyD} and all its applications, the operators $\rD_I$ and $\vdt$ both generate (twisted) translations in superspace, and the $\rD_I$'s anticommute with the supercharges and so are supersymmetry-covariant. Owing to the identity $\rD_{[I}\rD_{J]}=\inv2[\rD_I,\rD_J]$, these facts justify comparing the relation\eq{SomethingRandom} with the standard differential geometry definition\eq{e:TdR}. However, it is crucial that\eq{SomethingRandom} produces not a linear combination of translations and ``rotations'' as does\eq{e:TdR}, but a {\em\/composition\/} of the two.
 The higher even-order operators in the progression\eq{e:ValEnv} all produce similar types of transformations, and we dub this entire class of geometric effects {\em\/holoraumy\/}; we will specifically refer to ``quadratic holoraumy'' in distinction from the quartic and higher order operators as producing ``higher order holoraumy.''

 The identity $\rD_{[I}\rD_{J]}=\inv2[\rD_I,\rD_J]$ shows that the quadratic super-differential operator\eq{SomethingRandom} similarly compares the concatenation of two infinitesimal parallel translations---except that\eq{SomethingRandom} crucially uses the {\em\/wrong\/} type of bracket operation: The standard bracket operation comparing two anticommutative translations in superspace is the {\em\/anti\,}commutator, $\{\rD_I,\rD_J\}$---which must evaluate identically to $2i\d_{IJ}\vdt$ for every off-shell supermultiplet, by its very definition. It is thus the operators constructed using the ``wrong-type'' brackets,
\begin{subequations}
\label{e:4>2}
\begin{alignat}9
 \rD_{[I}\rD_{J]}
 &=\inv{2!}[\rD_I,\rD_J],\\
 \rD_{[I}\rD_J\rD_K\rD_{L]}
 &=\inv{4!}\Big(\big\{[\rD_I,\rD_J],[\rD_K,\rD_L]\big\}
                +\big\{[\rD_I,\rD_K],[\rD_L,\rD_J]\big\}
                 +\big\{[\rD_I,\rD_L],[\rD_J,\rD_K]\big\}\Big),
\end{alignat}
\end{subequations}
\etc, that elicit the non-trivial information. That is, only {\em\/jointly\/} do the standard bracket and the ``wrong-type'' bracket provide the complete information on the (un-symmetrized) concatenation of two (superspace-twisted) translations $\rD_I$ and $\rD_J$. Using the standard bracket relation\eq{e:SuSyD} implicitly, the holoraumy\eq{e:ValEnv} then exhibits this complete information.

The information elicited by the non-linear super-derivative operators\eq{e:ValEnv} is expected to strongly depend on the representation---which is precisely the feature that makes it useful. Indeed, this is not at all unexpected: In the familiar case of $\su(2)$, the defining relation $[\hat{J}_\a,\hat{J}_\b]=i\ve_{\a\b}{}^\g\hat{J}_\g$ holds identically in all representations. On the other hand, the results of the {\em\/wrong\/}-type bracket relations depend strongly on the representation:
 The familiar relation $\{\fc12\s_\a,\fc12\s_\b\}=\fc12\d_{\a\b}\Ione$ holds only for the spin-$\fRc12$ representation $\hat{J}_\a=\fc12\s_\a$; for larger spin, where $\hat{J}_\a$ are represented by matrices of larger size, the results of the {\em\/anti\kern1.5pt}commutators $\{\hat{J}_\a,\hat{J}_\b\}$ are no longer even linear combinations of the generators $\hat{J}_\a$ and the identity matrix, $\Ione$.

\paragraph{Quadratic Holoraumy Tensors:}
The realization that\eq{e:valise} is equivalent to\eq{e:Valise} where $\IGa_I$ generate the $\Cl(0,N)$ Clifford algebra\eq{e:Cl0N} also has another standard and just as important consequence\cite{rLM,rFRH}:
 Owing solely to the Clifford algebra defining relation\eq{e:Cl0N},
 the quadratic matrices $\IGa_{IJ}\Defl\inv2[\IGa_I,\IGa_J]$ canonically generate the $\Spin(0,N)$ group:
\begin{equation}
  \big[\,\IGa_{IJ}\,,\,\IGa_{KL}\,\big]
  = 2\d_{IL}\,\IGa_{JK} -2\d_{IK}\,\IGa_{JL} +2\d_{JK}\,\IGa_{IL} -2\d_{JL}\,\IGa_{IK}.
 \label{e:SON}
\end{equation}
This gives a special interpretation to the numerical matrices appearing in the quadratic holoraumy relation\eq{SomethingRandom}. Let us define:
\begin{equation}
 \sB_{IJ}\Defl \IL^{_{}}_{[I}\iL_{J]}\quad\text{and}\quad
 \sF_{IJ}\Defl \iL_{[I}\IL^{}_{J]},\qquad\text{so}\qquad
 \IGa_{IJ}=\bM{\sB_{IJ} & 0\\ 0 & \sF_{IJ}},
 \label{e:QHT}
\end{equation}
that is,
\begin{equation}
  \rD_{[I}\rD_{J]}\,\bm\F_i   = -i\,(\sB_{IJ})_i{}^k\,\vdt\bm\F_k
   \qquad\text{and}\qquad
  \rD_{[I}\rD_{J]}\,\bm\J_\hj = -i\,(\sF_{IJ})_\hj{}^\hl\,\vdt\bm\J_\hl.
 \label{e:DDH}
\end{equation}
The block-diagonal nature of the canonical result\eq{e:SON} and the definitions\eq{e:QHT} then imply that
\begin{subequations}
 \label{e:BFON}
\begin{align}
  \big[\,\sB_{IJ}\,,\,\sB_{KL}\,\big]
 &= 2\d_{IL}\,\sB_{JK} -2\d_{IK}\,\sB_{JL} +2\d_{JK}\,\sB_{IL} -2\d_{JL}\,\sB_{IK},
 \label{e:BON}\\
  \big[\,\sF_{IJ}\,,\,\sF_{KL}\,\big]
 &= 2\d_{IL}\,\sF_{JK} -2\d_{IK}\,\sF_{JL} +2\d_{JK}\,\sF_{IL} -2\d_{JL}\,\sF_{IK}.
 \label{e:FON}
\end{align}
\end{subequations}
That is, both (quadratic) holoraumy tensors $\sB_{IJ}$ and $\sF_{IJ}$ generate two separate Lie groups that are isomorphic to subgroups of $\Spin(N)$:
\begin{subequations}
 \label{e:HBF}
\begin{alignat}9
 \sB_{IJ}&:&~~
 \F_i&\to(\sB_{IJ})_i{}^k(i\vdt\F_i)
         =(\IL^{}_{[I})_i{}^\hj(\iL_{J]})_\hj{}^k\,(i\vdt\F_k), \label{e:BF}\\
 \sF_{IJ}&:&~~
 \J_\hj&\to(\sF_{IJ})_\hj{}^\hl(i\vdt\J_\hl)
         =(\iL_{[I})_\hj{}^i(\IL^{}_{J]})_i{}^\hl\,(i\vdt\J_\hl). \label{e:FJ}
\end{alignat}
\end{subequations}
That is, the $\sB_{IJ}$-tensors generate $\Spin(N)$-transformations upon the bosons $\F_i$, while the $\sF_{IJ}$-tensors generate the action of this group upon the fermions $\J_\hj$.
 In both cases, these field-space transformations are composed with a first order domain-space (here, world-line) translation.

\subsection{Quadratic Holoraumy Group and Characteristics}
Looking back at\eq{e:SuSyD}, we see that the extension of $\Spin(N)$ by the $\ZZ_2$-group of sign-changes $\rD_I\to-\rD_I$ forms the Lie group $\Pin(N)$---the maximal group of outer automorphisms, $\Aut(\SSp^{1|N})$, of the supersymmetry algebra\eq{e:SuSyD}. This then provides the geometric significance of the quadratic holoraumy tensors\eq{e:QHT} as generating the $\Aut(\SSp^{1|N})$-action on the component fields of the supermultiplet. Indeed, the higher even-order holoraumy tensors, such as in\eq{SomethingRandom3} and higher, all appear in the power expansion of the formal exponential group-elements (Eqs.\eq{e:DDH} imply that $\sB_{IJ}$ and $\sF_{IJ}$ are antihermitian)
\begin{equation}
 \exp\{\,\inv2\L^{[IJ]}\sB_{IJ}\,\}\in \sH_\sB,
  \quad\text{and}\quad
 \exp\{\,\inv2\hat\L^{[IJ]}\sF_{IJ}\,\}\in \sH_\sF,
 \label{e:HoloG}
\end{equation}
owing to the relations\eq{e:4>2}.
 Using again only\eq{e:Cl0N}, powers of $\IGa_{IJ}$ can always be expressed as their totally antisymmetrized products plus lower-order products; decomposing into blocks, the same follows for $\sB_{IJ}$ and of $\sF_{IJ}$.

\paragraph{Algebraic Invariants:}
On the other hand, the index structure in $(\sB_{IJ})_i{}^k$ and $(\sF_{IJ})_\hj{}^\hl$ reminds of the gauge algebra-valued field strengths in Yang-Mills theory. Indeed, we have shown above that $(\sB_{IJ})_i{}^k$ and $(\sF_{IJ})_\hj{}^\hl$ take values in the Lie (sub)algebra of $\Spin(N)$, given respectively in the matrix representations acting on the bosons $\F_i$ and the fermions $\J_\hj$.
 It is then reasonable to also consider evaluating and comparing various characteristics of these holoraumy tensors, obtained using the methods of Ref.\cite{rHMP-Inv}. Here, we list such monomial invariants that are up to quartic in $(\sB_{IJ})_i{}^j$:
\begin{subequations}
 \label{e:234}
\begin{alignat}9
  (\sB_I{}^J)_i{}^j\,(\sB_J{}^I)_j{}^i,&\quad
  (\sB_I{}^J)_i{}^j\,(\sB_J{}^K)_j{}^k\,(\sB_K{}^I)_k{}^i,\\[1mm]
  (\sB_I{}^J)_i{}^j\,(\sB_J{}^K)_j{}^k\,(\sB_K{}^L)_k{}^\ell\,(\sB_L{}^I)_\ell{}^i,&\quad
  (\sB_I{}^J)_i{}^j\,(\sB_J{}^K)_j{}^k\,(\sB_K{}^L)_\ell{}^i\,(\sB_L{}^I)_k{}^\ell,\\[1mm]
  (\sB_I{}^J)_i{}^j\,(\sB_J{}^K)_j{}^i\,(\sB_K{}^L)_k{}^\ell\,(\sB_L{}^I)_\ell{}^k,&\quad
  (\sB_I{}^J)_i{}^j\,(\sB_J{}^K)_k{}^\ell\,(\sB_K{}^L)_j{}^i\,(\sB_L{}^I)_\ell{}^k,\\[1mm]
  (\sB_I{}^J)_i{}^j\,(\sB_J{}^I)_j{}^i\,(\sB_K{}^L)_k{}^\ell\,(\sB_L{}^K)_\ell{}^k,&\quad
  (\sB_I{}^J)_i{}^j\,(\sB_J{}^I)_k{}^\ell\,(\sB_K{}^L)_j{}^i\,(\sB_L{}^K)_\ell{}^k,\\[1mm]
  (\sB_I{}^J)_i{}^j\,(\sB_J{}^I)_j{}^k\,(\sB_K{}^L)_k{}^\ell\,(\sB_L{}^K)_\ell{}^i,&\quad
  (\sB_I{}^J)_i{}^j\,(\sB_J{}^I)_k{}^\ell\,(\sB_K{}^L)_\ell{}^i\,(\sB_L{}^K)_j{}^k.
\end{alignat}
\end{subequations}
For $N=8$ even, there also exist ``volume'' invariants\eq{e:charB}:
\begin{subequations}
\begin{gather}
   \ve^{I_1I_2I_3I_4I_5I_6I_7I_8}\,
  (\sB_{I_1I_2})_{i_1}{}^{i_2}(\sB_{I_3I_4})_{i_2}{}^{i_3}
  (\sB_{I_5I_6})_{i_3}{}^{i_4}(\sB_{I_7I_8})_{i_4}{}^{i_1}, \label{e:v8}\\
   \ve^{I_1I_2I_3I_4I_5I_6I_7I_8}\,
  (\sB_{I_1I_2})_{i_1}{}^{i_2}(\sB_{I_3I_4})_{i_2}{}^{i_1}
  (\sB_{I_5I_6})_{i_3}{}^{i_4}(\sB_{I_7I_8})_{i_4}{}^{i_3},\\
   \ve^{I_1I_2I_3I_4I_5I_6I_7I_8}\,
  (\sB_{I_1I_2})_{i_1}{}^{i_3}(\sB_{I_3I_4})_{i_2}{}^{i_1}
  (\sB_{I_5I_6})_{i_3}{}^{i_4}(\sB_{I_7I_8})_{i_4}{}^{i_2},
\end{gather}
\end{subequations}
and so on. Besides the analogous invariants constructed from $(\sF_{IJ})_a{}^b$, there are also $\sB$-$\sF$ mixed invariants such as
\begin{subequations}
\begin{alignat}9
 (\sB_I{}^J)_i{}^j\,(\sB_J{}^K)_j{}^i\,(\sF_K{}^L)_a{}^b\,(\sF_L{}^I)_b{}^a,&\quad
 (\sB_I{}^J)_i{}^j\,(\sF_J{}^K)_a{}^b\,(\sB_K{}^L)_j{}^i\,(\sF_L{}^I)_b{}^a,\\
 (\sB_I{}^J)_i{}^j\,(\sB_J{}^I)_j{}^i\,(\sF_K{}^L)_a{}^b\,(\sF_L{}^K)_b{}^a,&\quad
 (\sB_I{}^J)_i{}^j\,(\sF_J{}^I)_a{}^b\,(\sB_K{}^L)_j{}^i\,(\sF_L{}^K)_b{}^a,
\end{alignat}
as well as
\begin{equation}
  \ve^{I_1I_2I_3I_4I_5I_6I_7I_8}\,
  (\sB_{I_1I_2})_i{}^j(\sB_{I_3I_4})_i{}^j\,
   (\sF_{I_5I_6})_a{}^b(\sF_{I_7I_8})_b{}^a,
 \label{e:vMix}
\end{equation}
\end{subequations}
and so on. Using these, one can construct ``polynomial holoraumy invariants,'' somewhat akin to the polynomial curvature invariants that provide useful information about the geometry of manifolds.

\paragraph{Geometric Invariants:}
The first of the ``volume'' invariants\eq{e:v8} in fact has a very suggestive geometric meaning as it may also be calculated as the characteristic fermionic integral,
\begin{equation}
  \int\rd^N\q~\det[\,\sB_{IJ}\q^I\q^J-\l\Ione\,]\big| \propto
   \ve^{I_1\cdots I_N}\Tr[\,\sB_{I_1I_2}\cdots\sB_{I_{N-1}I_N}\,].
 \label{e:charB}
\end{equation}
In fact, the ``characteristic polynomial'' superfield $\det[\,\sB_{IJ}\q^I\q^J-\l\Ione\,]$ may also have non-zero fermionic integrals over certain subsets of the $\q$'s, as is the case for chiral representations\ft{In this sense, the chiral superspace is akin to a homologically nontrivial sub-superspace inside the full standard superspace.}.
 Clearly, analogous characteristic quantities can just as well be constructed also from $\sF_{IJ}$, from the coefficients of the characteristic polynomials
\begin{equation}
  \det[\,\sB_{IJ}\q^I\q^J-\l\Ione\,]
   \quad\text{and}\quad
  \det[\,\sF_{IJ}\q^I\q^J-\m\Ione\,].
\end{equation}
It is curious to see the fermionic coordinates $\q^I$ in superspace here play the role usually reserved for the differentials of ordinary (commuting) local coordinates on a manifold in differential geometry.

\paragraph{Two Remarks:}
 First, although it is well known that the Riemann tensor in 3+1~dimensional spacetime has twenty independent degrees of freedom, it is not known how to construct a complete basis of twenty algebraically independent polynomial curvature invariants\cite{rHE}; the situation in higher dimensions is clearly only more complex. We therefore do not expect to be able to reduce the copious list\eqs{e:234}{e:charB} by any a priori methods. Rather, these invariants should be evaluated---presumably by computer-aided methods---for as many supermultiplets as possible, so as to determine which of their combinations, if any, provide unambiguous distinction between world-line dimensional reductions of supermultiplets.
 Second, it is known that polynomial curvature invariants are not sufficient to differentiate between physically distinct spacetimes even in 3+1~dimensions\cite{rHE}, and we do not expect polynomial holoraumy invariants built from monomials such as\eqs{e:234}{e:charB} to differentiate between the dimensional reductions of all distinct higher-dimensional supermultiplets.

Nevertheless, it would seem prudent to explore this resolving capability of the polynomial holoraumy invariants built from monomials such as\eqs{e:234}{e:charB}, and we hope to return to this task in a later, computer-aided effort.

\paragraph{Quadratic Holoraumy Recursions:}
The higher holoraumy tensors that occur in\eqs{SomethingRandom2}{SomethingRandom3} and further may be generated from the quadratic holoraumy tensors and the original $\IL$-matrices\eq{e:Valise}:
\begin{subequations}
 \label{e:RecH}
\begin{alignat}9
 \IL^{}_{[I}\iL_J\IL^{}_{K]}
 &~=~\sB_{[IJ}\IL_{K]}& &~=~\IL_{[I}\sF_{JK]},
 \label{e:Rec1}\\
 \iL_{[I}\IL^{}_J\iL_{K]}
 &~=~\sF^{}_{[IJ}\iL_{K]}& &~=~\iL_{[I}\sB^{}_{JK]}.
 \label{e:Rec2}
\end{alignat}
\end{subequations}
In turn, these recursive identities also provide relationships between the bosonic and the fermionic quadratic holoraumy tensors, $\sB_{IJ}$ and $\sF_{IJ}$---represented by the second equations\eq{e:RecH}.

In turn, another set of relationships between the $\sB_{IJ}$ and $\sF_{IJ}$ matrices emerges from the very definition of the matrix $\sB_{IJ}$,
\begin{alignat}9
 (\sB_{IJ})_i{}^k
 &\Defl (\IL^{}_{[I}\iL_{J]})_i{}^k
   \Defl \inv2(\IL^{}_I\iL_J)_i{}^k
         -\inv2(\IL^{}_J\iL_I)_i{}^k,
\intertext{upon multiplying the first term from the right and the second term from the left by
 $\Ione=\IL^{}_{\2I}\iL_{\2I}$ (for each $\2I$, $\iL_{\2I}$ is the matrix-inverse of $\IL_{\2I}$, whence there is no summation on $\2I$):}
 &= \inv2(\IL^{}_{\2I}\iL_J\IL^{}_{\2I}\iL_{\2I})_i{}^k
        -\inv2(\IL^{}_{\2I}\iL_{\2I}\IL^{}_J\iL_{\2I})_i{}^k
  =-(\IL^{}_{\2I})_i{}^\hj~
     \inv2(\iL_{\2I}\IL^{}_J-\iL_J\IL^{}_{\2I})_\hj{}^\hl~
          (\iL_{\2I})_\hl{}^k,\nn
\end{alignat}
from which we have:
\begin{equation}
  (\sB_{IJ})_i{}^k
 =-(\IL^{}_{\2I}\,\sF_{\2IJ}\,\iL_{\2I})_i{}^k.
\end{equation}
The analogous computations using $\Ione=\IL^{}_{\2J}\iL_{\2J}$ (no summation on $\2J$), as well as starting from $\sF_{IJ}=\iL_{[I}\IL^{}_{J]}$ then produces the $4\,{\times}\,\binom{N}2$ matrix relations:
\begin{subequations}
\begin{alignat}9
 (\sB_{IJ})_i{}^k
 &=-(\IL^{}_{\2I}\,\sF_{\2IJ}\,\iL_{\2I})_i{}^k,&\qquad
 (\sB_{IJ})_i{}^k
 &=-(\IL^{}_{\2J}\,\sF_{I\2J}\,\iL_{\2J})_i{}^k,\\
 (\sF_{IJ})_\hj{}^\hl
 &=-(\iL_{\2I}\,\sB_{\2IJ}\,\IL_{\2I})_\hj{}^\hl,&\qquad
 (\sF_{IJ})_i{}^k
 &=-(\iL_{\2J}\,\sB_{I\2J}\,\IL_{\2J})_\hj{}^\hl,
\end{alignat}
\end{subequations}
for each of the $\binom{N}2$ choices of $\2I,\2J=1,2,\cdots,N$, with no summation over either $\2I$ or $\2J$.

Finally, there exists also a sum-rule obtained by iterative use of the relations\eq{e:Cl0N}:
\begin{equation}
  \IGa_K\,\IGa_{IJ}\,\IGa_K
   = \inv2 \big(\IGa_K\,\IGa_I\IGa_J\,\IGa_K
     - \IGa_K\,\IGa_J\IGa_I\,\IGa_K\big)
   =(N{-}4)\IGa_{IJ}.
\end{equation}
Reading off the blocks of these matrices, this implies:
\begin{equation}
  \iL_K\,\sB^{_{}}_{IJ}\,\IL^{_{}}_K
  =(N{-}4)\sF_{IJ}
 \qquad\text{and}\qquad
  \IL^{_{}}_K\,\sF^{_{}}_{IJ}\,\iL_K
  =(N{-}4)\sB_{IJ},
\end{equation}
which provide additional useful relations between $\sF_{IJ}$ and $\sB_{IJ}$ for $N\neq4$, and a constraining sum-rule for $N=4$.

\subsection{The $N=4$ Valises}
\label{s:N4valises}
Motivated by the most familiar and most often employed framework of 3+1-dimensional simple (${\cal N}\,{=}\,1$) supersymmetry, we now focus on the case of its 0-brane dimensional reduction, the world-line $(N\,{=}\,4)$-supersymmetry. Furthermore, in view of the main theorem of Ref.\cite{rDHIL13} we focus on the minimal supermultiplets which have 4+4 components.
 The Dirac algebra\ft{Most importantly, this is {\em\/not\/} the familiar Dirac algebra used 3+1-dimensional spacetime physics for which we will use lower-case $\g$-matrices, but refers to the similar algebra of matrices defined in\eq{e:Cl0N}.}\eq{e:Dirac} then reduces to the sixteen matrices
\begin{equation}
  \Ione,\quad
  \IGa_I,\quad
  \IGa_{IJ},\quad
  \IGa_{IJK}\Defr\ve_{IJKL}\ha\IGa{}^L,\quad
  \IGa_{IJKL}\Defr\ve_{IJKL}\ha\IGa.
 \label{e:Dirac16}
\end{equation}
The minimal $N=4$ supermultiplets all have four bosonic and four fermionic components, so that the list\eq{e:Dirac} involves an $8\,{\times}\,8$-dimensional matrix representation of the $\IGa$-matrices\eq{e:Cl0N}---and in the block off-diagonal form\eq{e:Valise}. That is, the list\eq{e:Dirac16} is regenerated entirely from the ``half-sized,'' $4\,{\times}\,4$-dimensional, invertible $\IL$-matrices defined in\eq{e:Valise}. The rank-2 matrices $\IGa_{IJ}$ generate the $\spin(4)=\spin(3)_-\,{\oplus}\,\spin(3)_+$ algebra of the connected component of the group of automorphisms, $\Aut(\SSp^{1|4})=\Pin(4)$, of the world-line supersymmetry algebra\eq{e:SuSyD}.
Notice that the notation in\eq{e:Dirac16}---as well as throughout Section~\ref{s:HoloR}---is chosen to explicitly ``forget'' all (the?) structure inherited from the action of the higher-dimensional Lorentz symmetry; we will return to this below.

\paragraph{The Reference Algebra:}
Explicit calculation shows that the fifteen real, non-identity matrices\eq{e:Dirac16} in fact close an irreducible Lie algebra, which then must be $\spin(3,3)\approx\mathfrak{sl}(4,\IR)$\cite{rFRH,rWyb}.
 Together with $\Ione$, this provides a complete basis of real, $4\,{\times}\,4$, invertible matrices, for which we use the basis from Ref.\cite{rUMD09-1} and which are displayed in Table~\ref{t:16} in Appendix~\ref{a:M} for convenience.

The list of the fifteen non-identity $8\,{\times}\,8$-dimensional matrices\eq{e:Dirac16} is therefore constructed from the $4\,{\times}\,4$-dimensional matrix generators of $\spin(3,3)$ and their inverses, and so explicitly provide a real 8-dimensional representation\ft{Since $\spin(3,3)\approx\su(4)$, this real 8-dimensional representation is identified with the complex ${\bf4}$ of $\su(4)$.} of $\spin(3,3)$. These matrices may alternatively be regarded as spanning either of the algebras
\begin{equation}
 \Cl(1,3)\approx\Ione\,{\oplus}\,\spin(3,3),\qquad
  \spin(3,3)\approx\mathfrak{sl}(4,\IR)
             \approx\su(4).
 \label{e:Cl(1,3)}
\end{equation}
We are thus led to inquire how are the holoraumy tensors\eq{e:QHT}---and the two copies of the $\spin(N\,{=}\,4)=\spin(3)_-\,{\oplus}\spin(3)_+$ Lie algebra\eq{e:BFON} that they generate---embedded, respectively, in this $\spin(3,3)$ reference algebra\eq{e:Cl(1,3)}. That is to say, given two 4+4-component off-shell supermultiplets of $N\,{=}\,4$-extended world-line supersymmetry, we can construct the list of sixteen $4\,{\times}\,4$-dimensional matrices for both, and compare these two lists to discern the transformation required to relate the two supermultiplets.

The ``location'' of the $\IGa_{IJ}=(\sB_{IJ}\,{\oplus}\,\sF_{IJ})$ holoraumy tensors of one supermultiplet as compared to the like tensors of another supermultiplet---within any particular fixed basis (such as Table~\ref{t:16}) of $\Cl(1,3)\approx\Ione\,{\oplus}\,\spin(3,3)$---then provides valuable relative information about the two supermultiplets. This ``location'' can be made more precise by means of the diagram of maps:
\begin{equation}
 \begin{array}{rcr@{\,}c@{\,}l}
 && \sH_\sB\qquad\qquad\hbox{~} && \qquad\qquad~ \sH_\sF\\*[-1mm]
 \Aut(\SSp^{1|4})=\Pin(4) &\too{\text{(\ref{e:HBF})}}
  &\big(\Spin(3)_-\oplus\Spin(3)_+\big)_B &\times
  &\big(\Spin(3)_-\oplus\Spin(3)_+\big)_F\\*[1mm]
 && \h_{\sss B}\!\!\searrow\qquad\quad\hbox{~} && \qquad\quad\swarrow\!\h_{\sss F} \\*[1mm]
  &\supset&\Spin(3)_\rot&\times&\Spin(3)_\exR\\*[-1mm]
 \smash{\9{\text{Table~\ref{t:16}, Appendix~\ref{a:M}}}
          {\text{(reference)}~\Spin(3,3)}\left\{\rule{0pt}{4ex}\right.}
 &&\cap\qquad~~\hbox{~}&&\quad\cup\\*[-1mm]
  &\supset&\Spin(1,3)_{\text{Lorentz}}&\times&\rOp{U}(1)_R\\
 \end{array}
 \label{e:map}
\end{equation}
where $\h_{\sss B}$ and $\h_{\sss F}$ map the holoraumy groups $\sH_\sB$ and $\sH_\sF$, respectively, to $\Spin(3)_\rot\,{\times}\,\Spin(3)_\exR\subset\Spin(3,3)$ and so specify the ``location'' of $\sB_{IJ}$ and $\sF_{IJ}$ respectively in the reference algebra $\spin(3,3)$.
 The bottom part of the mapping diagram\eq{e:map} may be visualized by realizing the hermitian generators of $\Spin(3,3)$ as $6\,{\times}\,6$ antisymmetric real matrices:
\begin{equation}
 \vC{\begin{picture}(50,26)
   \put(0,2){\includegraphics[width=30mm]{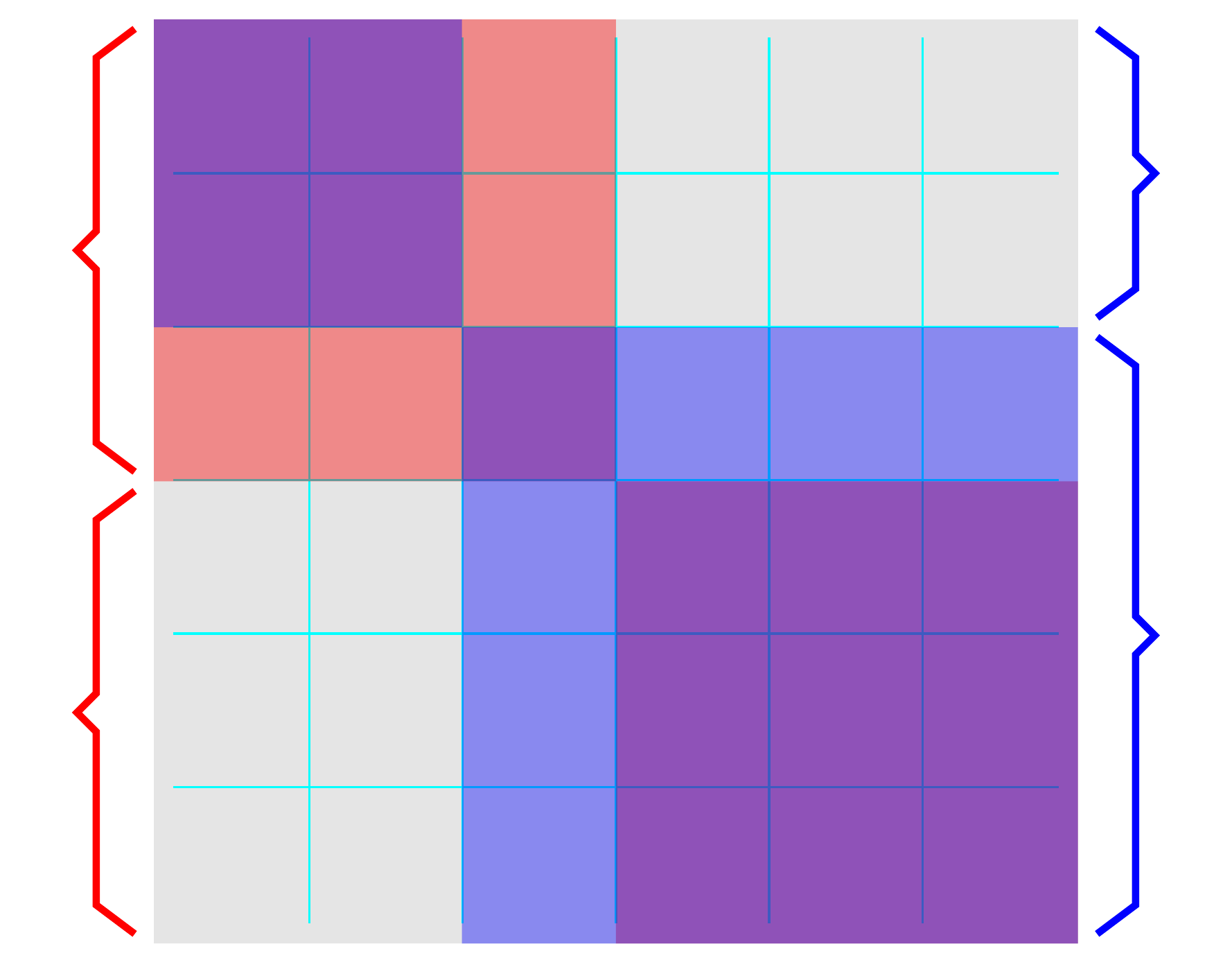}}
     \put(0,20){\Rx{\C1{$\spin(3)_\exR=\spin(3,0)$}}}
     \put(0,15){\Rx{\C1{$\{\g^0,\,\g^{123},\,\g^{0123}\}$\quad}}}
     \put(0,8){\Rx{\C1{$\spin(3)_\rot=\spin(0,3)$}}}
     \put(0,3){\Rx{\C1{$\{\g^{12},\,\g^{23},\,\g^{31}\}$\quad}}}
     \put(30,20){\Lx{\C3{$\spin(2,0)\to\rOp{U}(1)_R:~\{\g^5=i\g^{0123}\}$}}}
     \put(30,11){\Lx{\C3{$\spin(1,3)\to\Spin(1,3)_{\text{Lorentz}}$}}}
     \put(30,6){\Lx{\C3{$\{\g^{\m\n},\,\m,\n=0,1,2,3\}$}}}
     \put(5,5){\rotatebox{45}{\large\boldmath$\Spin(3,3)$}}
 \end{picture}}
\end{equation}
The corresponding $\Spin(3)_\rot$ group is both the maximal compact subgroup of the 3+1-dimensional Lorentz group $\Spin(1,3)=\SL(2,\IC)$ and also one of the two factors in the maximal compact connected subgroup of $\Aut(\SSp^{1|4})=\Pin(4)$. The complementary $\Spin(3)_\exR$ factor in turn contains the well-known 3+1-dimensional $U(1)$ $R$-symmetry generated by the $\g^5$-matrix.
 To disambiguate between the several $\spin(3)$ algebras, we summarize:
\begin{subequations}
 \label{e:3333}
\begin{alignat}9
 \spin(3)_-&\oplus\spin(3)_+&&\subset\spin(4)\to\Pin(4)=\Aut(\SSp^{1|4}),
 &&\quad\text{automorphisms of\eq{e:SuSyD}};\\
 \spin(3)_\rot&\oplus\spin(3)_\exR&&=\spin(0,3)\oplus\spin(3,0)\subset\spin(3,3),
 &&\quad\text{refers to Table~\ref{t:16}}.
\end{alignat}
\end{subequations}

\paragraph{World-Line Equivalence:}
The $\IL_I$-matrices\eq{e:Valise} are specified with respect to a chosen basis of component (super)fields, and are therefore subject to a basis change of the bosonic and fermionic component fields, $\cal X$ and $\cal Y$ respectively:
\begin{subequations}
 \label{e:XY}
\begin{align}
   \cX\,{\oplus}\,\cY:&~~\IL_I \to \cX\,\IL_I\,\cY^{-1},\quad\text{so that}
 \label{e:XLY}\\
   \cX\,{\oplus}\,\cY:&~~
    \sB_{IJ} \to \cX\,\sB_{IJ}\,\cX^{-1}~~\text{and}~~
     \sF_{IJ} \to \cY\,\sF_{IJ}\,\cY^{-1}~~&\text{for}~~I,J&=1,2,3,4.
\end{align}
\end{subequations}
Clearly, the quadratic holoraumy tensors $\sB_{IJ}$ and $\sF_{IJ}$ transform separately by bosonic and fermionic component (super)field redefinitions $\cal X$ and $\cal Y$, respectively.

Now, let $\sB^{\sss(A)}_{IJ},\sF^{\sss(A)}_{IJ}$ denote the quadratic holoraumy tensors of the $A^\text{th}$ supermultiplet. If the two supermultiplets are equivalent, there must exist a component (super)field basis change that relates their $\IL_I$-matrices:
\begin{subequations}
 \label{e:XY=}
\begin{alignat}9
   \cX\,{\oplus}\,\cY:&~~\IL_I^{\sss(2)} = \cX\,\IL_I^{\sss(1)}\,\cY^{-1},~~
    \text{for each}~~I=1,2,3,4~~\text{so that} \label{e:XY=2}\\
   \cX\,{\oplus}\,\cY:&~~
    \sB_{IJ}^{\sss(2)} = \cX\,\sB_{IJ}^{\sss(1)}\,\cX^{-1}~~\text{and}~~
     \sF_{IJ}^{\sss(2)} = \cY\,\sF_{IJ}^{\sss(1)}\,\cY^{-1}~~\text{for each}~~I,J=1,2,3,4.
     \label{e:XY=3}
\end{alignat}
\end{subequations}
As standard in field theory, $\cX,\cY$ and their inverses must be local field redefinitions. Such local bijective transformations\eq{e:XY=} then define {\em\/world-line supermultiplet equivalences\/}.

Since the maximal compact subgroup of $\Spin(3,3)$ is $\Spin(3,0)\,{\times}\,\Spin(0,3)\approx\Spin(4)$, it is tempting to identify this abstract subgroup with the connected part of the $\Aut(\SSp^{1|4})=\Pin(4)$ group generated by the $\IGa_{IJ}$ from among\eq{e:Dirac}, which combines in a block-diagonal form the two copies of $\Spin(4)$ generated by $\sB_{IJ},\sF_{IJ}$ in\eq{e:BFON}.
 In turn, it is a standard result in Lie group theory that every regular subgroup $H\subset G$ of a simple Lie group $G$ in fact has a continuum of distinct embeddings in $G$, but that they are all equivalent by $G$-conjugation\cite{rWyb,rHall,rPR-GTh}.
 This makes it tempting to conclude that {\em\/every\/} two possible holoraumy $\Spin(4)$ subgroups\eq{e:SON} of $\Spin(3,3)$, generated by\eq{e:Cl(1,3)} are isomorphic.
 The same would then seem to follow also for\eq{e:BFON}, \ie, that the holoraumy groups computed for any two 4+4-component supermultiplets are isomorphic to each other by way of\eq{e:XY=}, and so cannot distinguish between inequivalent supermultiplets.

Explicit calculations in Section~\ref{s:3LPs} however show that this does not hold, and that concrete results for $\IGa_{IJ}=\sB_{[IJ]}\,{\oplus}\,\sF_{IJ}$ can---and indeed do distinguish between several world-line supermultiplets that were dimensionally reduced from higher-dimensional spacetime and for which we have carried out the computations explicitly. Reverse-engineering this dimensional reduction, we are then able to exhibit some necessary conditions (\ie, obstructions, conversely) for dimensionally extending a given world-line supermultiplet into a desired 3+1-dimensional supermultiplet. Both in concept and in practice then, the present results on holoraumy expand on the 1\,$\to$\,2-dimensional extension results of Ref.\cite{rGH-obs} and complements the analysis of Refs.\cite{rFIL,rFL}.

\paragraph{Results:}
While we have so far not been able to provide a mathematically rigorous proof of the precise extent to which holoraumy tensors can be used to distinguish off-shell supermultiplets, this preliminary study and the explicit examples in Section~\ref{s:3LPs} do demonstrate its potential. For example, we will show below in detail that the holoraumy---defined and calculated solely from world-line physics---faithfully discerns between the inequivalent dash-chromotopologies\cite{r6-3,r6-3.1,r6-1.2} and also ``twisting''\cite{rTwSJG0,rGHR}, originally defined for supersymmetry in 1+1-dimensional spacetime. In addition, however, it also signals the existence of a complex structure, which is necessary in a chiral supermultiplet but obstructs the extension to a vector supermultiplet, for example. This also clears up a minor conundrum, provided by the fact that the so-called {\em\/twisted-chiral\/} superfield is manifestly complex, albeit being a dimensional reduction of the 3+1-dimensional vector superfield\cite{rTwSJG0,rGHR}, which is manifestly real.
 Also, a finer distinctive property of the holoraumy tensors correlates with the dimensional extension to different real 3+1-dimensional supermultiplets, such as the vector {\em\/vs\/}.\ the tensor supermultiplet (both in the Wess-Zumino gauge).

 The possible reasons that---contrary to the above-cited suspicion---certain of the possible holoraumy subgroups of $\Spin(4)\subset\Spin(3,3)$ are {\em\/not\/} isomorphic to each other, and so {\em\/can\/} distinguish between 4+4-component supermultiplets include:
\begin{enumerate}\itemsep=-3pt\vspace{-2mm}
 \item The detailed comparison of the holoraumy tensors and where they are within the $\spin(3,3)$ algebra\eq{e:ValEnv} crucially depends on the details of the real forms of the respective groups.
 \item The maximal compact subgroup of $\Spin(3,0)\,{\times}\,\Spin(0,3)\subset\Spin(3,3)$ is singled out by the signature of the metrics that these groups preserve.
 \item The main theorem of Ref.\cite{rDHIL13} permits restricting to monomial\ft{A matrix is {\em\/monomial\/} if it has a single nonzero entry in every row and every column.} $4\,{\times}\,4$ $\IL_I$-matrices\eq{e:Valise}; to preserve this, the component field redefinitions\eq{e:XY=2} must be significantly restricted.
\end{enumerate}
Needless to say, the definitive determination (and rigorous proof) of precisely how ``resolving'' the holoraumy concept is in reconstructing the higher-dimensional spacetime symmetry structures from the world-line dimensional reduction of supermultiplets hinges on a complete understanding of the precise relationship between the maximal group of outer automorphisms, $\Aut(\SSp^{1|N})\approx\Pin(0,N)$---which is what features prominently throughout this work, and the full Poincar\'e group in the intended higher-dimensional spacetime. This ultimate goal of uncovering the precise workings of an as yet conjectured ``supersymmetry holography'' is quite beyond our present scope. However, the subsequent skein of sample supermultiplets should provide a good starting point for such a more complete study.

\section{Several Examples}
\label{s:3LPs}
We now turn to exhibit a few well known 3+1-dimensional supermultiplets, dimensionally reduced to the world-line in Refs.\cite{rUMD09-1,rGRS13-RADIO}, for which we calculate and analyze the holoraumy tensors\eq{e:QHT}.

\subsection{The Chiral Supermultiplet Valise}
 \label{s:CS}
We begin with the familiar example of the chiral supermultiplet, of which the 0-brane dimensional reduction to the coordinate time world-line and in terms of real component superfields $(A,B,F,G|\j_a)$ is given by the\eq{e:Valise}-like super-differential system\cite{rUMD09-1}
\begin{subequations}
 \label{e:vCS1D}
\begin{alignat}9
  \rD_a A
 &=\j_a,&\qquad
  \rD_a B
 &= i (\g^5)_a{}^b \, \j_b,\mkern100mu \label{e:vCS1DAB} \\*
  \rD_a F
  &= (\g^0)_a{}^b \,\j_b,&\qquad
  \rD_a G
  &= i \,(\g^5\g^0){}_a{}^b \,\j_b, \label{e:vCS1DFG}\\*[1mm]
  \rD_a \j_b
  &=\makebox[0pt][l]{$
    i (\g^0)_{a \,b}\,\vdt A - (\g^5\g^0)_{a \,b}\,\vdt B
    -i C_{a\,b}\,\vdt F +(\g^5)_{ a \, b}\,\vdt G,$} \label{e:vCS1Dj}
\end{alignat}
where the component superfields $A,B,F,G$ all have the same engineering dimension, $\inv2$ lower than the fermions $\j_a$, and are all real, as are their lowest component fields obtained by standard projection\cite{r1001,rBK}.

 Using the matrices in Tables~\ref{t:16} and~\ref{t:160}, the system\eqs{e:vCS1DAB}{e:vCS1Dj} may also be tabulated as
\begin{equation}
  \begin{array}{@{} c|cccc|cccc @{}}
\text{\bsf vCS} &\bm A &\bm B &\bm F &\bm G &\bm{\j_1} &\bm{\j_2} &\bm{\j_3} &\bm{\j_4} \\ 
    \toprule
\C1{\bm{\rD_1}} & \j_1 &-\j_4 & \j_2 &-\j_3
                & i\vdt{A} & i\vdt{F} &-i\vdt{G} &-i\vdt{B} \\ 
\C2{\bm{\rD_2}} & \j_2 & \j_3 &-\j_1 &-\j_4
                &-i\vdt{F} & i\vdt{A} & i\vdt{B} &-i\vdt{G} \\ 
\C3{\bm{\rD_3}} & \j_3 &-\j_2 &-\j_4 & \j_1
                & i\vdt{G} &-i\vdt{B} & i\vdt{A} &-i\vdt{F} \\ 
\C4{\bm{\rD_4}} & \j_4 & \j_1 & \j_3 & \j_2
                & i\vdt{B} & i\vdt{G} & i\vdt{F} & i\vdt{A} \\ 
    \bottomrule
  \end{array}
 \label{e:vCS1Dt}
\end{equation}
\end{subequations}
This makes it clear that the system\eq{e:vCS1D} is {\em\/monomial\/}: Appearances in\eq{e:vCS1Dj} to the contrary, the result of applying each super-derivative to each single component superfield is again a single component superfield or its derivative, not a linear combination of such terms. Such supermultiplets are faithfully depicted by graphs called {\em\/Adinkras\/}\cite{r6-1}. In particular, the system\eq{e:vCS1D} is then depicted as
\begin{equation}
 \vC{\begin{picture}(60,28)
   \put(0,0){\includegraphics[width=60mm]{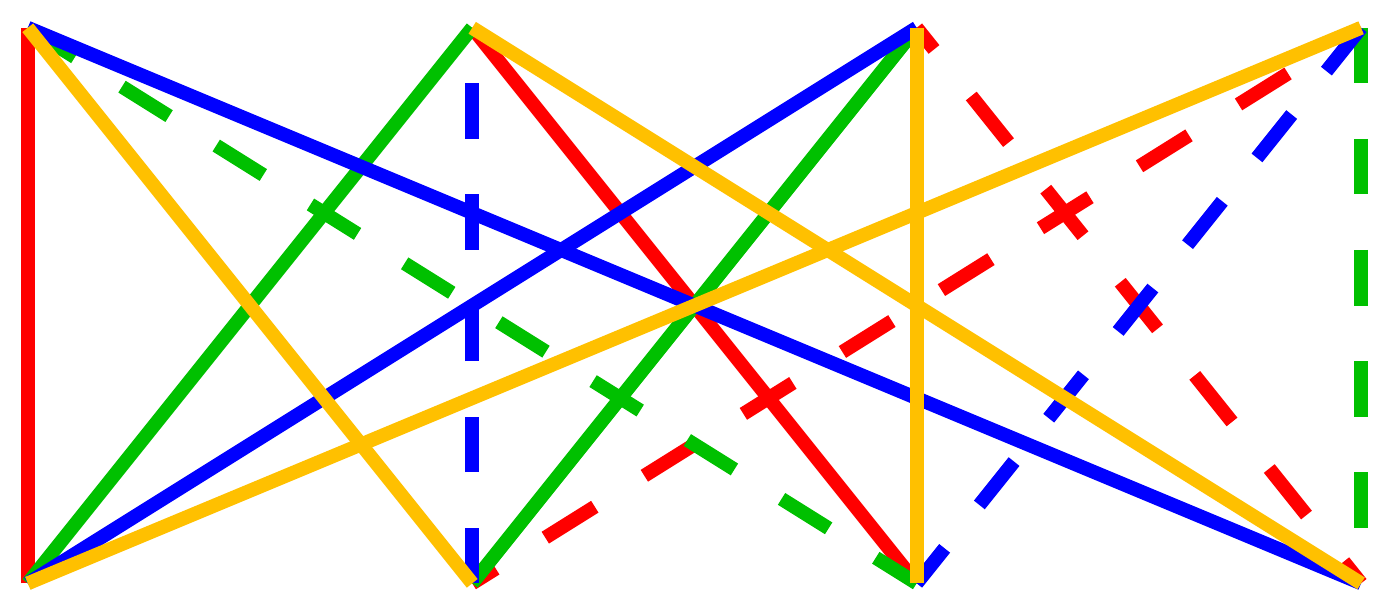}}
    \put(1,0){\cB{$A$}}
    \put(20.5,0){\cB{$B$}}
    \put(39.5,0){\cB{$F$}}
    \put(59,0){\cB{$G$}}
    \put(1,24){\bB{$\j_1$}}
    \put(20.5,24){\bB{$\j_2$}}
    \put(39.5,24){\bB{$\j_3$}}
    \put(59,24){\bB{$\j_4$}}
 \end{picture}}
 \label{e:vCS-A}
\end{equation}
where the nodes depict the component superfields, are drawn at the height proportional to their engineering dimension and where the edges depict the $\rD_a$-relations in\eq{e:vCS1D}.
The corresponding monomial $\IL$-matrices, as defined in\eq{e:valise}, are:
\begin{alignat}9
  \C1{\IL_1^{\!\!\sss\text{CS}}}&=\!\bM{1&0&0&0\\ 0&0&0&\!-1\\ 0&1&0&0\\ 0&0&\!-1&0\\},&~~
  \C2{\IL_2^{\!\!\sss\text{CS}}}&=\!\bM{0&1&0&0\\ 0&0&1&0\\ \!-1&0&0&0\\ 0&0&0&\!-1\\},&~~
  \C3{\IL_3^{\!\!\sss\text{CS}}}&=\!\bM{0&0&1&0\\ 0&\!-1&0&0\\ 0&0&0&\!-1\\ 1&0&0&0\\},&~~
  \C4{\IL_4^{\!\!\sss\text{CS}}}&=\!\bM{0&0&0&1\\ 1&0&0&0\\ 0&0&1&0\\ 0&1&0&0\\}.
 \label{e:vCS-L}
\end{alignat}

With the supermultiplet thus specified both in the tensorial representation\eq{e:vCS1D} and in terms of its $\IL$-matrices\eq{e:vCS-L}, we can compute straightforwardly the quadratic holoraumy tensors both ways, for a comparative illustration.

\paragraph{Tensorial Computation:}
Iterating\eq{e:vCS1D}, we obtain\ft{Unlike in the general definition\eq{e:FJ}, both the fermions and the super-derivatives are here counted by indices of the same type by virtue of the fact that their numbers equal in the chiral supermultiplet.}:
\begin{alignat}9
 \rD_{[a}\rD_{b]}\,\j_c
  &=\inv2[\,\rD_a\,,\,\rD_b\,]\,\j_c
   =-i(\sF_{ab}^{\sss\text{(CS)}})_c{}^d\,(\vdt\j_d), \label{e:FabCSj}
\intertext{where the quadratic fermionic holoraumy tensor $(\sF_{ab})_c{}^d$ may be written, utilizing a series of Fierz identities, as:}
 (\sF_{ab}^{\sss\text{(CS)}})^{}_c{}^d
  &=\inv2(\g^0\g_{mn})_{ab}\,(\g^{mn})_c{}^d,\qquad
  m,n,p=1,2,3. \label{e:CSFH3}
\end{alignat}
The fermionic holoraumy is thus generated by $\g^{12},\, \g^{23}$ and $\g^{31}$, the generators of the (spatial) rotation subgroup $\Spin(3)_\rot=\Spin(0,3)\subset\Spin(1,3)$ of the Lorentz group, and ``rotates'' the ``direction'' of $(\vdt\j)$ relative to the original ``direction'' of $\j$. As stated above for the general case, the holoraumy\eq{e:FabCSj} indeed composes $\t$-translations with homogeneous linear transformations in the $\j$-space---the latter of which in fact really are spatial rotations within the Lorentz group acting on the fermions $\j$. That is to say, the fermionic quadratic holoraumy tensor $\sF_{ab}^{\text{(vCS)}}$ is a $\Spin(3)_\rot$-valued 2-form in the 4-component space of fermions.

The bosonic holoraumy tensor is computed in the same way---again utilizing a series of Fierz identities, producing:
\begin{equation}
  (\sB_{ab}^{\sss\text{(CS)}})^{}_i{}^j
   =C_{ab}\,(\g^{12})_i{}^j +i(\g^5)_{ab}\,(\g^{23})_i{}^j
    +i(\g^5\g^0)_{ab}\,(\g^{13})_i{}^j, \label{e:CSBH3}
\end{equation}
To summarize, the mapping\eq{e:map} locates:
\begin{equation}
  \sB_{ab}^{\sss\text{(CS)}} \5\h\into \spin(3)_\rot,
   \qquad\text{and}\qquad
  \sF_{ab}^{\sss\text{(CS)}} \5\h\into \spin(3)_\rot,
 \label{e:CSrotrot}
\end{equation}
with the concrete results\eq{e:CSFH3} and\eq{e:CSBH3} specifying the details of these maps.

Having exhibited the first concrete example, we are in position to draw the Reader's attention to an important feature: The computations resulting in\eq{e:CSFH3} and\eq{e:CSBH3} strongly depend on a particular series of 3+1-dimensional Fierz identities for the $\g$-matrices, the structure of which in turn strongly depends on the dimension and signature of the higher-dimensional spacetime for which the used $\g$-matrices have been defined.

In the subsequent ``matrix computation'' below, we will recover the algebraic structure of these results from the purely world-line information\eqs{e:vCS-A}{e:vCS-L}. As mentioned in the comment after Eq.\eq{e:Dirac16}, this computational framework and notation were chosen to explicitly (try to) ``forget'' the structure of the action of the higher-dimensional Lorentz symmetry.
 The very fact that the algebraic structure of the results\eq{e:CSFH3} and\eq{e:CSBH3} {\em\/can\/} be recovered from the subsequent ``matrix computation'' implies that this purely world-line representation of the 0-brane dimensional reduced supermultiplet nevertheless does retain---holographically---the structure of the higher-dimensional Lorentz symmetry action within the supermultiplet from which\eqs{e:vCS-A}{e:vCS-L} was obtained. In the rest of the article, we will show that the same holography persists throughout all of the examples.

\paragraph{Matrix Computation:}
In turn, with the explicit $\IL$-matrices\eq{e:vCS-L} identified, it is a straightforward matter to compute:
\begin{subequations}
 \label{e:vCS-}
\begin{alignat}9
  \sB_{12}^{\sss\text{(CS)}}&=-\sB_{34}^{\sss\text{(CS)}}&&=-\g^{12},&\quad
  \sB_{23}^{\sss\text{(CS)}}&=-\sB_{14}^{\sss\text{(CS)}}&&=+\g^{13},&\quad
  \sB_{31}^{\sss\text{(CS)}}&=-\sB_{24}^{\sss\text{(CS)}}&&=-\g^{23}, \label{e:vCS-B} \\
  \sF_{12}^{\sss\text{(CS)}}&=+\sF_{34}^{\sss\text{(CS)}}&&=-\g^{13},&\quad
  \sF_{23}^{\sss\text{(CS)}}&=+\sF_{14}^{\sss\text{(CS)}}&&=+\g^{23},&\quad
  \sF_{31}^{\sss\text{(CS)}}&=+\sF_{24}^{\sss\text{(CS)}}&&=-\g^{12}, \label{e:vCS-F}
\end{alignat}
\end{subequations}
where we have identified the results in terms of the reference matrices as given in Table~\ref{t:16}.

The first of each of these triple equalities produce the basis-independent results
\begin{equation}
  \sB_{ab+}^{\sss\text{(CS)}}\Defl
  \sB_{ab}^{\sss\text{(CS)}}+\inv2\ve_{ab}{}^{cd}\,\sB_{cd}^{\sss\text{(CS)}}=0
   \qquad\text{and}\qquad
  \sF_{ab-}^{\sss\text{(CS)}}\Defl
  \sF_{ab}^{\sss\text{(CS)}}-\inv2\ve_{ab}{}^{cd}\,\sF_{cd}^{\sss\text{(CS)}}=0,
 \label{e:vCS-pr}
\end{equation}
which are a consequence of the fact that the chiral superfield is annihilated by a pair of complex combinations of the super-derivatives tantamount to the result\eq{e:DCS=0}, the product of which yields the annihilating quasi-projection operators\cite{r6-1.2,rH-TSS}
\begin{equation}
  \big[\,\rD_{[a}\rD_{b]} + \inv2\ve_{ab}{}^{cd}\,\rD_{[c}\rD_{d]}\,\big].
 \label{e:CSDD}
\end{equation}
In turn, annihilation by these operators is equivalent to the Adinkra\eq{e:vCS-A} exhibiting closed (\C1{red}-\C2{green}-\C3{blue}-\C4{orange}) 4-color cycles with the cycle parity $\textsl{CP}(\text{\ref{e:vCS-A}})=+1$; for a precise definition, see Ref.\cite{rH-TSS}. Finally, for all $N\,{=}\,4$ supermultiplets with $4+4$ components, this cycle parity equals the ``chromocharacter'' $\c_o$, as defined in Ref.\cite{rUMD09-1}.

By the general result\eq{e:BFON}, $\sB_{ab}^{\sss\text{(CS)}}$ and $\sF_{ab}^{\sss\text{(CS)}}$ generate rotations in the two $\IR^4$-like sectors of the field-space, $(A,B,F,G)$ and $(\j_1,\j_2,\j_3,\j_4)$, respectively. The relations\eq{e:vCS-pr} reduce these rotations
\begin{equation}
 \begin{array}{rcl}
  \spin(4)_{\sss B} &\too{\text{(\ref{e:vCS-pr})}}& \spin(3)_-\\
  \spin(4)_{\sss F} &\too{\text{(\ref{e:vCS-pr})}}& \spin(3)_+\\
 \end{array}\bigg\}\subset\spin(3)_-\oplus\spin(3)_+=\spin(4)
 \label{e:RedHCS}
\end{equation}
to complementary parts of the algebra of $\Pin(4)=\Aut(\SSp^{1|4})$.
 A similar reduction holds all minimal supermultiplets\cite{r6-1.2,rH-TSS} and will be referred as the minimality condition.

The results\eq{e:vCS-} further show that not only do both $\sB_{ab}^{\sss\text{(CS)}}$ and $\sF_{ab}^{\sss\text{(CS)}}$ generate only a $\Spin(3)\subset\Pin(4)$, but have the {\em\/same\/} $\h$-images within our reference $\spin(3,3)$. Eqs.\eqs{e:vCS-B}{e:vCS-F} clearly imply that
\begin{equation}
  \vp:~\h
   \{\sF^{\sss\text{(CS)}}_{12},\sF^{\sss\text{(CS)}}_{23},\sF^{\sss\text{(CS)}}_{31}\}
  = \h
     \{-\sB^{\sss\text{(CS)}}_{23},-\sB^{\sss\text{(CS)}}_{31},\sB^{\sss\text{(CS)}}_{12}\}
  ~\supset~\spin(3)_\rot,
    \label{e:vCS-B=F}
\end{equation}
which is an even ($\det\,{=}\,{+}1$) signed permutation of the generators, and does not depend on the concrete matrix realizations\eq{e:vCS-}. $\h$ denotes the mapping\eq{e:map} that identifies the elements of $\sB_{ab}$ and $\sF_{ab}$, acting on different spaces, with elements from the reference algebra, $\spin(3,3)$ and its maximal compact subalgebra $\spin(3)_\rot\oplus\spin(3)_\exR$.
 Summarizing the composition of the result\eq{e:CSrotrot} following\eq{e:RedHCS},
\begin{equation}
  \big(\h_{\sss B};\h_{\sss F}\big)^{\sss\text{(CS)}}
  :~~\bigg\{
   \begin{array}{r@{~\in~}ccl}
    \sB^{\sss\text{(CS)}}_{ab} & \spin(3)_- &\too{\Ione} & \spin(3)_\rot,\\[1mm]
    \sF^{\sss\text{(CS)}}_{ab} & \spin(3)_+ &\too{\vp}   & \spin(3)_\rot,\\
   \end{array}
 \label{e:CSmap}
\end{equation}
where $\vp=\big({{\SSS12},\,{\SSS23},\,{\SSS31}\atop\7{\SSS23},\,\7{\SSS31},\,{\SSS12}}\big)$ is the even ($\det=+1$) relative signed permutation\eq{e:vCS-B=F} relating the images of the fermionic and bosonic holonomy within the reference algebra $\spin(3,3)$.

\subsection{The Vector Supermultiplet Valise}
 \label{s:VS}
The 3+1-dimensional vector supermultiplet, given in the Wess-Zumino gauge and our Majorana (real component) framework and dimensionally reduced to the coordinate time world-line, is specified by the super-differential system\cite{rUMD09-1}:
\begin{subequations}
\label{e:vVS1D}
\begin{alignat}9
 \rD_a\,A_m
 &= (\g_m)_a{}^b\,\l_b,\qquad
 \rD_a\,d
  = i(\g^5\g^0)_a{}^b\,\l_b = -(\g^{123})_a{}^b\,\l_b,\\
 \rD_a\,\l_b
 &= -i(\g^0\g^n)_{ab}\,(\vdt\,A_n) + (\g^5)_{ab}\,(\vdt d),
\end{alignat}
where the $\vdt d$ is to be identified with the usual auxiliary field in the vector supermultiplet.
 In tabular format, using the matrices in Tables~\ref{t:16} and~\ref{t:160}, these produce:
\begin{equation}
  \begin{array}{@{} c|cccc|cccc @{}}
\text{\bsf vVS} &\bm{A_1} &\bm{A_2} &\bm{A_3} &\bm{d}
  &\bm{\l_1} &\bm{\l_2} &\bm{\l_3} &\bm{\l_4} \\ 
    \toprule
\bm{\C1{\rD_1}} & \l_2 &-\l_4 & \l_1 &-\l_3 & i\vdt A_3 & i\vdt A_1 &-i\vdt d &-i\vdt A_2 \\ 
\bm{\C2{\rD_2}} & \l_1 & \l_3 &-\l_2 &-\l_4 & i\vdt A_1 &-i\vdt A_3 & i\vdt A_2 &-i\vdt d \\ 
\bm{\C3{\rD_3}} & \l_4 & \l_2 & \l_3 & \l_1 & i\vdt d & i\vdt A_2 & i\vdt A_3 & i\vdt A_1 \\ 
\bm{\C4{\rD_4}} & \l_3 &-\l_1 &-\l_4 & \l_2 &-i\vdt A_2 & i\vdt d & i\vdt A_1 &-i\vdt A_3 \\ 
    \bottomrule
  \end{array}
 \label{e:vVS1Dt}
\end{equation}
\end{subequations}
and are depicted as:
\begin{equation}
 \vC{\begin{picture}(80,28)
   \put(0,0){\includegraphics[width=60mm]{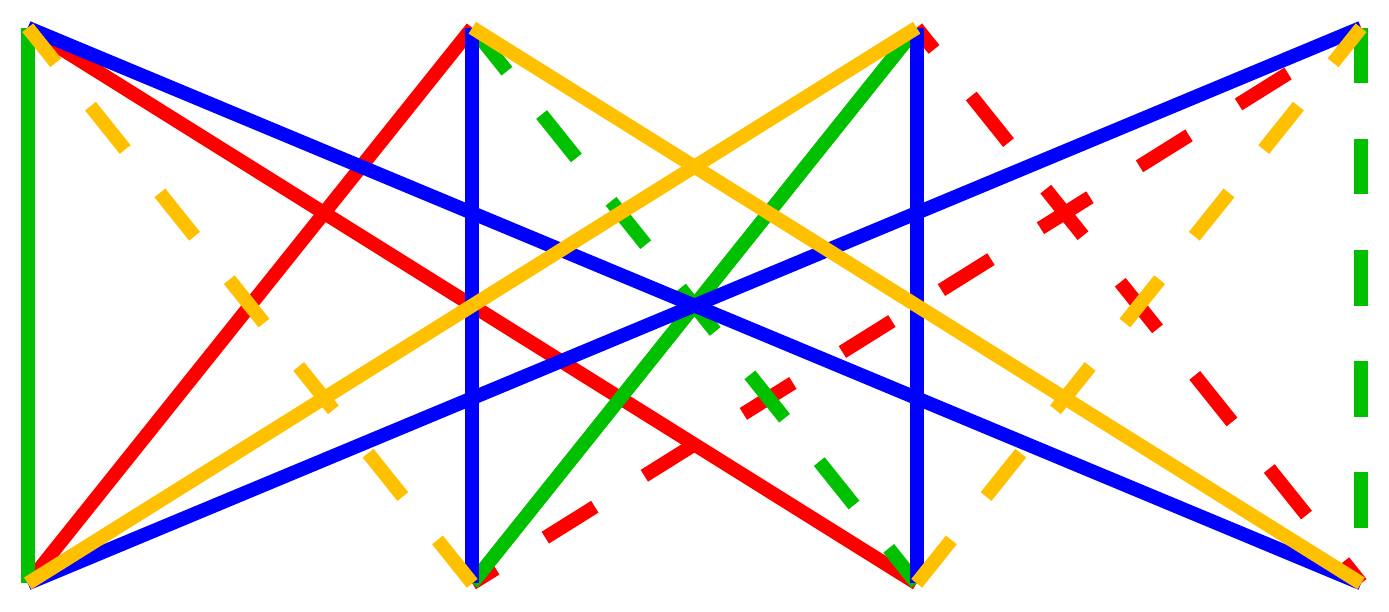}}
    \put(1,0){\cB{$A_1$}}
    \put(20.5,0){\cB{$A_2$}}
    \put(39.5,0){\cB{$A_3$}}
    \put(59,0){\cB{$d$}}
    \put(1,24){\bB{$\l_1$}}
    \put(20.5,24){\bB{$\l_2$}}
    \put(39.5,24){\bB{$\l_3$}}
    \put(59,24){\bB{$\l_4$}}
 \end{picture}}
 \label{e:vVS-A}
\end{equation}
and we read off the $\IL$-matrices:
\begin{alignat}9
  \C1{\IL_1^{\!\!\sss\text{VS}}}&=\!\bM{0&1&0&0\\ 0&0&0&\!\!-1\\ 1&0&0&0\\ 0&0&\!\!-1&0\\},&~~
  \C2{\IL_2^{\!\!\sss\text{VS}}}&=\!\bM{1&0&0&0\\ 0&0&1&0\\ 0&\!\!-1&0&0\\ 0&0&0&\!\!-1\\},&~~
  \C3{\IL_3^{\!\!\sss\text{VS}}}&=\!\bM{0&0&0&1\\ 0&1&0&0\\ 0&0&1&0\\ 1&0&0&0\\},&~~
  \C4{\IL_4^{\!\!\sss\text{VS}}}&=\!\bM{0&0&1&0\\ \!-1&0&0&0\\ 0&0&0&\!\!-1\\ 0&1&0&0\\}.
 \label{e:vVS-L}
\end{alignat}

With the supermultiplet now given both in the tensorial representation\eq{e:vVS1D} and in terms of its $\IL$-matrices\eq{e:vVS-L}, we can compute straightforwardly the quadratic holoraumy tensors both ways, for a comparative illustration.

\paragraph{Tensorial Computation:}
Iterating\eq{e:vVS1D}, we compute straightforwardly:
\begin{alignat}9
 \rD_{[a}\rD_{b]}\,\l_c
  &=\inv2[\,\rD_a\,,\,\rD_b\,]\,\l_c
   =-i(\sF_{ab}^{\sss\text{(VS)}})_c{}^d\,(\vdt\l_d), \label{e:FabVSj}
\intertext{where the quadratic fermionic holoraumy tensor $(\sF_{ab})_c{}^d$ may be written, utilizing again a series of Fierz identities, as:}
 (\sF_{ab}^{\sss\text{(VS)}})^{}_c{}^d
  &= C_{ab}\,(\g^0)_c{}^d
    +(\g^5)_{ab}\,(\g^5\g^0)_c{}^d
    +(\g^5\g^0)_{ab}\,(\g^5)_c{}^d. \label{e:VSFH3}
\end{alignat}
As in the case of the chiral supermultiplet, the fermionic quadratic holoraumy tensor $\sF_{ab}^{\sss\text{(VS)}}$ is again a $\Spin(3)$-valued 2-form in the 4-component space of fermions. However, this time the Lie group is generated by $\g^0,\g^{123}$ and $\g^{0123}$, was dubbed the ``extended $R$-symmetry'' in Ref.\cite{rGHS-HoloSuSy}, and we denote it $\Spin(3)_\exR$. The fermionic holoraumy (generated by $\g^0$, $\g^{123}=-i\g^5\g^0$ and $\g^{0123}=-i\g^5$), thus ``rotates'' the ``direction'' of $(\vdt\l)$ relative to the original ``direction'' of $\l$ by means of a $\Spin(3)_\exR$-action within the fermionic $\IR^4$-like field-space.

Similarly computed, the bosonic holoraumy tensor is:
\begin{equation}
 (\sB_{ab}^{\sss\text{(VS)}})^{}_i{}^j
  =-i(\g^5\g^1)_{ab}\,(\g^{23})_i{}^j
   +i(\g^5\g^2)_{ab}\,(\g^{12})_i{}^j
   +i(\g^5\g^3)_{ab}\,(\g^{13})_i{}^j. \label{e:VSBH3}
\end{equation}
To summarize, the mapping\eq{e:map} locates:
\begin{equation}
  \sB_{ab}^{\sss\text{(VS)}} \5\h\into \spin(3)_\rot,
   \qquad\text{and}\qquad
  \sF_{ab}^{\sss\text{(VS)}} \5\h\into \spin(3)_\exR,
\end{equation}
with the concrete results\eq{e:VSFH3} and\eq{e:VSBH3} again specifying the details of these maps. The algebraic structure of these results is shown below to be encoded just as well in the purely world-line description of the 0-brain dimensional reduction\eqs{e:vVS-A}{e:vVS-L} of this supermultiplet.

\paragraph{Matrix Computation:}
Given the explicit $\IL$-matrices\eq{e:vVS-L}, straightforward matrix algebra produces the quadratic holonomy matrices:
\begin{subequations}
 \label{e:vVS-BF}
\begin{alignat}9
  \sB_{12}^{\sss\text{(VS)}}&=+\sB_{34}^{\sss\text{(VS)}}&&=+\g^{12},&\quad
  \sB_{23}^{\sss\text{(VS)}}&=+\sB_{14}^{\sss\text{(VS)}}&&=+\g^{23},&\quad
  \sB_{31}^{\sss\text{(VS)}}&=+\sB_{24}^{\sss\text{(VS)}}&&=+\g^{13}, \label{e:vVS-B} \\
  \sF_{12}^{\sss\text{(VS)}}&=-\sF_{34}^{\sss\text{(VS)}}&&=-\g^0,&\quad
  \sF_{23}^{\sss\text{(VS)}}&=-\sF_{14}^{\sss\text{(VS)}}&&=+\g^{0123},&\quad
  \sF_{31}^{\sss\text{(VS)}}&=-\sF_{24}^{\sss\text{(VS)}}&&=-\g^{123}, \label{e:vVS-F}
\end{alignat}
\end{subequations}
where we have again identified the results in terms of the reference matrices as given in Table~\ref{t:16}.

The first of these equalities produce the basis-independent results
\begin{equation}
  \sB_{ab-}^{\sss\text{(VS)}}\Defl
  \sB_{ab}^{\sss\text{(VS)}}-\inv2\ve_{ab}{}^{cd}\,\sB_{cd}^{\sss\text{(VS)}}=0
   \qquad\text{and}\qquad
  \sF_{ab+}^{\sss\text{(VS)}}\Defl
  \sF_{ab}^{\sss\text{(VS)}}+\inv2\ve_{ab}{}^{cd}\,\sF_{cd}^{\sss\text{(VS)}}=0,
 \label{e:vVS-pr}
\end{equation}
which are a consequence of the fact that the vector superfield in the Wess-Zumino gauge is annihilated by quasi-projection operators\cite{r6-1.2,rH-TSS}
\begin{equation}
  \big[\rD_{[a}\rD_{b]} - \inv2\ve_{ab}{}^{cd}\,\rD_{[c}\rD_{d]}\big],
 \label{e:VSDD}
\end{equation}
which has the relative sign opposite to the chiral supermultiplet result\eq{e:CSDD}.
 Equivalently, the Adinkra\eq{e:TS-A} exhibits closed (\C1{red}-\C2{green}-\C3{blue}-\C4{orange}) 4-color cycles with $\textsl{CP}(\text{\ref{e:vVS-A}})=-1=\c_o$\cite{rH-TSS,rUMD09-1}. That\eq{e:VSDD} has the relative sign opposite to that one in\eq{e:CSDD} follows from the following facts:
 ({\small\bf1})~The vector superfield in the Wess-Zumino gauge is the complement within a real intact superfield $\IU$ of the gauged-away formal imaginary part of a chiral superfield.
 ({\small\bf2})~Cycle parity is additive: $\textsl{CP}(\IU)=0$,
 $\textsl{CP}(\bm\L)=+1=\textsl{CP}(\bm\L^\dag)$, and
 $\textsl{CP}\big(\IU/\Imm(\bm\L)\big)=0-1=-1$.

The general result\eq{e:BFON} of course again holds, so that $\sB_{ab}^{\sss\text{(VS)}}$ and $\sF_{ab}^{\sss\text{(VS)}}$ generate rotations in the $\IR^4$-like sectors of the field-space, respectively $(A_1,A_2,A_3,d)$ and $(\l_1,\l_2,\l_3,\l_4)$.
The minimality relations\eqs{e:vVS-pr}{e:VSDD} reduce these rotations
\begin{equation}
 \begin{array}{rcl}
  \spin(4)_{\sss B} &\too{\text{(\ref{e:vVS-pr})}}& \spin(3)_+\\
  \spin(4)_{\sss F} &\too{\text{(\ref{e:vVS-pr})}}& \spin(3)_-\\
 \end{array}\bigg\}\subset\spin(3)_-\oplus\spin(3)_+=\spin(4)
 \label{e:RedHVS}
\end{equation}
to complementary parts of the algebra of $\Pin(4)=\Aut(\SSp^{1|4})$, but in a way opposite of that in the chiral supermultiplet; see\eq{e:RedHCS}.

Also unlike the case of the chiral supermultiplet, the relations\eq{e:vVS-BF} do not reduce the direct sum of these rotations from a full $\spin(4)\subset\spin(3,3)$: the $\sB_{ab}^{\sss\text{(VS)}}$ and $\sF_{ab}^{\sss\text{(VS)}}$ tensors are each valued in a separate and mutually commuting $\Spin(3)$ subgroup of $\Spin(4)$. In particular, the mapping\eq{e:map} locates:
\begin{equation}
  \sB_{ab+}^{\sss\text{(VS)}} \5\h\into \spin(3)_\rot,
   \quad\text{while}\quad
  \sF_{ab-}^{\sss\text{(VS)}} \5\h\into \spin(3)_\exR,
 \label{e:VSrotexR}
\end{equation}
where $\spin(3)_\rot$ is the rotational subgroup of the Lorentz group $\Spin(1,3)$, and $\spin(3)_\exR$ is the ``extended $R$-symmetry'' algebra generated by $\g^0$, $\g^{123}$ and $\g^{0123}$. Referring again to the diagram\eq{e:map}, we summarize the results\eq{e:RedHVS} and\eq{e:VSrotexR} as
\begin{equation}
  \big(\h_{\sss B};\h_{\sss F}\big)^{\sss\text{(VS)}}
  :~~\bigg\{
   \begin{array}{r@{~\in~}ccl}
    \sB^{\sss\text{(VS)}}_{ab} & \spin(3)_+ &\too{\Ione} & \spin(3)_\rot,\\[1mm]
    \sF^{\sss\text{(VS)}}_{ab} & \spin(3)_- &\too{\Ione} & \spin(3)_\exR.\\
   \end{array}
 \label{e:VSmap}
\end{equation}
Since $\sB_{ab+}$ and $\sF_{ab-}$ map to mutually commuting factors of the reference algebra $\spin(3,3)$, there is no relative permutation---unlike the case of the chiral supermultiplet\eq{e:CSmap}.

\subsection{Complex Structure vs.\ Absence Thereof}
\label{s:cpxS}
This comparison between the chiral and the vector supermultiplet is best highlighted by recalling that the chiral supermultiplet is complex, while the vector supermultiplet is inherently real.

To this end, note that the components\eq{e:vCS1D} are also found in the superspace expansion of the ({\em\/complex\/}) chiral superfield $\bm\F$, which is defined to satisfy the super-differential constraints $\bar{D}_{\!\dt\a}\,\bm\F=0$. Since the Weyl spinor super-derivatives $\bar{D}_{\!\dt\a}$ may be identified with the Majorana super-derivative expressions $\inv2([\Ione{+}\g^5]\rD){}_a$\cite{r1001,rWB,rGHR,rBK}, the lowest components of the condition $\bar{D}_{\!\dt\a}\,\bm\F=0$ translate into
\begin{equation}
  \big(\inv2[\Ione +\g^5]\rD\big){}_a\,(A+iB) ~=~
  \inv2[\d_a{}^b +i(\g^{0123})_a{}^b]\rD_b\,(A+iB) ~=0.
 \label{e:DCS=0}
\end{equation}
Applying the conjugate super-derivatives $\inv2[\Ione{-}\g^5]\rD$ repeatedly on $(A{+}iB)$ then generates the supersymmetric completion of this complex pairing throughout the rest of the supermultiplet:
\begin{equation}
  \Big\{ \big(A{+}iB,\,F{-}iG\,\big|\,\j_1{-}i\j_4,\,\j_2{+}i\j_3\big)
   ~;~[\C1{\rD_1}-i\C4{\rD_4}],[\C2{\rD_2}+i\C3{\rD_3}] \,\Big\},
 \label{e:cpxvCS}
\end{equation}
which is therefore a complex supermultiplet with respect to the complex supersymmetry action generated by $\inv2([\Ione{-}\g^5]\rD)_a\simeq D_\a$ and $[\Ione{-}\g^5]\j\simeq\j_\a$.
 Since $\g^5\Defl i\g^{0123}$ is purely imaginary, it defines the supersymmetric complex structure on all fermions, as indicated in\eq{e:cpxvCS}. Also, the combination $\vdt(F{-}iG)$ may be identified with the usually auxiliary component of $\bm\F$.

Thus, the complex structure of the whole supermultiplet is completely determined by the initial choices $A{+}iB$ and $D_\a=\inv2([\Ione{-}\g^5]\rD)_a$: the super-differential system\eq{e:vCS1D} then determines the rest of\eq{e:cpxvCS}. In general, we define:
\begin{defn}
 A supermultiplet has a {\em\/supersymmetric complex structure\/} if
 the action of a fixed complex pairing of all super-derivatives (and supercharges) 
 consistently combines all component (super)fields into complex pairs
 throughout the supermultiplet.
\end{defn}
In any $\s$-model, the dynamical bosons provide local coordinates in the target manifold, while the dynamical fermions span the local tangent space.
 The simultaneous and coinciding (aligned) reduction $\spin(4)_B\,{\oplus}\,\spin(4)_F\too\h(\spin(3)\approx\su(2))\subset\spin(3,3)$ of these holoraumic rotations in both the fermionic (tangential) and the bosonic (coordinate) target-space of the chiral supermultiplet is consistent with the chiral supermultiplet admitting a supersymmetric complex structure\eq{e:cpxvCS}.

This mirrors the maximal holonomy group of a (real $2n$-dimensional) complex K\"ahler $n$-fold being $\SU(n)\subset\SO(2n)$, rather than the full $\SO(2n)$ of an orientable real $2n$-dimensional manifold: the existence of a complex structure reduces the holonomy group. Much the same, the existence of the supersymmetric complex structure\eq{e:cpxvCS} reduces the holoraumy group\eq{e:HoloG} from $\Spin(4)$ to the $\Spin(3)_\rot\approx\SU(2)_\rot$ subgroup of field-space ``rotations,'' composed with $\t$-translations.

The careful Reader will have noticed that it is the existence of a K\"ahler metric on a complex $n$-dimensional manifold that reduces the holonomy $\Spin(2n)\to\rOp{U}(n)$, and the Ricci-flatness of such a K\"ahler metric that further reduces the holonomy $\rOp{U}(n)\to\SU(n)$.
 The type and characteristics of the target-space metric is ultimately determined by the action of the $\s$-model built from this supermultiplet. The above considerations refer only to the supermultiplet itself, and to the local target space patch spanned by its components  and equipped with the canonical flat supersymmetric metric; see appendix~\ref{s:metric} for a proof and construction. As the flat metric is both K\"ahler and Ricci-flat, the holoraumy reduction
\begin{equation}
 \big(\Spin(2n)_B\,{\times}\,\Spin_F(2n)\big)\circ\IR^1_{\vdt}\too\h\SU(n)\circ\IR^1_{\vdt}
\end{equation}
is indeed a consequence of only the complex structure. Generalizing this, we have:
\begin{conc}\label{C:SUn}
In a valise supermultiplet with $2n$ real bosons and $2n$ real fermions, an isomorphism of the bosonic and the fermionic holoraumy groups to the same $\SU(n)$ subgroup within any reference framework implies the existence of a supersymmetric complex structure in the entire supermultiplet.
\end{conc}

Compare now the corresponding characteristics of the chiral and vector supermultiplet:
\begin{description}\itemsep=-3pt\vspace*{-1mm}
 \item[\bsf Chiral supermultiplet:]
 The {\em\/valise\/} version\eq{e:vCS1D} of this supermultiplet has a $(\IC^2_{\sss B}|\IC^2_{\sss F})$-like field-space, spanned by the complex components indicated in\eq{e:cpxvCS}.
  \begin{enumerate}\itemsep=-3pt\vspace*{-1mm}
   \item By virtue of the supermultiplet minimality relations\eq{e:vCS-pr}, $\sB_{ab}^{\sss\text{(CS)}}$- and $\sF_{ab}^{\sss\text{(CS)}}$-rotations both span $\SU(2)\approx\Spin(3)$ subgroups of the maximal field-space rotations.
   \item The images of these two {\em\/a priori\/} separate $Spin(3)$-groups in $\Spin(3,3)$ coincide, and span the same $\SU(2)_\rot\approx\Spin(3)_\rot$, which is the common subgroup of the Lorentz group $\Spin(1,3)$ and the maximal connected component $\Spin(4)\subset\Aut(\SSp^{1|4})$.
   \item This coincidence implies that the complex structures on $\IC^2_{\sss B}$ and  $\IC^2_{\sss F}$ are related by supersymmetry, and provide a {\em\/supersymmetric complex structure\/}. Indeed, the two complex structures are {\em\/generated\/} by supersymmetry from $(A,B)\to(A{+}iB)$.
  \end{enumerate}\vspace*{-2mm}

 \item[\bsf Vector supermultiplet \rm(in the Wess-Zumino gauge):]
 The {\em\/valise\/} version\eq{e:vVS1D} of this supermultiplet has an $(\IR^4_{\sss B}|\IR^4_{\sss F})$-like field-space, spanned by $(A_1,A_2,A_3,d|\l_1,l_2,\l_3,\l_4)$.
  \begin{enumerate}\itemsep=-3pt\vspace*{-1mm}
   \item By virtue of the supermultiplet minimality relations\eq{e:vVS-pr}, $\sB_{ab}^{\sss\text{(VS)}}$- and $\sF_{ab}^{\sss\text{(VS)}}$-rotations both span $\SU(2)\approx\Spin(3)$ subgroups of the maximal field-space rotations.
   \item The images of these two {\em\/a priori\/} separate $Spin(3)$-groups in $\Spin(3,3)$ mutually commute, and jointly span its maximal compact subgroup $\Spin(3)_\rot\times\Spin(3)_\exR$, isomorphic to $\Spin(4)$, the maximal connected component of $\Aut(\SSp^{1|4})$, the outer group of automorphisms of the supersymmetry algebra\eq{e:SuSyD}.
   \item The perfect {\em\/mis-alignment\/} and mutual commutation of $\h(\sB_{ab}^{\sss\text{(VS)}})$- and $\h(\sF_{ab}^{\sss\text{(VS)}})$-trans\-for\-mations within the reference $\Spin(3,3)$ correlates with the {\em\/absence\/} of a supersymmetric complex structure in the vector supermultiplet.
  \end{enumerate}\vspace*{-2mm}
\end{description}\vspace*{-2mm}
That is, although the 4-plets $(A_1,A_2,A_3,d)$, $(\l_1,\l_2,\l_3,\l_4)$ and $(\rD_1,\rD_2,\rD_3,\rD_4)$ separately can be combined into complex pairs, no such choice gives the field-space the $(\IC^2_{\sss B}|\IC^2_{\sss F})$ structure compatible with the supersymmetry\eq{e:vVS1D}; see also Eqs.\eqs{e:l=j}{e:cpxv2tCS} below.

\subsection{The Tensor Supermultiplet}
 \label{s:TS}
The 3+1-dimensional tensor supermultiplet, again in the Wess-Zumino gauge, our Majorana (real component) framework and dimensionally reduced to the coordinate time world-line, is specified by the super-differential system\cite{rUMD09-1}:
\begin{subequations}
\label{e:TS1D}
\begin{alignat}9
 \rD_a\,\vf
 &= \c_a,\qquad
 \rD_a\,B_{mn}
  = -\inv2(\g_{mn})_a{}^b\,\c_b,\qquad m,n=1,2,3,\\*
 \rD_a\,\c_b
 &= i(\g^0)_{ab}\,\vdt\vf
   -(\g^5\g^m)_{ab}\,\ve_m{}^{rs}\,\vdt B_{rs}.
\end{alignat}
Using the matrices in Tables~\ref{t:16} and~\ref{t:160}, we tabulate these results:
\begin{equation}
  \begin{array}{@{} c|cccc|cccc @{}}
\text{\bsf TS} &\bm{\vf} &2\bm{B_{12}} &2\bm{B_{23}} &2\bm{B_{31}}
  &\bm{\c_1} &\bm{\c_2} &\bm{\c_3} &\bm{\c_4} \\ 
    \toprule
\bm{\C1{\rD_1}} & \c_1 &-\c_3 &-\c_4 &-\c_2
                &  i\vdt\vf     &-2i\vdt B_{31} &-2i\vdt B_{12} &-2i\vdt B_{23} \\ 
\bm{\C2{\rD_2}} & \c_2 & \c_4 &-\c_3 & \c_1
                & 2i\vdt B_{31} & i\vdt\vf      &-2i\vdt B_{23} & 2i\vdt B_{12} \\ 
\bm{\C3{\rD_3}} & \c_3 & \c_1 & \c_2 &-\c_4
                & 2i\vdt B_{12} & 2i\vdt B_{23} & i\vdt\vf      &-2i\vdt B_{31} \\ 
\bm{\C4{\rD_4}} & \c_4 &-\c_2 & \c_1 & \c_3
                & 2i\vdt B_{23} &-2i\vdt B_{12} & 2i\vdt B_{31} & i\vdt\vf \\ 
    \bottomrule
  \end{array}
 \label{e:1DTSt}
\end{equation}
\end{subequations}
and verify that the transformations are again monomial, whereby the supermultiplet can be depicted by the Adinkra:
\begin{equation}
 \vC{\begin{picture}(80,35)(0,-3)
   \put(0,-1){\includegraphics[width=70mm]{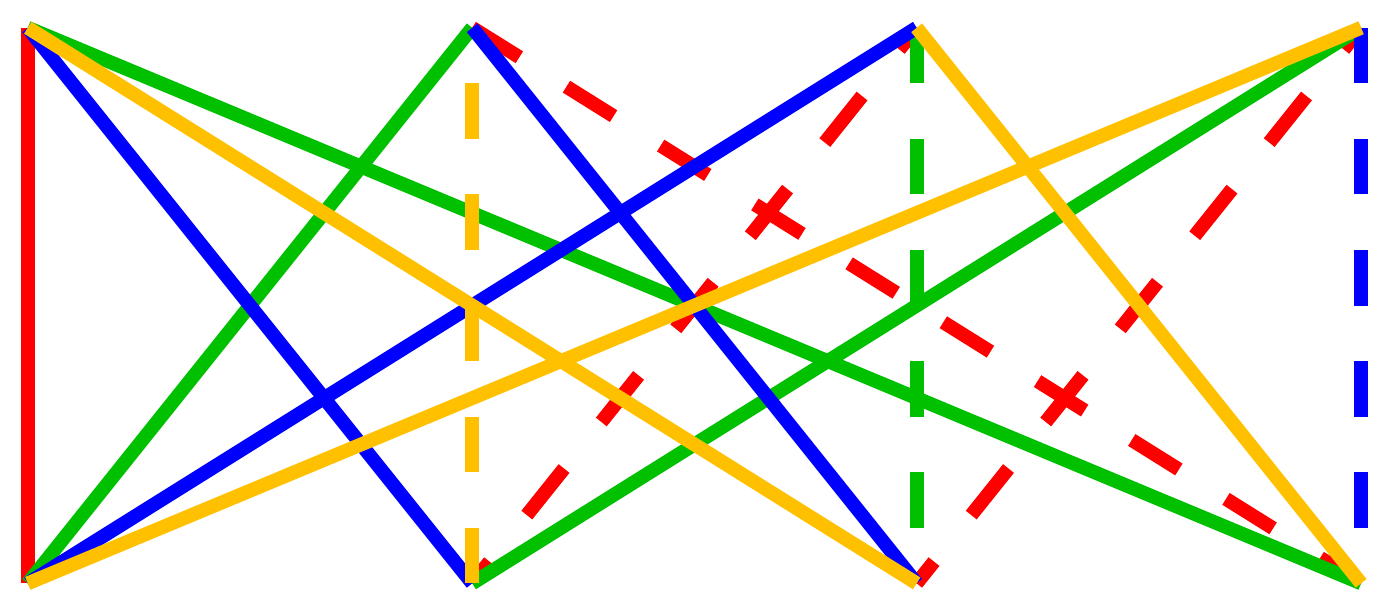}}
    \put(2,0){\cB{$\vf$}}
    \put(24,0){\cB{$2B_{12}$}}
    \put(46,0){\cB{$2B_{23}$}}
    \put(68,0){\cB{$2B_{31}$}}
    \put(2,28){\bB{$\c_1$}}
    \put(24,28){\bB{$\c_2$}}
    \put(46,28){\bB{$\c_3$}}
    \put(68,28){\bB{$\c_4$}}
 \end{picture}}
 \label{e:TS-A}
\end{equation}
and we read off the $\IL$-matrices:
\begin{alignat}9
  \C1{\IL_1^{\!\!\sss\text{TS}}}&=\!\bM{1&0&0&0\\0&0&\!\!-1&0\\0&0&0&\!\!-1\\0&\!\!-1&0&0\\}, &~~
  \C2{\IL_2^{\!\!\sss\text{TS}}}&=\!\bM{0&1&0&0\\ 0&0&0&1\\ 0&0&\!\!-1&0\\ 1&0&0&0\\},&~~
  \C3{\IL_3^{\!\!\sss\text{TS}}}&=\!\bM{0&0&1&0\\ 1&0&0&0\\ 0&1&0&0\\ 0&0&0&\!\!-1\\},&~~
  \C4{\IL_4^{\!\!\sss\text{TS}}}&=\!\bM{0&0&0&1\\ 0&\!\!-1&0&0\\ 1&0&0&0\\ 0&0&1&0\\}.
 \label{e:TS-L}
\end{alignat}

With the supermultiplet now given both in the tensorial representation\eq{e:TS1D} and in terms of its $\IL$-matrices\eq{e:TS-L}, we can compute straightforwardly the quadratic holoraumy tensors both ways, for a comparative illustration.

\paragraph{Tensorial Computation:}
Iterating\eq{e:vVS1D}, we compute straightforwardly:
\begin{alignat}9
 \rD_{[a}\rD_{b]}\,\l_c
  &=\inv2[\,\rD_a\,,\,\rD_b\,]\,\l_c
   =-i(\sF_{ab}^{\sss\text{(TS)}})_c{}^d\,(\vdt\l_d), \label{e:FabTSj}
\intertext{where the quadratic fermionic holoraumy tensor $(\sF_{ab})_c{}^d$ may be written, utilizing again a series of Fierz identities, as:}
 (\sF_{ab}^{\sss\text{(TS)}})^{}_c{}^d
  &=-C_{ab}\,(\g^0)_c{}^d
    +(\g^5)_{ab}\,(\g^5\g^0)_c{}^d
    -(\g^5\g^0)_{ab}\,(\g^5)_c{}^d. \label{e:TSFH3}
\end{alignat}
As in the case of the chiral and the vector supermultiplet, the fermionic quadratic holoraumy tensor $\sF_{ab}^{\sss\text{(TS)}}$ is again a $\Spin(3)$-valued 2-form in the 4-component space of fermions. Just as for the vector supermultiplet, $\{\sF_{ab}^{\sss\text{(TS)}}\}\in\spin(3)_\exR$. Similarly computed, the bosonic holoraumy tensor is now:
\begin{equation}
 (\sB_{ab}^{\sss\text{(TS)}})^{}_i{}^j
  =-i(\g^5\g^1)_{ab}\,(\g^{12})_i{}^j
   -i(\g^5\g^2)_{ab}\,(\g^{23})_i{}^j
   +i(\g^5\g^3)_{ab}\,(\g^{13})_i{}^j. \label{e:TSBH3}
\end{equation}
To summarize,
\begin{equation}
  \sB_{ab}^{\sss\text{(TS)}} \5\h\into \spin(3)_\rot,
   \qquad\text{and}\qquad
  \sF_{ab}^{\sss\text{(TS)}} \5\h\into \spin(3)_\exR,
\end{equation}
with the concrete results\eq{e:TSFH3} and\eq{e:TSBH3} again specifying the details of these maps. The algebraic structure of these results is shown below to again be encoded just as well in the purely world-line description of the 0-brain dimensional reduction\eqs{e:TS-A}{e:TS-L} of this supermultiplet.

\paragraph{Matrix Computation:}
Given the explicit $\IL$-matrices\eq{e:TS-L}, straightforward matrix algebra produces the quadratic holonomy matrices:
\begin{subequations}
 \label{e:vTS-BF}
\begin{alignat}9
  \sB_{12}^{\sss\text{(TS)}}&=+\sB_{34}^{\sss\text{(TS)}}&&=+\g^{23},&\quad
  \sB_{23}^{\sss\text{(TS)}}&=+\sB_{14}^{\sss\text{(TS)}}&&=+\g^{12},&\quad
  \sB_{31}^{\sss\text{(TS)}}&=+\sB_{24}^{\sss\text{(TS)}}&&=+\g^{13}, \label{e:vTS-B} \\
  \sF_{12}^{\sss\text{(TS)}}&=-\sF_{34}^{\sss\text{(TS)}}&&=+\g^0,&\quad
  \sF_{23}^{\sss\text{(TS)}}&=-\sF_{14}^{\sss\text{(TS)}}&&=-\g^{0123},&\quad
  \sF_{31}^{\sss\text{(TS)}}&=-\sF_{24}^{\sss\text{(TS)}}&&=-\g^{123}, \label{e:vTS-F}
\end{alignat}
\end{subequations}
where we have again identified the results in terms of the reference matrices as given in Table~\ref{t:16}.

The first of these equalities produce the basis-independent results
\begin{equation}
  \sB_{ab-}^{\sss\text{(TS)}}\Defl
  \sB_{ab}^{\sss\text{(TS)}}-\inv2\ve_{ab}{}^{cd}\,\sB_{cd}^{\sss\text{(TS)}}=0
   \qquad\text{and}\qquad
  \sF_{ab+}^{\sss\text{(TS)}}\Defl
  \sF_{ab}^{\sss\text{(TS)}}+\inv2\ve_{ab}{}^{cd}\,\sF_{cd}^{\sss\text{(TS)}}=0,
 \label{e:vTS-pr}
\end{equation}
which are a consequence of the fact that the tensor superfield in the Wess-Zumino gauge is annihilated by quasi-projection operators\cite{r6-1.2,rH-TSS}
\begin{equation}
  \big[\rD_{[a}\rD_{b]} - \inv2\ve_{ab}{}^{cd}\,\rD_{[c}\rD_{d]}\big],
 \label{e:TSDD}
\end{equation}
where the relative sign is opposite from the one in\eq{e:CSDD} and same as the one in\eq{e:VSDD}.
 Equivalently, the Adinkra\eq{e:TS-A} exhibits closed (\C1{red}-\C2{green}-\C3{blue}-\C4{orange}) 4-color cycles with $\textsl{CP}(\text{\ref{e:TS-A}})=-1=\c_o$\cite{rH-TSS,rUMD09-1}. 

Again, $\sB_{ab}^{\sss\text{(TS)}}$ and $\sF_{ab}^{\sss\text{(TS)}}$ generate rotations in the $\IR^4$-like sectors of the field-space, respectively $(\vf,2B_{12},2B_{23},2B_{31})$ and $(\c_1,\c_2,\c_3,\c_4)$.
The minimality relations\eq{e:vTS-pr} again reduce these rotations
\begin{equation}
 \begin{array}{rcl}
  \spin(4)_{\sss B} &\too{\text{(\ref{e:vTS-pr})}}& \spin(3)_+\\
  \spin(4)_{\sss F} &\too{\text{(\ref{e:vTS-pr})}}& \spin(3)_-\\
 \end{array}\bigg\}\subset\spin(3)_-\oplus\spin(3)_+=\spin(4)
 \label{e:RedHTS}
\end{equation}
Unlike the case of the chiral supermultiplet and just as in the vector supermultiplet, the relations\eq{e:vTS-pr} do not reduce these rotations from $\Spin(4)$: the $\sB_{ab}^{\sss\text{(TS)}}$ and $\sF_{ab}^{\sss\text{(TS)}}$ tensors are each valued in a separate and mutually commuting $\Spin(3)$ subgroup of $\Spin(4)$. In particular,
\begin{equation}
  \sB_{ab+}^{\sss\text{(TS)}} \5\h\into \spin(3)_\rot,
   \quad\text{while}\quad
  \sF_{ab-}^{\sss\text{(TS)}} \5\h\into \spin(3)_\exR,
 \label{e:TSrotexR}
\end{equation}
and referring again to the diagram\eq{e:map}, we summarize the results\eq{e:RedHTS} and\eq{e:TSrotexR} as
\begin{equation}
  \big(\h_{\sss B};\h_{\sss F}\big)^{\sss\text{(TS)}}
  :~~\bigg\{
   \begin{array}{r@{~\in~}ccl}
    \sB^{\sss\text{(TS)}}_{ab} & \spin(3)_+ &\too{\Ione} & \spin(3)_\rot,\\[1mm]
    \sF^{\sss\text{(TS)}}_{ab} & \spin(3)_- &\too{\Ione} & \spin(3)_\exR,\\
   \end{array}
 \label{e:TSmap}
\end{equation}
precisely as is the case\eq{e:VSmap} with the vector supermultiplet.

Motivated by this holoraumy isomorphism, $(\h_{\sss B};\h_{\sss F})^{\sss\text{(TS)}}\simeq(\h_{\sss B};\h_{\sss F})^{\sss\text{(VS)}}$, we find the component field identification
\begin{subequations}
 \label{e:VS>TS}
\begin{equation}
  (\vf,2B_{12},2B_{23},2B_{31}|\c_1,\c_2,\c_3,\c_4)
  \cong (A_1,A_2,d,-A_3|\l_2,\l_1,\l_4,\l_3),
\end{equation}
which may be written, in terms of\eq{e:XY=}, as
\begin{equation}
 \bM{\vf\\2B_{12}\\2B_{23}\\2B_{31}\\}
  =\underbrace{\bM{1&0&0&0\\ 0&1&0&0\\ 0&0&0&1\\ 0&0&\!\!-1&0\\}}
    _{=:\,{\cal X}^{^\text{(TS)}}\!\!_{\text{(VS)}}}
    \bM{A_1\\A_2\\A_3\\d\\},\quad
 \bM{\c_1\\\c_2\\\c_3\\\c_4\\}
  =\underbrace{\bM{0&1&0&0\\ 1&0&0&0\\ 0&0&0&1\\ 0&0&1&0\\}}
    _{=:\,{\cal Y}^{^\text{(TS)}}\!\!_{\text{(VS)}}}
    \bM{\l_1\\\l_2\\\l_3\\\l_4\\},\quad
 \bigg\{\!
       \begin{array}{r@{\>}l}
          \det[{\cal X}^{^\text{(TS)}}\!\!_{\sss\text{(VS)}}]&=+1,\\*[2mm]
          \det[{\cal Y}^{^\text{(TS)}}\!\!_{\sss\text{(VS)}}]&=+1,
       \end{array}
\end{equation}
\end{subequations}
Since $d=\int\!\rd\t\,\sD$ is a non-local redefinition of the auxiliary component $\sD$ in the vector supermultiplet, the component field identification\eq{e:VS>TS} is a {\em\/non-local\/} relationship between the world-line dimensional reductions of the tensor and the vector supermultiplets,\eq{e:TS1D} and\eq{e:VS1D}, respectively. 

\subsection{The Twisted-Chiral Supermultiplet Valise}
 \label{s:tCSv}
The twisted-chiral supermultiplet was first constructed in Ref.\cite{rTwSJG0}, by dimensionally reducing the vector supermultiplet to 1+1-dimensional spacetime. For completeness, this derivation is re-traced in Appendix~\ref{a:tCS}, dimensionally reducing however straight to 1-dimensional the world-line. Here, we cite the end-result:
\begin{subequations}
\label{e:vtCS1D}
\begin{equation}
  \begin{array}{@{} c|cccc|cccc @{}}
 \text{\bsf vtCS} &\bm{\Tw A} &\bm{\Tw B} &\bm{\Tw F} &\bm{\Tw G} 
  &\bm{\Tw\j_1} &\bm{\Tw\j_2} &\bm{\Tw\j_3} &\bm{\Tw\j_4} \\ 
    \toprule
\bm{\C1{\rD_1}} & \Tw\j_1 & -\Tw\j_4 & \Tw\j_2 & \Tw\j_3
    &i\vdt \Tw{A} &i\vdt\Tw{F} & i\vdt \Tw{G} &-i\vdt \Tw{B} \\ 
\bm{\C2{\rD_2}} & \Tw\j_2 &-\Tw\j_3 &-\Tw\j_1 & -\Tw\j_4
    &-i\vdt\Tw{F} & i\vdt \Tw{A} &-i\vdt \Tw{B} &-i\vdt \Tw{G} \\ 
\bm{\C3{\rD_3}} & \Tw\j_3 & \Tw\j_2 & \Tw\j_4 & -\Tw\j_1
    &-i\vdt \Tw{G} & i\vdt \Tw{B} & i\vdt \Tw{A} &i\vdt\Tw{F} \\ 
\bm{\C4{\rD_4}} &\Tw\j_4 &\Tw\j_1 &-\Tw\j_3 & \Tw\j_2
    &i\vdt \Tw{B} & i\vdt \Tw{G} &-i\vdt\Tw{F} & i\vdt \Tw{A} \\
    \bottomrule
  \end{array}
 \label{e:1DVSt3}
\end{equation}
and in tensorial notation:
\begin{align}
  \rD_a\,\Tw{A}&=\Tw\j_a,\quad
  \rD_a\,\Tw{B}=-(\g^{23})_a{}^b\,\Tw\j_b,\quad
  \rD_a\,\Tw{F}=-(\g^{13})_a{}^b\,\Tw\j_b,\quad
  \rD_a\,\Tw{G}=(\g^{12})_a{}^b\,\Tw\j_b, \label{e:1tCVS1}\\*
  \rD_a\,\Tw\j_b
            &=i(\g^0)_{ab}\,(\vdt\,\Tw{A}) -i(\g^{023})_{ab}\,(\vdt\,\Tw{B}) 
              -i(\g^{013})_{ab}\,(\vdt\,\Tw{F}) +i(\g^{012})_{ab}\,(\vdt\,\Tw{G}).
  \label{e:1tCVS2}
\end{align}
\end{subequations}
The transformation rules\eq{e:vtCS1D} specify the world-line valise twisted-chiral supermultiplet that differs from the original definition\cite{rTwSJG0} only through the component (super)field redefinitions\eq{e:l=j} and\eq{e:ABFG} detailed in the Appendix~\ref{a:tCS}, both of which have $\det=+1$.

For a comparison with\eq{e:vCS1Dt} however, we perform one additional redefinition:
\begin{equation}
  \Tw\j_3\to\ha\j_3\Defl-\Tw\j_3,
 \label{e:-}
\end{equation}
and obtain:
\begin{equation}
  \begin{array}{@{} c|cccc|cccc @{}}
 \text{\bsf v}\ha{\text{\bsf tCS}} &\bm{\Tw A} &\bm{\Tw B} &\bm{\Tw F} &\bm{\Tw G} 
  &\bm{\Tw\j_1} &\bm{\Tw\j_2} &\bm{\ha\j_3} &\bm{\Tw\j_4} \\ 
    \toprule
\bm{\C1{\rD_1}} & \Tw\j_1 & -\Tw\j_4 & \Tw\j_2 & -\ha\j_3
    &i\vdt \Tw{A} & i\vdt\Tw{F} &-i\vdt \Tw{G} &-i\vdt \Tw{B} \\ 
\bm{\C2{\rD_2}} & \Tw\j_2 & \ha\j_3 &-\Tw\j_1 & -\Tw\j_4
    &-i\vdt\Tw{F} & i\vdt \Tw{A} & i\vdt \Tw{B} &-i\vdt \Tw{G} \\ 
\bm{\C3{\rD_3}} &-\ha\j_3 & \Tw\j_2 & \Tw\j_4 & -\Tw\j_1
    &-i\vdt \Tw{G} & i\vdt \Tw{B} &-i\vdt \Tw{A} & i\vdt\Tw{F} \\ 
\bm{\C4{\rD_4}} &\Tw\j_4 &\Tw\j_1 & \ha\j_3 & \Tw\j_2
    &i\vdt \Tw{B} & i\vdt \Tw{G} & i\vdt\Tw{F} & i\vdt \Tw{A} \\
    \bottomrule
  \end{array}
 \label{e:vtCSt1D}
\end{equation}
which now differs from the valise version of the world-line dimensionally reduced chiral supermultiplet supersymmetry transformation pattern in\eq{e:vCS1Dt} in---{\em\/and only in\/}---the signs of each resulting term in the $\C3{\rD_3}$-row, and so is the world-line twisted-chiral supermultiplet as defined in Ref.\cite{r6-3,r6-3.1,rFIL}.

Quite clearly, the ``one additional redefinition'' of the fermions\eq{e:-} has a negative determinant, and so does not belong to the {\em\/connected component\/} $\Spin(4)\subset\Aut(\SSp^{1|4})$, but its negative-determinant complement within the full $\Aut(\SSp^{1|4})=\Pin(4)$. It should be noted that the classification of world-line supermultiplets in Refs.\cite{rPT,rT01,rT01a,rCRT,rKRT,r6-1,rKT07,r6-3,r6-3.2,r6-1.2,rGKT10,r6-3.1,rDHIL13} generally proceeds up to general component field redefinitions---which then include negative-determinant component field transformations such as\eq{e:-}. While perfectly reasonable within the framework of purely world-line physics, we will show that it behooves us to trace these characteristics when inquiring whether a given world-line supermultiplet is the dimensional reduction of a supermultiplet from higher-dimensional spacetime.

The tabular representations of the super-differential relations\eq{e:vtCS1D} and\eq{e:vtCSt1D} make it clear that both versions of this system are also monomial, and may be depicted, respectively, by the Adinkras:
\begin{equation}
 \vC{\begin{picture}(62,32)(0,-2)
   \put(0,0){\includegraphics[width=60mm]{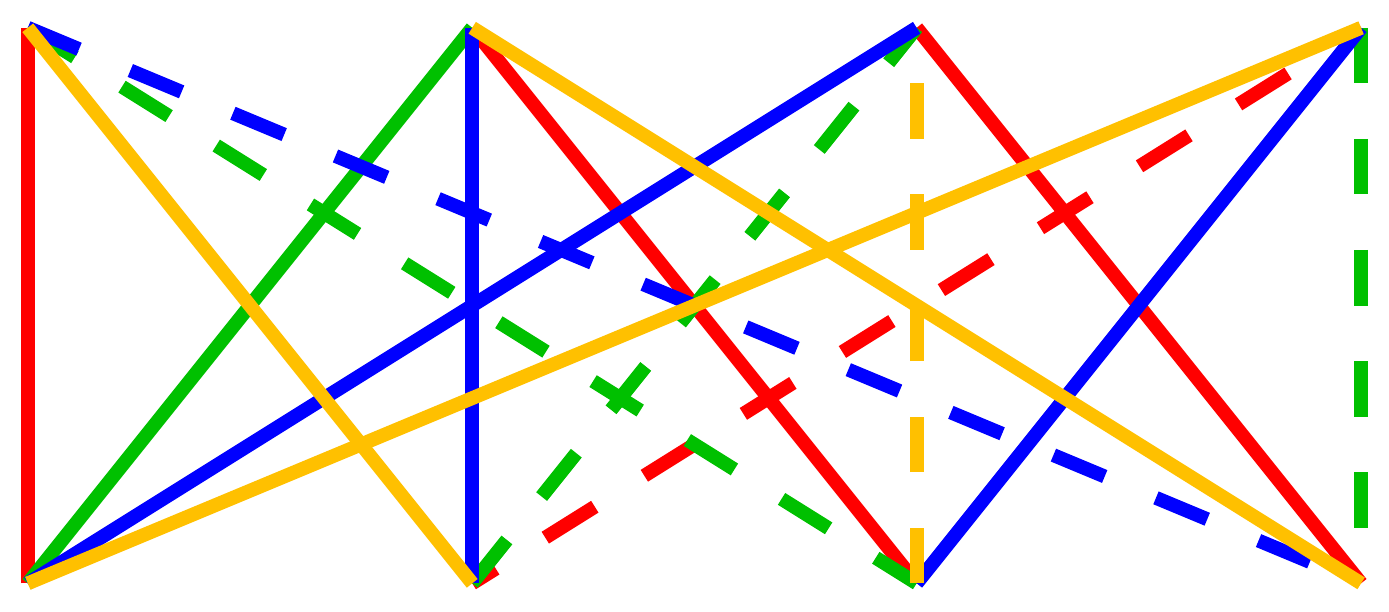}}
    \put(-10,15){\bsf vtCS}
    \put(-11,9){(\ref{e:vtCS1D})}
    \put(1,0){\cB{$\Tw{A}$}}
    \put(20.5,0){\cB{$\Tw{B}$}}
    \put(39.5,0){\cB{$\Tw{F}$}}
    \put(59,0){\cB{$\Tw{G}$}}
    \put(1,24){\bB{$\Tw\j_1$}}
    \put(20.5,24){\bB{$\Tw\j_2$}}
    \put(39.5,24){\bB{$\Tw\j_3$}}
    \put(59,24){\bB{$\Tw\j_4$}}
 \end{picture}}
  \qquad\qquad
 \vC{\begin{picture}(62,32)(0,-2)
   \put(0,0){\includegraphics[width=60mm]{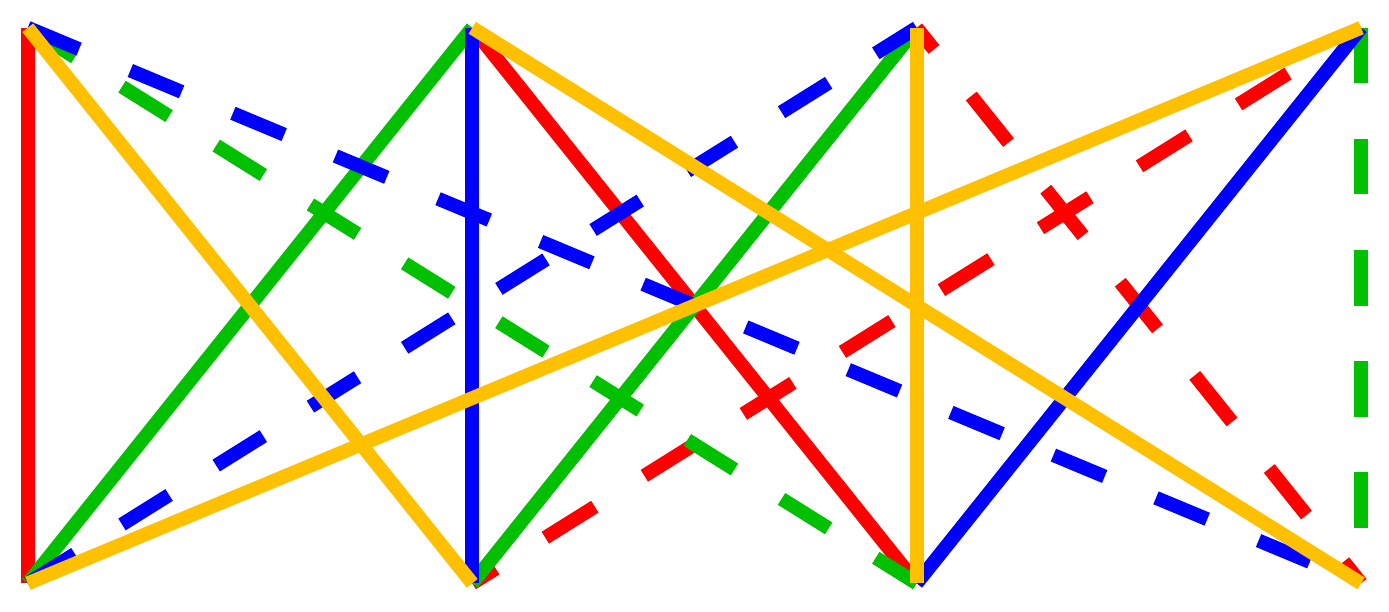}}
    \put(1,0){\cB{$\Tw{A}$}}
    \put(20.5,0){\cB{$\Tw{B}$}}
    \put(39.5,0){\cB{$\Tw{F}$}}
    \put(59,0){\cB{$\Tw{G}$}}
    \put(1,24){\bB{$\Tw\j_1$}}
    \put(20.5,24){\bB{$\Tw\j_2$}}
    \put(39.5,24){\bB{$\ha\j_3$}}
    \put(59,24){\bB{$\Tw\j_4$}}
    \put(61,15){\bsf v$\ha{\text{\bsf tCS}}$}
    \put(61,9){(\ref{e:vtCSt1D})}
 \end{picture}}
 \label{e:vtCS-A}
\end{equation}
the only difference between them being dashedness of the edges adjacent to $\Tw\j_3$, \ie, $\ha\j_3$, as implied by the ``one additional redefinition''\eq{e:-}.
The corresponding monomial $\IL$-matrices, as defined in\eq{e:valise}, are:
\begin{subequations}
 \label{e:vtCS-L}
\begin{alignat}9
 \mkern-10mu
  \C1{\IL_1^{\!\!\sss\text{tCS}}}
           &=\!\bM{1&0&0&0\\ 0&0&0&\!\!-1\\ 0&1&0&0\\ 0&0&1&0\\},&~
  \C2{\IL_2^{\!\!\sss\text{tCS}}}
           &=\!\bM{0&1&0&0\\ 0&0&\!\!-1&0\\ \!-1&0&0&0\\ 0&0&0&\!\!-1\\},&~
  \C3{\IL_3^{\!\!\sss\text{tCS}}}
           &=\!\bM{0&0&1&0\\ 0&1&0&0\\ 0&0&0&1\\ \!-1&0&0&0\\},&~
  \C4{\IL_4^{\!\!\sss\text{tCS}}}
           &=\!\bM{0&0&0&1\\ 1&0&0&0\\ 0&0&\!\!-1&0\\ 0&1&0&0\\},
 \label{e:vtCS-L1}
\intertext{for\eq{e:vtCS1D}, whereas\eq{e:vtCSt1D} produces:}
 \mkern-10mu
  \C1{\IL_1^{\!\!\sss\hCS}}
           &=\!\bM{1&0&0&0\\ 0&0&0&\!\!-1\\ 0&1&0&0\\ 0&0&\!\!-1&0\\},&~
  \C2{\IL_2^{\!\!\sss\hCS}}
           &=\!\bM{0&1&0&0\\ 0&0&1&0\\ \!-1&0&0&0\\ 0&0&0&\!\!-1\\},&~
  \C3{\IL_3^{\!\!\sss\hCS}}
           &=\!\bM{0&0&\!\!-1&0\\ 0&1&0&0\\ 0&0&0&1\\ \!-1&0&0&0\\},&~
  \C4{\IL_4^{\!\!\sss\hCS}}
           &=\!\bM{0&0&0&1\\ 1&0&0&0\\ 0&0&1&0\\ 0&1&0&0\\}.
 \label{e:vtCS-L2}
\end{alignat}
\end{subequations}

\paragraph{Complex Structures:}
Given the almost perfect identity between the super-differential system\eq{e:CS1D} and\eq{e:vtCSt1D}, where $\C3{\rD_3^{\sss\text{(}\hCS\text{)}}}=-\C3{\rD_3^{\sss\text{(CS)}}}$ being the only difference, we readily conclude that
\begin{subequations}
 \label{e:cpxv2tCS}
\begin{alignat}9
  \text{\bsf vtCS}&:\quad
  \big(\, (\Tw{A}{+}i\Tw{B}), (\Tw{F}{-}i\Tw{G})
      ~|~ (\Tw\j_1{-}i\Tw\j_4),~ (\Tw\j_2{-}i\Tw\j_3)
   ~;~[\C1{\rD_1}-i\C4{\rD_4}],[\C2{\rD_2}-i\C3{\rD_3}] \,\big),
 \label{e:cpxv2tCSa} \\
  \text{\bsf v}\ha{\text{\bsf tCS}}&:\quad
  \big(\, (\Tw{A}{+}i\Tw{B}), (\Tw{F}{-}i\Tw{G})
      ~|~ (\Tw\j_1{-}i\Tw\j_4),~ (\Tw\j_2{+}i\ha\j_3)
   ~;~[\C1{\rD_1}-i\C4{\rD_4}],[\C2{\rD_2}-i\C3{\rD_3}] \,\big),
 \label{e:cpxv2tCSb}
\end{alignat}
\end{subequations}
form two versions of the complex valise world-line twisted-chiral supermultiplets\eq{e:vtCS1D} and\eq{e:vtCSt1D}.

The geometric meaning of this is as follows:
\begin{enumerate}\itemsep=-3pt\vspace*{-3mm}
 \item As in the complex chiral supermultiplet\eq{e:cpxvCS}, so also in both versions of the twisted-chiral supermultiplet\eq{e:cpxv2tCSa} and\eq{e:cpxv2tCSb}, the four bosons span a $\IC^2$-like target-space in like complex combinations: $(A{+}iB),(F{-}iG)$ and $(\Tw{A}{+}i\Tw{B}),(\Tw{F}{-}i\Tw{G})$.
 \item Both in the complex chiral supermultiplet\eq{e:cpxvCS}, and in the ``redefined''\cite{r6-3,r6-3.1,rFIL} twisted-chiral supermultiplet\eq{e:cpxv2tCSa}, the four fermions jointly span a $\IC^2$-like tangent space, all four in like complex combinations:
 $(\j_1{-}i\j_4)\sim(\Tw\j_1{-}i\Tw\j_4)$, as well as
 $(\j_2{+}i\j_3)\sim(\Tw\j_2{+}i\Tw\j_3)$, respectively.
 \item In the ``original''\cite{rTwSJG0} twisted-chiral supermultiplet\eq{e:cpxv2tCSa} however, the four fermions span a $\IC\,{\oplus}\,\IC^*$-like (holomorphic+antiholomorphic) tangent space in the opposite complex combinations:
 $(\j_1{-}i\j_4)\sim(\Tw\j_1{-}i\Tw\j_4)$, but
 $(\j_2{+}i\j_3)\5*\sim(\Tw\j_2{-}i\Tw\j_3)$, respectively.
\end{enumerate}\vspace*{-2mm}
Indeed, the ``one additional redefinition''\eq{e:-} serves to map the fermionic space $\IC\,{\oplus}\,\IC^*\to\IC^2$, and so ``attune'' the component field complex structures in the manner of the complex chiral supermultiplet.

Having obtained both the tensorial and the $\IL$-matrix representation\eqs{e:vtCS1D}{e:vtCS-L}, we proceed computing the quadratic holoraumy tensors.

\paragraph{Tensorial Computation:}
Iterating\eqs{e:1tCVS1}{e:1tCVS2}, we compute
\begin{alignat}9
 \rD_{[a}\rD_{b]}\,\Tw\j_c
  &=\inv2[\,\rD_a\,,\,\rD_b\,]\,\Tw\j_c
   =-i(\sF_{ab}^{\sss\text{(tCS)}})_c{}^d\,(\vdt\Tw\j_d), \label{e:FabtCSj}
\intertext{where}
 (\sF_{ab}^{\sss\text{(tCS)}})^{}_c{}^d
  &=-C_{ab}\,(\g^0)_c{}^d
    +(\g^5)_{ab}\,(\g^5\g^0)_c{}^d
    -(\g^5\g^0)_{ab}\,(\g^5)_c{}^d. \label{e:tCSFH3}
\intertext{Similarly computed, the bosonic holoraumy tensor is:}
 (\sB_{ab}^{\sss\text{(tCS)}})^{}_i{}^j
 &=+i(\g^5\g^1)_{ab}\,(\g^{13})_i{}^j
   +i(\g^5\g^2)_{ab}\,(\g^{12})_i{}^j
   +i(\g^5\g^3)_{ab}\,(\g^{23})_i{}^j. \label{e:tCSBH3}
\end{alignat}
To summarize,
\begin{equation}
  \sB_{ab}^{\sss\text{(tCS)}} \5\h\into \spin(3)_\rot,
   \qquad\text{and}\qquad
  \sF_{ab}^{\sss\text{(tCS)}} \5\h\into \spin(3)_\exR,
\end{equation}
with the concrete results\eq{e:tCSFH3} and\eq{e:tCSBH3} again specifying the details of these maps. The algebraic structure of these results is shown below to again be encoded just as well in the purely world-line description of the 0-brain dimensional reduction\eqs{e:vtCS-A}{e:vtCS-L1} of this supermultiplet.

\paragraph{Matrix Computation:}
Given the $\IL$-matrices\eq{e:vtCS-L1}, straightforward matrix algebra produces:
\begin{subequations}
 \label{e:tCS-BF}
\begin{alignat}9
  \sB_{12}^{\sss\text{(tCS)}}&=+\sB_{34}^{\sss\text{(tCS)}}&&=-\g^{12},&\quad
  \sB_{23}^{\sss\text{(tCS)}}&=+\sB_{14}^{\sss\text{(tCS)}}&&=-\g^{13},&\quad
  \sB_{31}^{\sss\text{(tCS)}}&=+\sB_{24}^{\sss\text{(tCS)}}&&=+\g^{23}, \label{e:tCS-B} \\
  \sF_{12}^{\sss\text{(tCS)}}&=-\sF_{34}^{\sss\text{(tCS)}}&&=+\g^0,&\quad
  \sF_{23}^{\sss\text{(tCS)}}&=-\sF_{14}^{\sss\text{(tCS)}}&&=+\g^{0123},&\quad
  \sF_{31}^{\sss\text{(tCS)}}&=-\sF_{24}^{\sss\text{(tCS)}}&&=-\g^{123}, \label{e:tCS-F}
\end{alignat}
\end{subequations}
where we have again identified the results in terms of the reference matrices as given in Table~\ref{t:16}.

The first of these equalities produce the basis-independent results
\begin{equation}
  \sB_{ab-}^{\sss\text{(tCS)}}\Defl
  \sB_{ab}^{\sss\text{(tCS)}}-\inv2\ve_{ab}{}^{cd}\,\sB_{cd}^{\sss\text{(tCS)}}=0
   \qquad\text{and}\qquad
  \sF_{ab+}^{\sss\text{(tCS)}}\Defl
  \sF_{ab}^{\sss\text{(tCS)}}+\inv2\ve_{ab}{}^{cd}\,\sF_{cd}^{\sss\text{(tCS)}}=0,
 \label{e:tCS-pr}
\end{equation}
which are a consequence of the fact that the twisted-chiral supermultiplet is annihilated by quasi-projection operators\cite{r6-1.2,rH-TSS}
\begin{equation}
  \big[\rD_{[a}\rD_{b]} - \inv2\ve_{ab}{}^{cd}\,\rD_{[c}\rD_{d]}\big],
 \label{e:tCSDD}
\end{equation}
where the relative sign is opposite from the one in\eq{e:CSDD} and same as the one in\eq{e:TSDD}, as well as the one in\eq{e:VSDD} from which\eq{e:1DVSt3} was derived.
 Equivalently, the Adinkras\eq{e:vtCS-A} exhibit closed (\C1{red}-\C2{green}-\C3{blue}-\C4{orange}) 4-color cycles with $\textsl{CP}(\text{\ref{e:vtCS-A}})=-1=\c_o$\cite{rH-TSS,rUMD09-1}. 

Again, $\sB_{ab}^{\sss\text{(tCS)}}$ and $\sF_{ab}^{\sss\text{(tCS)}}$ generate rotations in the $\IR^4$-like sectors of the field-space, respectively $(\Tw{A},\Tw{B},\Tw{F},\Tw{B})$ and $(\Tw\j_1,\Tw\j_2,\Tw\j_3,\Tw\j_4)^{\strut}$.
The minimality relations\eq{e:tCS-pr} again reduce these rotations
\begin{equation}
 \begin{array}{rcl}
  \spin(4)_{\sss B} &\too{\text{(\ref{e:tCS-pr})}}& \spin(3)_+\\
  \spin(4)_{\sss F} &\too{\text{(\ref{e:tCS-pr})}}& \spin(3)_-\\
 \end{array}\bigg\}\subset\spin(3)_-\oplus\spin(3)_+=\spin(4)
 \label{e:RedHtCS}
\end{equation}
Unlike the case of the chiral supermultiplet and just as in the vector supermultiplet, the relations\eq{e:tCS-pr} do not reduce the direct sum of these rotations from $\Spin(4)$: the $\sB_{ab}^{\sss\text{(tCS)}}$ and $\sF_{ab}^{\sss\text{(tCS)}}$ tensors are each valued in a separate and mutually commuting $\Spin(3)$ subgroup of $\Spin(4)$. In particular,
\begin{equation}
  \sB_{ab+}^{\sss\text{(tCS)}} \5\h\into \spin(3)_\rot,
   \quad\text{while}\quad
  \sF_{ab-}^{\sss\text{(tCS)}} \5\h\into \spin(3)_\exR,
 \label{e:tCSrotexR}
\end{equation}
and referring again to the diagram\eq{e:map}, we summarize the results\eq{e:RedHtCS} and\eq{e:tCSrotexR} as
\begin{equation}
  \big(\h_{\sss B};\h_{\sss F}\big)^{\sss\text{(tCS)}}
  :~~\bigg\{
   \begin{array}{r@{~\in~}ccl}
    \sB^{\sss\text{(tCS)}}_{ab} & \spin(3)_+ &\too{\Ione} & \spin(3)_\rot,\\[1mm]
    \sF^{\sss\text{(tCS)}}_{ab} & \spin(3)_- &\too{\Ione} & \spin(3)_\exR,\\
   \end{array}
 \label{e:tCSmap}
\end{equation}
precisely as is the case\eq{e:VSmap} with the vector supermultiplet and\eq{e:TSmap} with the tensor supermultiplet.

However, using the $\IL$-matrices\eq{e:vtCS-L2} produces instead:
\begin{subequations}
 \label{e:tCSt-BF}
\begin{alignat}9
  \sB_{12}^{\sss\HCS}&=+\sB_{34}^{\sss\HCS}&&=-\g^{12},&\quad
  \sB_{23}^{\sss\HCS}&=+\sB_{14}^{\sss\HCS}&&=-\g^{13},&\quad
  \sB_{31}^{\sss\HCS}&=+\sB_{24}^{\sss\HCS}&&=+\g^{23}, \label{e:tCSt-B} \\
  \sF_{12}^{\sss\HCS}&=-\sF_{34}^{\sss\HCS}&&=-\g^{13},&\quad
  \sF_{23}^{\sss\HCS}&=-\sF_{14}^{\sss\HCS}&&=-\g^{23},&\quad
  \sF_{31}^{\sss\HCS}&=-\sF_{24}^{\sss\HCS}&&=+\g^{12}. \label{e:tCSt-F}
\end{alignat}
\end{subequations}
The first of these equalities produce the basis-independent results
\begin{equation}
  \sB_{ab-}^{\sss\HCS}\Defl
  \sB_{ab}^{\sss\HCS}-\inv2\ve_{ab}{}^{cd}\,\sB_{cd}^{\sss\HCS}=0
   \qquad\text{and}\qquad
  \sF_{ab+}^{\sss\HCS}\Defl
  \sF_{ab}^{\sss\HCS}+\inv2\ve_{ab}{}^{cd}\,\sF_{cd}^{\sss\HCS}=0,
 \label{e:tCSt-pr}
\end{equation}
just like\eq{e:tCSt-pr}, since\eq{e:1DVSt3} and\eq{e:vtCSt1D} differ only by a component field redefinition\eq{e:-}, and so are both annihilated by the same quasi-projection operators\cite{r6-1.2,rH-TSS}
\begin{equation}
  \big[\rD_{[a}\rD_{b]} - \inv2\ve_{ab}{}^{cd}\,\rD_{[c}\rD_{d]}\big].
 \label{e:tCStDD}
\end{equation}
The relative sign is opposite from the one in\eq{e:CSDD}, and same as the one in\eq{e:TSDD}, \eq{e:VSDD} and\eq{e:1DVSt3}, thus implying  $\textsl{CP}(\text{\ref{e:vtCS-A}})=-1=\c_o$\cite{rH-TSS,rUMD09-1}.

Again, $\sB_{ab}^{\sss\HCS}$ and $\sF_{ab}^{\sss\HCS}$ generate rotations in the $\IR^4$-like sectors of the field-space, respectively $(\Tw{A},\Tw{B},\Tw{F},\Tw{B})$ and $(\Tw\j_1,\Tw\j_2,\ha\j_3,\Tw\j_4)^{\strut}$.
The relations\eq{e:tCSt-pr} reduce these rotations
\begin{equation}
 \begin{array}{rcl}
  \spin(4)_{\sss B} &\too{\text{(\ref{e:tCSt-pr})}}& \spin(3)_+\\
  \spin(4)_{\sss F} &\too{\text{(\ref{e:tCSt-pr})}}& \spin(3)_-\\
 \end{array}\bigg\}\subset\spin(3)_-\oplus\spin(3)_+=\spin(4)
 \label{e:RedHtCSt}
\end{equation}
Unlike the case of the vector, tensor and ``original'' twisted-chiral supermultiplet and just as in the case of the chiral supermultiplet, the relations\eq{e:tCSt-BF} do reduce these rotations from $\Spin(4)$: the $\sB_{ab}^{\sss\HCS}$ and $\sF_{ab}^{\sss\HCS}$ tensors are both valued in the same $\Spin(3)$ subgroup of $\Spin(4)$. In particular,
\begin{equation}
  \sB_{ab+}^{\sss\HCS} \5\h\into \spin(3)_\rot,
   \quad\text{while}\quad
  \sF_{ab-}^{\sss\HCS} \5\h\into \spin(3)_\rot,
 \label{e:tCStrotrot}
\end{equation}
and referring again to the diagram\eq{e:map}, we summarize the results\eq{e:RedHtCSt} and\eq{e:tCStrotrot} as
\begin{equation}
  \big(\h_{\sss B};\h_{\sss F}\big)^{\sss\HCS}
  :~~\bigg\{
   \begin{array}{r@{~\in~}ccl}
    \sB^{\sss\HCS}_{ab} & \spin(3)_+ &\too{\Ione} & \spin(3)_\rot,\\[1mm]
    \sF^{\sss\HCS}_{ab} & \spin(3)_- &\too{\vp'} & \spin(3)_\rot,\\
   \end{array}
 \label{e:tCStmap}
\end{equation}
just as in the case\eq{e:CSmap} of the chiral supermultiplet, but with the roles of $\spin(3)_-$ and $\spin(3)_+$ swapped, and where
 $\vp'=\big({{\SSS12},\,{\SSS23},\,{\SSS31}\atop{\SSS23},\,\7{\SSS31},\,\7{\SSS12}}\big)$ is the even ($\det=+1$) relative signed permutation
\begin{equation}
  \vp':~\h\{\sF^{\HCS}_{12},\sF^{\HCS}_{23},\sF^{\HCS}_{31}\}
  = \h\{\sB^{\HCS}_{23},-\sB^{\HCS}_{31},-\sB^{\HCS}_{12}\}
  ~\supset~\spin(3)_\rot,
    \label{e:vCSt-B=F}
\end{equation}
relating the images of the fermionic and bosonic holonomy in the reference algebra $\spin(3,3)$. This permutation is the same as\eq{e:vCS-B=F}, up to an even number of sign-changes in the generators, which is an outer automorphism of all $\spin(3)$ algebras.

\section{Non-Valise Supermultiplets}
 \label{s:nvCS}
The straightforward dimensional reduction of higher-dimensional supermultiplets to the coordinate time world-line in fact contains more information than can be discerned from the valise supermultiplets discussed in Section~\ref{s:3LPs}. Those particular example valises have been obtained upon the following non-local field redefinitions:
\begin{subequations}
 \label{e:nodeL}
\begin{alignat}9
 \text{chiral\eq{e:vCS1D}}:&&~~
 (A,B|\j_a|\cF,\cG) &\to (A,B,F,G|\j_a),&~~
 (F,G)&\Defl\big(\int\!\rd\t\,\cF,\int\!\rd\t\,\cG\big);\\
 \text{vector\eq{e:vVS1D}}:&&~~
 (A_1,A_2,A_3|\l_a|\sD) &\to (A_1,A_2,A_3,d|\j_a),&~~
 d&\Defl\int\!\rd\t~\sD;\\
 \9{\text{(\ref{e:vtCS1D}) \&\eq{e:ABFG}}}
 {\text{twisted-chiral}}:&&~~
 (A_1,A_2,A_3|\l|\sD)   &\to (\Tw{A},\Tw{B},\Tw{F},\Tw{G}|\Tw\j_a),&~~
 \Tw{F}&\Defl-\int\!\rd\t~\sD.
\end{alignat}
\end{subequations}
In the supermultiplets displayed in the left-hand side of the field redefinition maps\eq{e:nodeL}, the action of $\rD_{[I}\rD_{J]}$ still produces a spin-preserving transformation on the fields, but does not always reduce to a uniform factorization of a field-space rotation and domain-space (time) translation. To indicate this distinction, we will label the resulting operators:
\begin{equation}
  \rD_{[I}\rD_{J]}(\text{bosons})\Defr-i\ha\sB_{IJ}(\text{bosons})
   \quad\text{and}\quad
  \rD_{[I}\rD_{J]}(\text{fermions})\Defr-i\ha\sF_{IJ}(\text{fermions}),
\end{equation}
and examine these results in turn.

\subsection{Chiral supermultiplet}
Restoring the original component superfields $\cF=\vdt F$ and $\cG=\vdt G$ and drawing their nodes above the fermionic ones to indicate their relative engineering dimensions, $[\cF]=[\cG]=[\j_a]{+}\inv2=[A]{+}1=[B]{+}1$, the system\eq{e:vCS1Dt} becomes:
\begin{equation}
 \begin{array}{@{} c|cccc|cccc @{}}
\text{\bsf CS} &\bm A &\bm B &\bm{\cF} &\bm{\cG} &\bm{\j_1} &\bm{\j_2} &\bm{\j_3} &\bm{\j_4} \\ 
    \toprule
\C1{\bm{\rD_1}} & \j_1 &-\j_4 & \vdt\j_2 &-\vdt\j_3 & i\vdt{A} & i\cF &-i\cG &-i\vdt{B} \\ 
\C2{\bm{\rD_2}} & \j_2 & \j_3 &-\vdt\j_1 &-\vdt\j_4 &-i\cF & i\vdt{A} & i\vdt{B} &-i\cG \\ 
\C3{\bm{\rD_3}} & \j_3 &-\j_2 &-\vdt\j_4 & \vdt\j_1 & i\cG &-i\vdt{B} & i\vdt{A} &-i\cF \\ 
\C4{\bm{\rD_4}} & \j_4 & \j_1 & \vdt\j_3 & \vdt\j_2 & i\vdt{B} & i\cG & i\cF & i\vdt{A} \\ 
    \bottomrule
 \end{array}
 \qquad
 \vC{\begin{picture}(40,35)(0,-2)
   \put(0,0){\includegraphics[height=31mm]{CScN.pdf}}
    \put(16,0){\cB{$A$}}
    \put(25,0){\cB{$B$}}
    \put(1,14){\bB{$\j_1$}}
    \put(11,14){\bB{$-\j_4$}}
    \put(30,14){\bB{$\j_2$}}
    \put(39,14){\bB{$\j_3$}}
    \put(16,29){\cB{$\cF$}}
    \put(25,29){\cB{$-\cG$}}
 \end{picture}}
 \label{e:CS1D}
\end{equation}
showing the complex combinations
\begin{equation}
   \big(\, (A{+}iB) \,\big|\, (\j_1{-}i\j_4),\, (\j_2{+}i\j_3) \,\big|\,(\cF{-}i\cG)
   \,:\, [\C1{\rD_1}{-}i\C4{\rD_4}],[\C2{\rD_2}{+}i\C3{\rD_3}] \,\big)
 \label{e:cpxCS}
\end{equation}
of our real components that are in precise correspondence with usual the complex and Weyl components\cite{r1001,rWB,rBK} given the like complex combinations of the super-derivatives; see Eqs.\eq{e:cpxvCS}, as induced by\eq{e:DCS=0}.

Two of the bosonic holoraumy generators $\ha\sB_{IJ}$ now become nontrivial compositions with world-line translations. 
In particular, we obtain for\eq{e:CS1D}:
\begin{subequations}
 \label{e:CSR[t]}
\begin{equation}
  \ha\sB_{12}^{\sss\text{(CS)}} =-\ha\sB_{34}^{\sss\text{(CS)}}
  =\bM{0&0&\!\!-1&0\\ 0&0&0&1\\ \vdt^2&0&0&0\\ 0&\!\!-\vdt^2&0&0\\ },\qquad
  \ha\sB_{31}^{\sss\text{(CS)}} =-\ha\sB_{24}^{\sss\text{(CS)}}
  =\bM{0&0&0&\!\!-1\\ 0&0&\!\!-1&0\\ 0&\vdt^2&0&0\\ \vdt^2&0&0&0\\ },
 \label{e:R[t]CS}
\end{equation}
while
\begin{equation}
  \ha\sB_{23}^{\sss\text{(CS)}} =-\ha\sB_{14}^{\sss\text{(CS)}}
  =\bM{0&\!\!-1&0&0\\ 1&0&0&0\\ 0&0&0&\!\!-1\\ 0&0&1&0\\ }\circ\vdt
 \label{e:R[0]CS}
\end{equation}
\end{subequations}
remains a uniform composition of field-space (simultaneous, double) rotations and domain-space translations:
\begin{equation}
  \ha\sB_{23}^{\sss\text{(CS)}}(A,B|\cF,\cG)
  =\vdt(-B,A|-\cG,\cF).
\end{equation}
In turn, all the generators $\ha\sF_{IJ}^{\sss\text{(CS)}}=\sF_{IJ}^{\sss\text{(CS)}}\circ\vdt$ remain a uniform composition. This ``algebraic subset'' of the holoraumy transformations then is generated by $\sB_{23}^{\sss\text{(CS)}}$ and $\sF_{IJ}^{\sss\text{(CS)}}$, and generates the group $\rOp{U}(1)_{\sss B}\times\SU(2)_{\sss F}$. This type of effective (dynamical) symmetry reduction, shown in the central column of the summary:
\begin{equation}
  \begin{array}{@{} r@{}lr@{\>\to\>}l@{\>\too\h\>}l @{}}
 \text{\bsf Chiral}& &\MC2l{\text{\bsf Algebraic Subset of Holoraumy}}
                     &\text{\bsf Table~\ref{t:16}} \\ 
    \toprule
 \text{original}&\eq{e:CS1D}
 &\SO(4)_{\sss B}\,{\times}\,\SO(4)_{\sss F}
 &\rOp{U}(1)_{\sss B}\,{\times}\,\SU(2)_{\sss F}
 &\SU(2)_\rot\\ 
 \text{valise}&\eq{e:vCS1D}
 &\SO(4)_{\sss B}\,{\times}\,\SO(4)_{\sss F}
 &\SU(2)_{\sss B}\,{\times}\,\SU(2)_{\sss F}
 &\SU(2)_\rot\\ 
    \bottomrule
  \end{array}
 \label{e:CST}
\end{equation}
has been systematically explored for the $N\,{=}\,8$ ultra-multiplet in Ref.\cite{rFGH}, so that our present results extend that work both conceptually in Section~\ref{s:HoloR}, and also to $N=4$ supermultiplets.

The generators\eq{e:R[t]CS} can no longer form a matrix group in the usual sense with\eq{e:R[0]CS}, but they do form an $\IR[\vdt]$-module. While this general discussion of holoraumy is well beyond our present scope, we note in passing that when acting on functions expanded in plane-waves, such operators become matrix functions of plane-wave frequencies, and so provide an algebraic (Fourier-dual) approach to these differential operators as bundles over the energy-momentum space.

\subsection{Vector Supermultiplet}
The coordinate time world-line dimensional reduction of the standard vector supermultiplet in the Wess-Zumino gauge is given by the super-differential system\cite{rUMD09-1}:
\begin{subequations}
\label{e:VS1D}
\begin{alignat}9
 \rD_a\,A_m
 &= (\g_m)_a{}^b\,\l_b,\qquad m=1,2,3,\quad \g_m=\h_{mn}\g^n=\g^m,\\
 \rD_a\,\l_b
 &= -i(\g^0\g^n)_{ab}\,(\vdt\,A_n) + (\g^5)_{ab}\,\sD,\\
 \rD_a\,\sD
 &= i(\g^5\g^0)_a{}^b\,(\vdt\,\l_b).
\end{alignat}
Using the matrices in Tables~\ref{t:16} and~\ref{t:160}, we tabulate these results:
\begin{equation}
  \begin{array}{@{} c|cccc|cccc @{}}
\text{\bsf VS} &\bm{A_1} &\bm{A_2} &\bm{A_3} &\pmb{\sD}
  &\bm{\l_1} &\bm{\l_2} &\bm{\l_3} &\bm{\l_4} \\ 
    \toprule
\bm{\C1{\rD_1}} & \l_2 &-\l_4 & \l_1 &-\vdt\l_3 & i\vdt A_3 & i\vdt A_1 &-i\sD &-i\vdt A_2 \\ 
\bm{\C2{\rD_2}} & \l_1 & \l_3 &-\l_2 &-\vdt\l_4 & i\vdt A_1 &-i\vdt A_3 & i\vdt A_2 &-i\sD \\ 
\bm{\C3{\rD_3}} & \l_4 & \l_2 & \l_3 & \vdt\l_1 & i\sD & i\vdt A_2 & i\vdt A_3 & i\vdt A_1 \\ 
\bm{\C4{\rD_4}} & \l_3 &-\l_1 &-\l_4 & \vdt\l_2 &-i\vdt A_2 & i\sD & i\vdt A_1 &-i\vdt A_3 \\ 
    \bottomrule
  \end{array}
 \qquad
 \vC{\begin{picture}(35,35)(0,-2)
   \put(-.5,0){\includegraphics[height=31mm]{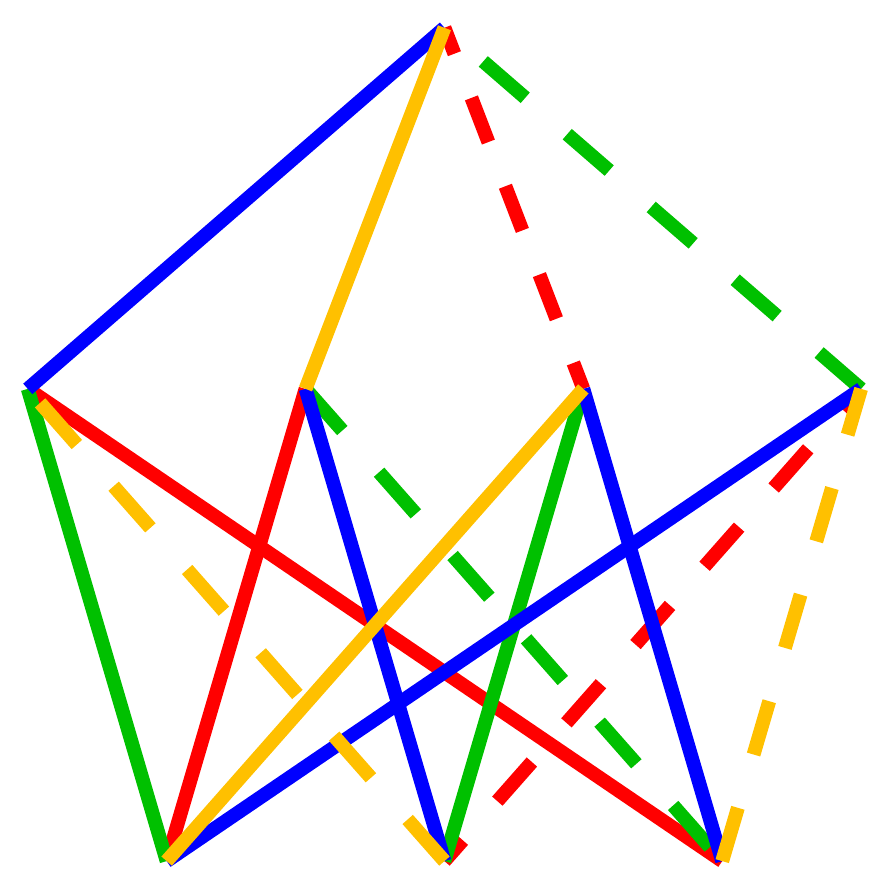}}
    \put(5,0){\cB{$A_1$}}
    \put(15,0){\cB{$A_2$}}
    \put(25,0){\cB{$A_3$}}
    \put(0,16){\bB{$\l_1$}}
    \put(10,16){\bB{$\l_2$}}
    \put(20,16){\bB{$\l_3$}}
    \put(30,16){\bB{$\l_4$}}
    \put(15,29){\cB{$\sD$}}
 \end{picture}}
 \label{e:1DVSt}
\end{equation}
\end{subequations}
The holoraumy generators for the vector supermultiplet (in the Wess-Zumino) gauge\eq{e:vVS-B} then become:
\begin{subequations}
 \label{e:VSR[t]}
\begin{gather}
  \ha\sB_{12}^{\sss\text{(VS)}} =+\ha\sB_{34}^{\sss\text{(VS)}}
  =\bM{0&0&\vdt&0\\ 0&0&0&\!\!-1\\ \!-\vdt&0&0&0\\ 0&\vdt^2&0&0\\ },\qquad
  \ha\sB_{31}^{\sss\text{(VS)}} =+\ha\sB_{24}^{\sss\text{(VS)}}
  =\bM{0&\!\!-\vdt&0&0\\ \vdt&0&0&0\\ 0&0&0&-1\\ 0&0&\vdt^2&0\\ },\\
  \ha\sB_{23}^{\sss\text{(VS)}} =+\ha\sB_{14}^{\sss\text{(VS)}}
  =\bM{0&0&0&1\\ 0&0&\vdt&0\\ 0&\!\!-\vdt&0&0\\ \!-\vdt^2&0&0&0\\ },
 \label{e:R[0]VS}
\end{gather}
\end{subequations}
none of which form a uniform composition of an algebraic matrix and $\vdt$. In contrast, $\sF_{IJ}^{\sss\text{(VS)}}$ of course remain algebraic. This is then summarized as:
\begin{equation}
  \begin{array}{@{} r@{}lr@{\>\to\>}l@{\>\too\h\>}l @{}}
 \text{\bsf vector}& &\MC2l{\text{\bsf Algebraic Subset of Holoraumy}}
                     &\text{\bsf Table~\ref{t:16}} \\ 
    \toprule
 \text{original}&\eq{e:VS1D}
 &\SO(4)_{\sss B}\,{\times}\,\SO(4)_{\sss F} 
 &\Ione_{\sss B}\,{\times}\,\SU(2)_{\sss F}
 &\SU(2)_\exR\\ 
 \text{valise}&\eq{e:vVS1D}
 &\SO(4)_{\sss B}\,{\times}\,\SO(4)_{\sss F} 
 &\SU(2)_{\sss B}\,{\times}\,\SU(2)_{\sss F} 
 &\SU(2)_\rot\,{\times}\,\SU(2)_\exR\\ 
    \bottomrule
  \end{array}
 \label{e:VST}
\end{equation}

\subsection{Twisted-Chiral Supermultiplet}
\label{s:tCS}
Restoring the original component superfields $\Tw\cF=\vdt\Tw{F}=-\sD$ and $\Tw\cG=\vdt\Tw{G}=\vdt A_1$, the system\eq{e:1DVSt3} becomes:
\begin{equation}\mkern-20mu
  \begin{array}{@{} c|cccc|cccc @{}}
 \text{\bsf tCS} &\bm{\Tw A} &\bm{\Tw B} &\bm{\Tw\cF} &\bm{\Tw\cG} 
  &\bm{\Tw\j_1} &\bm{\Tw\j_2} &\bm{\Tw\j_3} &\bm{\Tw\j_4} \\ 
    \toprule
\bm{\C1{\rD_1}} & \Tw\j_1 & -\Tw\j_4 & \vdt\Tw\j_2 & \vdt\Tw\j_3
    &i\vdt \Tw{A} &i\Tw\cF & i\Tw\cG &-i\vdt \Tw{B} \\ 
\bm{\C2{\rD_2}} & \Tw\j_2 &-\Tw\j_3 &-\vdt\Tw\j_1 & -\vdt\Tw\j_4
    &-i\Tw\cF & i\vdt \Tw{A} &-i\vdt \Tw{B} &-i\Tw\cG \\ 
\bm{\C3{\rD_3}} & \Tw\j_3 & \Tw\j_2 & \vdt\Tw\j_4 & -\vdt\Tw\j_1
    &-i\Tw\cG & i\vdt \Tw{B} & i\vdt \Tw{A} &i\Tw\cF \\ 
\bm{\C4{\rD_4}} &\Tw\j_4 &\Tw\j_1 &-\vdt\Tw\j_3 & \vdt\Tw\j_2
    &i\vdt \Tw{B} & i\Tw\cG &-i\Tw\cF & i\vdt \Tw{A} \\
    \bottomrule
  \end{array}
 \qquad
 \vC{\begin{picture}(40,35)(0,-2)
   \put(0,0){\includegraphics[height=31mm]{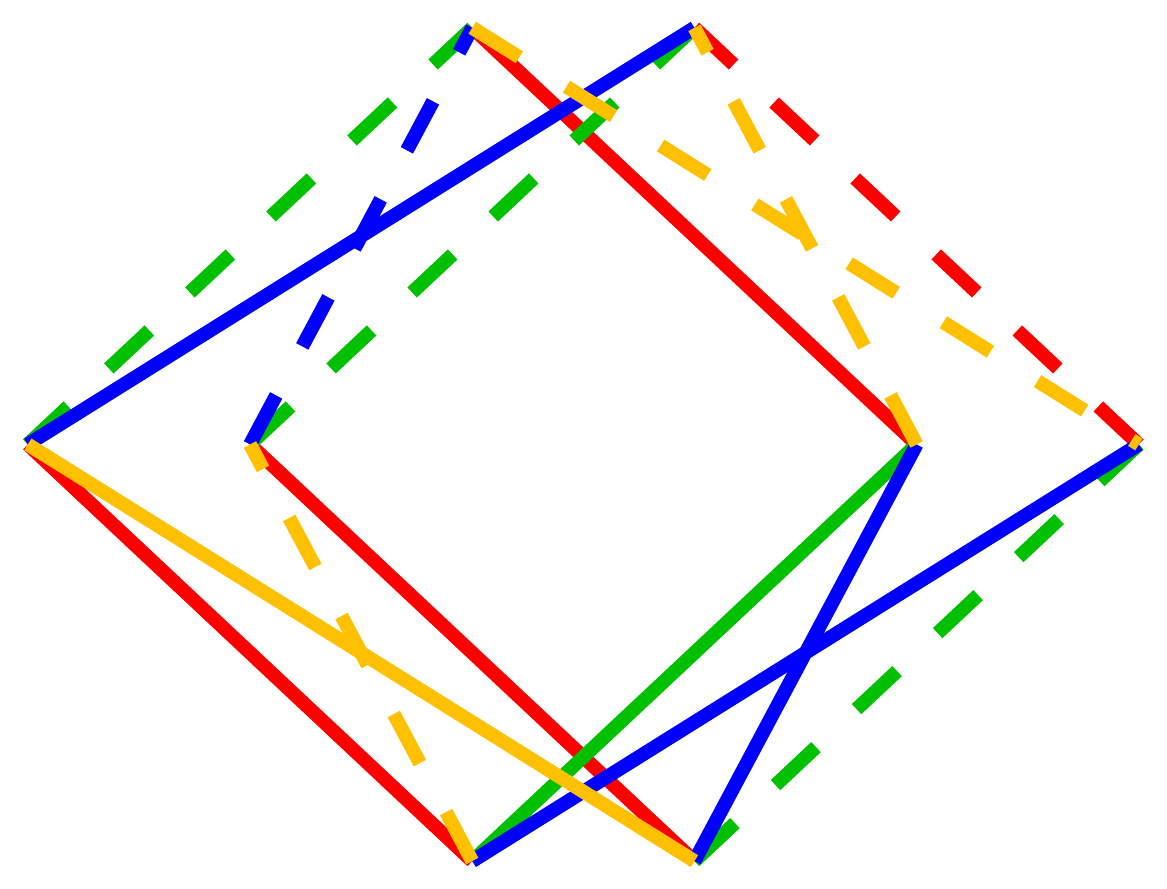}}
    \put(16,0){\cB{$\Tw A$}}
    \put(25,0){\cB{$\Tw B$}}
    \put(1,14){\bB{$\Tw\j_1$}}
    \put(11,14){\bB{$-\Tw\j_4$}}
    \put(30,14){\bB{$\Tw\j_2$}}
    \put(39,14){\bB{$\Tw\j_3$}}
    \put(16,29){\cB{$\Tw\cF$}}
    \put(25,29){\cB{$-\Tw\cG$}}
 \end{picture}}
 \label{e:tCS1D}
\end{equation}
whereas the\eq{e:-}-``redefined'' twisted-chiral supermultiplet is:
\begin{equation}\mkern-20mu
  \begin{array}{@{} c|cccc|cccc @{}}
\ha{\text{\bsf tCS}} &\bm{\Tw A} &\bm{\Tw B} &\bm{\Tw\cF} &\bm{\Tw\cG}
                &\bm{\Tw\j_1} &\bm{\Tw\j_2} &\bm{\ha\j_3} &\bm{\Tw\j_4} \\ 
    \toprule
\C1{\bm{\rD_1}} & \Tw\j_1 &-\Tw\j_4 & \vdt\Tw\j_2 &-\vdt\ha\j_3
                & i\vdt\Tw{A} & i\Tw\cF &-i\Tw\cG &-i\vdt\Tw{B} \\ 
\C2{\bm{\rD_2}} & \Tw\j_2 & \ha\j_3 &-\vdt\Tw\j_1 &-\vdt\Tw\j_4
                &-i\Tw\cF & i\vdt\Tw{A} & i\vdt\Tw{B} &-i\Tw\cG \\ 
\C3{\bm{\rD_3}} &-\ha\j_3 & \Tw\j_2 & \vdt\Tw\j_4 &-\vdt\Tw\j_1
                &-i\Tw\cG & i\vdt\Tw{B} &-i\vdt\Tw{A} & i\Tw\cF \\ 
\C4{\bm{\rD_4}} & \Tw\j_4 & \Tw\j_1 & \vdt\ha\j_3 & \vdt\Tw\j_2
                & i\vdt\Tw{B} & i\Tw\cG & i\Tw\cF & i\vdt\Tw{A} \\ 
    \bottomrule
  \end{array}
 \qquad
 \vC{\begin{picture}(40,35)(0,-2)
   \put(0,0){\includegraphics[height=31mm]{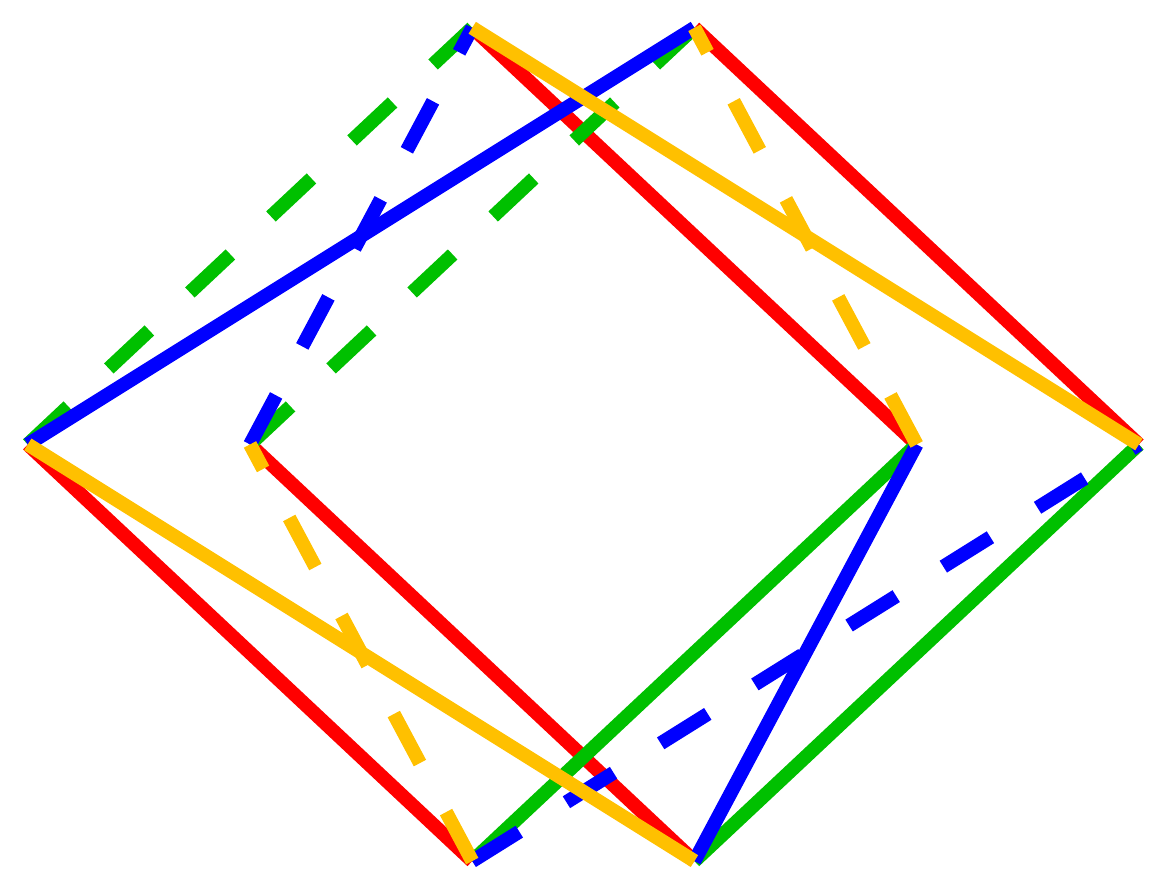}}
    \put(16,0){\cB{$\Tw A$}}
    \put(25,0){\cB{$\Tw B$}}
    \put(1,14){\bB{$\Tw\j_1$}}
    \put(11,14){\bB{$-\Tw\j_4$}}
    \put(30,14){\bB{$\Tw\j_2$}}
    \put(39,14){\bB{$\ha\j_3$}}
    \put(16,29){\cB{$\Tw\cF$}}
    \put(25,29){\cB{$-\Tw\cG$}}
 \end{picture}}
 \label{e:tCSt1D}
\end{equation}
Perhaps more so that comparing the two side-by-side Adinkras in\eq{e:vtCS-A}, a comparison of\eq{e:tCSt1D} with\eq{e:tCS1D} and\eq{e:CS1D} 

The bosonic holoraumy generators for both versions of the twisted-chiral supermultiplet,\eq{e:tCS1D} and\eq{e:tCSt1D}, are no longer all uniform compositions with $\vdt$:
\begin{subequations}
 \label{e:tCSR[t]}
\begin{gather}
  \ha\sB_{12}^{\sss\text{(tCS)}} =+\ha\sB_{34}^{\sss\text{(tCS)}}
  =\bM{0&0&\!\!-1&0\\ 0&0&0&1\\ \vdt^2&0&0&0\\ 0&\!\!-\vdt^2&0&0\\ },\qquad
  \ha\sB_{31}^{\sss\text{(tCS)}} =+\ha\sB_{24}^{\sss\text{(tCS)}}
  =\bM{0&0&0&1\\ 0&0&1&0\\ 0&\!\!-\vdt^2&0&0\\ \!-\vdt^2&0&0&0\\ },
\intertext{while}
  \ha\sB_{23}^{\sss\text{(tCS)}} =+\ha\sB_{14}^{\sss\text{(tCS)}}
  =\bM{0&1&0&0\\ \!-1&0&0&0\\ 0&0&0&1\\ 0&0&\!\!-1&0\\ }\circ\vdt,
 \label{e:R[0]tCS}
\end{gather}
\end{subequations}
remains a uniform composition with $\vdt$. In turn, the fermionic holoraumy generators remain uniform compositions, and $\sF_{IJ}^{\sss\text{(tCS)}}$ equal\eq{e:tCS-F} while $\sF_{IJ}^{\sss\HCS}$ equal\eq{e:tCSt-F}. To summarize:
\begin{equation}
  \begin{array}{@{} rr@{\>\to\>}l@{\>\too\h\>}l @{}}
 \text{\bsf Twisted-Chiral} &\MC2l{\text{\bsf Algebraic Subset of Holoraumy}}
                     &\text{\bsf cf.\ Table~\ref{t:16}} \\ 
    \toprule
 \text{original\eq{e:tCS1D}}
 &\SO(4)_{\sss B}\,{\times}\,\SO(4)_{\sss F} 
 &\rOp{U}(1)_{\sss B}\,{\times}\,\SU(2)_{\sss F}
 &\rOp{U}(1)_\rot\,{\times}\,\SU(2)_\exR\\ 
 \text{valise\eq{e:vtCS1D}}
 &\SO(4)_{\sss B}\,{\times}\,\SO(4)_{\sss F} 
 &\SU(2)_{\sss B}\,{\times}\,\SU(2)_{\sss F} 
 &\SU(2)_\rot\,{\times}\,\SU(2)_\exR\\ 
    \midrule
 \text{original\eq{e:tCSt1D}}
 &\SO(4)_{\sss B}\,{\times}\,\SO(4)_{\sss F} 
 &\rOp{U}(1)_{\sss B}\,{\times}\,\SU(2)_{\sss F}
 &\SU(2)_\rot\\ 
 \text{valise\eq{e:vtCSt1D}}
 &\SO(4)_{\sss B}\,{\times}\,\SO(4)_{\sss F} 
 &\SU(2)_{\sss B}\,{\times}\,\SU(2)_{\sss F} 
 &\SU(2)_\rot\\ 
    \bottomrule
  \end{array}
 \label{e:tCS1Dc}
\end{equation}
This clearly distinguishes the ``original''\cite{rTwSJG0} twisted-chiral supermultiplet\eq{e:cpxv2tCSa} and its ``redefined''\cite{r6-3,r6-3.1,rFIL} twisted-chiral supermultiplet\eq{e:cpxv2tCSa}. A comparison with\eq{e:CST} shows that the chiral supermultiplet\eq{e:CS1D} and the ``redefined'' twisted-chiral supermultiplet\eq{e:tCSt1D} have isomorphic holoraumy groups, but of course are differentiated by cycle parity, \ie, $\c_o$; compare the discussions after Eqs.\eq{e:CSDD} and\eq{e:tCSDD}.

These two supermultiplets are transformed into each other by means of a fermion redefinition such as\eq{e:-} that has a negative determinant. Such a redefinition of fermions is compatible with Lorentz transformations only on the world-line and on the world-sheet, where the Lorentz groups are abelian, $\ZZ_2$ and $\Spin(1,1)\approx\IR^*$, respectively\cite{rFRH}. It follows that this feature may be used to distinguish world-sheet reductions of supermultiplets from $n{+}1$-dimensional spacetime with $n\geqslant2$: Starting with $2{+}1$-dimensional spacetime, minimal spinors have an even number of real components so that Lorentz-compatible sign change transformations cannot have negative determinants.

\section{Higher-Dimensional Holoraumy}
\label{s:HoloSS}
Superspace methods\cite{r1001,rBK} are applicable without loss of generality\cite{rHTSSp08} in all discussions of supersymmetry and in spacetimes of all dimensions, albeit the formalism and especially the off-shell representations are completely understood only for a limited number of supersymmetries, $N\leqslant8$\cite{rGLPR}. It is nevertheless possible to extend the definitions of holoraumy from Section~\ref{s:HoloR} to higher spacetime dimensions, as follows:
\begin{enumerate}\itemsep=-3pt\vspace*{-3mm}
 \item Start with the simple fact that every supermultiplet consists of bosonic and fermionic components, and represent each of them by a Salam-Strathdee intact superfield the lowest component of which equals the particular component field.
 \item The supermultiplet is then represented by a first order super-differential coupled system
\begin{equation}
 \begin{aligned}
    D_\a (\text{bosons})   &= (\text{fermions + derivatives thereof}),\\
    D_\a (\text{fermions}) &= (\text{bosons + derivatives thereof})
 \end{aligned}
\end{equation}
where the expressions on the right-hand side of course contain also ordinary spacetime derivatives of the indicated components, and the ``$D_\a$'' symbolically denote the super-differential operators closing the desired supersymmetry algebra of the chosen type (Poincar\'e, super-conformal, with central charges, \etc), and in the spacetime of chosen dimensions and signature.
 \item Construct the enveloping super-differential system as in\eq{e:ValEnv}, and define:
\begin{equation}
 \ha\sB_{\a\b} (\text{bosons})
  \Defl [\,D_\a\,,\,D_\b\,] (\text{bosons}), \qquad
 \ha\sF_{\a\b} (\text{fermions})
  \Defl [\,D_\a\,,\,D_\b\,] (\text{fermions}),
\end{equation}
and so on, obtaining thus the complete action of $\wedge^*\Span(D_\a)$ on the chosen supermultiplet. 
\end{enumerate}\vspace*{-2mm}
This specifies the action of the complete hierarchy of holoraumy operators on the considered supermultiplet. In full generality, these operators do not form a group, but an $\IR[\N]$ (or $\IC[\N]$) module, where $\N$ stands for the list of (bosonic) spacetime derivatives, including all types of gauge-covariant derivatives, as needed.

It is straightforward that the action of so-defined quadratic holoraumy operators $\ha\sB_{\a\b}$ and $\ha\sF_{\a\b}$ (as well as their higher-order analogues) generally take the form of non-uniform {\em\/compositions\/} of field-space homogeneous linear transformations and domain-space transformations, as is the case in\eq{e:CSR[t]},\eq{e:VSR[t]} and\eq{e:tCSR[t]}. The domain-space transformations may well include the inhomogeneous transformations of spacetime translations, but possibly also the other bosonic generators from the considered supersymmetry algebra.

\paragraph{Holoraumy Invariants:}
Since the holoraumy operators $\ha\sB_{\a\b}$ and $\ha\sF_{\a\b}$ are generally non-uniform compositions of field-space and domain-space transformations, it is not possible to factorize them into a uniform composition of tensors such as $\sB_{IJ}$ and $\sF_{IJ}$ in\eqs{e:QHT}{e:HBF} with a purely field-space action on one hand, and domain-space differential operators on the other. They naturally generate diverse elements in the $\IR[\N]$ (or $\IC[\N]$) modules mentioned above.

However---{\em\/on a case-by-case basis\/} and in a {\em\/representation-dependent way\/}:
  \begin{enumerate}\itemsep=-3pt\vspace*{-3mm}
    \item Each supermultiplet will have a characteristic restriction of the holoraumy operators to appropriate subsets of components (\eg, those of the same engineering dimension), which factorizes each so-restricted subset of holoraumy operators uniformly, as is the case within\eq{e:R[0]CS} within\eq{e:CSR[t]}, \eq{e:R[0]VS} within\eq{e:VSR[t]} and \eq{e:R[0]tCS} within\eq{e:tCSR[t]}.
    \item The reduction of the holoraumy operators to the (on-shell) physical degrees of freedom within every supermultiplet will factorize uniformly, for dimensional reasons.
    \item The so-restricted (so-reduced) holoraumy operators will span an algebraic structure that is constrained by the process of restriction (reduction), and form substructures of the $\IR[\N]$ (or $\IC[\N]$) modules mentioned above.
  \end{enumerate}\vspace*{-2mm}
The {\em\/representation-dependent\/} structure obtained in this way then provides the foundation for constructing {\em\/representation-dependent\/} invariants generalizing\eqs{e:234}{e:charB}. Within the restricted (reduced) subset of holoraumy invariants that {\em\/do\/} factorize, these invariants are still numerical. When computed with the full, un-restricted (un-reduced) holoraumy operators, these invariants are specific spacetime operators, characteristic of the particular representation for which they were calculated.

It is this {\em\/representation-dependent\/} and highly hierarchical holoraumy structure and its computable invariants that we hope will facilitate in classifying and identifying supermultiplets.

\paragraph{SuSy-Holography:} This brings us to an important topic, having to do with specifying which component (super)field redefinitions are regarded as equivalence relations when assessing whether or not two supermultiplets are to be regarded as {\em\/inequivalent\/} or {\em\/equivalent.\/}
 In turn, this is related to the question whether or not a particular supermultiplet from a spacetime of one dimension is the dimensional reduction of a supermultiplet from a spacetime of a larger dimension.

For example, the particular field redefinition\eq{e:-} may certainly be regarded as an equivalence relation when discussing world-line supermultiplets within the context of world-line models all by themselves. This is indeed the framework of Refs.\cite{rPT,rT01,rT01a,rCRT,rKRT,r6-1,rKT07,r6-3,r6-3.2,r6-1.2,rGKT10,r6-3.1,rDHIL13}, wherein {\em\/real,\/} finite-dimensional, unitary and off-shell supermultiplets are classified up to all component field redefinitions. This includes the negative-determinant ones that include\eq{e:-}, which are---within this purely world-line framework---also to be regarded as equivalence relations. 

However, the determination whether or not two supermultiplets are to be regarded as (in)equi\-valent changes considerably once additional structure is included, and is showcased already within the simple examples of Section~\ref{s:3LPs} as summarized in Table~\ref{t:4+4}.
\begin{table}[ht]
\small
$$
  \begin{array}{@{} c|c|c@{~}c@{~}c|c@{~}c@{~}c|c|c|c @{}}
 &\bm{\c_o} &\bm{\sB_{12}} &\bm{\sB_{23}} &\bm{\sB_{31}}
            &\bm{\sF_{\!12}} &\bm{\sF_{\!23}} &\bm{\sF_{\!31}}
 & \text{\bsf field-space} & \bm{\sB_{\!IJ}\simeq\sF_{\!IJ}}
 & \bm{\h(\sB_{\!IJ}\,{\oplus}\,\sF_{\!IJ})}\\ 
    \toprule
 \text{\bsf vCS}  &+1 &-\g^{12} &+\g^{13} &-\g^{23} &-\g^{13} &+\g^{23}   &-\g^{12}
 & (\IC^2|\IC^2)
 & \big({{\SSS12},\,{\SSS23},\,{\SSS31}\atop\7{\SSS23},\,\7{\SSS31},\,{\SSS12}}\big)
 & \su(2)=\spin(3)_\rot \\*[1mm] 
 \text{\bsf v}\ha{\text{\bsf tCS}}
                  &-1 &-\g^{12} &-\g^{13} &+\g^{23} &-\g^{13} &-\g^{23}   &+\g^{12}
 & (\IC^2|\IC^2)
 & \big({{\SSS12},\,{\SSS23},\,{\SSS31}\atop{\SSS23},\,\7{\SSS31},\,\7{\SSS12}}\big)
 & \su(2)=\spin(3)_\rot \\ 
    \midrule
 \text{\bsf vtCS} &-1 &-\g^{12} &-\g^{13} &+\g^{23} &+\g^{0}  &+\g^{0123} &-\g^{123}
 & (\IC^2|\IC{\oplus}\IC^*) & \text{---}
 & \spin(3)^{\sss B}_\rot{\oplus}\spin(3)^{\sss F}_\exR \\*[1mm]
 \text{\bsf vVS}  &-1 &+\g^{12} &+\g^{23} &+\g^{13} &-\g^{0}  &+\g^{0123} &-\g^{123}
 & (\IR^4|\IR^4) & \text{---}
 & \spin(3)^{\sss B}_\rot{\oplus}\spin(3)^{\sss F}_\exR \\*[1mm]
 \text{\bsf vTS}  &-1 &+\g^{23} &+\g^{12} &+\g^{13} &+\g^{0}  &-\g^{0123} &-\g^{123}
 & (\IR^4|\IR^4) & \text{---}
 & \spin(3)^{\sss B}_\rot{\oplus}\spin(3)^{\sss F}_\exR \\
    \bottomrule
  \end{array}
$$
\caption{Some of the key computational results from Section~\ref{s:3LPs}}
\label{t:4+4}
\end{table}
 
For one, the middle column in Table~\ref{t:4+4} clearly indicates the difference in the supersymmetric complex structures resulting from the field redefinition\eq{e:-} transforming the twisted-chiral supermultiplet, ``{\bsf vtCS}''\eq{e:1DVSt3} into its redefined form, ``$\text{\bsf v}\ha{\text{\bsf tCS}}$''\eq{e:vtCSt1D}; see also Sections~\ref{s:cpxS}, \ref{s:tCSv} and~\ref{s:tCS}. Thus, ``{\bsf vtCS}''\eq{e:1DVSt3} and ``$\text{\bsf v}\ha{\text{\bsf tCS}}$''\eq{e:vtCSt1D} are inequivalent {\em\/complex\/} supermultiplets.

This distinction of course remains when the component fields of these valise supermultiplets are redefined so as to avoid the obstruction to dimensional extension to 1+1-dimensional world-sheet\cite{rGH-obs}, as was done in\eq{e:tCS1D} and\eq{e:tCSt1D}, respectively. The holoraumy structure of the resulting supermultiplets continues to track this distinction in the supersymmetric complex structures of the two supermultiplets; see\eq{e:tCS1Dc} and Table~\ref{t:4+4}. Thus, ``{\bsf tCS}''\eq{e:tCS1D} and ``$\ha{\text{\bsf tCS}}$''\eq{e:tCSt1D} are inequivalent {\em\/complex\/} supermultiplets, both on the world-line and on the world-sheet.

In addition, in spacetimes where the total number of dimensions is at least three, all fermionic representations of the Lorentz group have an even number of components. Therefore, Lorentz-compatible fermionic field redefinitions cannot have negative determinants, and fermionic component field transformations such as\eq{e:-} cannot possibly be regarded as equivalence relations in such higher-dimensional spacetimes.

Denote by $\vr_0$ the coordinate world-line dimensional reduction, and let $M_i$ be the $i^\text{th}$ supermultiplet in any fixed spacetime $X$ of at least three dimensions and any (fixed) signature. Then, if $\vr_0(S_1)$ and $\vr_0(M_2)$ differ by a negative-determinant fermionic field redefinition, $M_1$ and $M_2$ are inequivalent in the higher-dimensional spacetime.
 In fact, if $\n_{\sss\sF}$ is a negative-determinant fermionic field redefinition, the world-line supermultiplet $\n_{\sss\sF}[\vr_0(M_1)]$ need not be the dimensional reduction of any supermultiplet in $X$.

We finally turn to an even subtler distinction that is detected by holoraumy:
 The vector supermultiplet\eq{e:vVS1Dt}, the tensor supermultiplet\eq{e:1DTSt} and the ``original'' twisted-chiral supermultiplet\eq{e:1DVSt3} have isomorphic holoraumy algebras.
The precise nature of the isomorphisms between these three pairs of holoraumy groups and algebras may be read off from the bottom three rows of the left-hand half of Table~\ref{t:4+4}. The distinctions are exhibited in Table~\ref{t:dets}, which specifies the isomorphisms in matrix form, with respect to the reference basis in Table~\ref{t:16}.
\begin{table}[htdp]
$$
  \begin{array}{@{} cccc @{}}
 & \pmb{\sB\too{b}\sB} & \pmb{\sF\too{f}\sF} & \bm{\big(\det(b),\det(f)\big)} \\ 
    \toprule
 \text{\bsf vVS\,$\to$\,vTS}:
  &\bM{0&1&0\\1&0&0\\0&0&1} &\bM{-1&0&0\\0&-1&0\\0&0&1} & (-1,+1) \\ 
 \text{\bsf vVS\,$\to$\,vtCS}:
  &\bM{-1&0&0\\0&0&-1\\0&1&0} &\bM{-1&0&0\\0&1&0\\0&0&1} & (-1,-1) \\ 
 \text{\bsf vTS\,$\to$\,vtCS}:
  &\bM{0&-1&0\\0&0&-1\\1&0&0} &\bM{1&0&0\\0&-1&0\\0&0&1} & (+1,-1) \\ 
    \bottomrule
  \end{array}
$$
 \caption{The isomorphism transformations $b\in\Aut(\sH_\sB)$ and$f\in\Aut(\sH_\sF)$}
 \label{t:dets}
\end{table}%
All of these isomorphisms involve odd (negative-determinant) signed permutations, although they are achieved by means of positive-determinant component field redefinitions; see\eq{e:VS>TS} and Appendix~\ref{a:tCS}.

\section{Conclusions}
 \label{s:Coda}
To summarize, we have defined, exhibited and discussed a novel super-differential algebraic structure inherent in all supermultiplets, which we dub the {\em\/holoraumy\/}. In full generality, it is a nontrivial composition of:
\begin{itemize}\itemsep=-3pt\vspace*{-3mm}
 \item homogeneous, linear transformations in the field-space
 \item inhomogeneous transformations in the domain space
\end{itemize}\vspace*{-2mm}
as realized on the component superfields spanning the specified representation of supersymmetry. This structure is revealed and is computable within the enveloping system of the system of super-differential equations that specify the supersymmetry transformations within the supermultiplet.

As a proof of concept, we have explored this structure within the simple framework of world-line $N$-extended supersymmetry without central charges, and have studied it in the illustrative cases of several well-known 3+1-dimensional supermultiplets, dimensionally reduced to the coordinate time world-line. The results may be gleaned from Tables~\ref{t:4+4} and~\ref{t:dets}.

The holoraumy tensors, their algebras and the groups they generate detect the following properties of supermultiplets:
\begin{enumerate}\itemsep=-3pt\vspace*{-3mm}
 \item The ``(un)twistedness'' of minimal supermultiplets (agreeing with the characteristic $\c_o$\cite{rUMD09-1} and the cycle parity\cite{rH-TSS}), {\em\/via\/} the ``minimality relations,'' such as\eq{e:vCS-pr}, (\ref{e:vVS-pr}) and\eq{e:vTS-pr}.
 \item The supersymmetric complex structure, by means of the even (positive-determinant) signed permutation isomorphism $\h(\sB_{\!IJ})\approx\h(\sF_{\!IJ})$---and lack thereof; see Table~\ref{t:4+4}.
 \item The subtler distinction between the valise versions of the world-line dimensional reductions of all three of the ({\it a})~vector and ({\it b})~tensor supermultiplets in Wess-Zumino gauge and ({\it c})~the ``original'' twisted-chiral supermultiplet of Ref.\cite{rTwSJG0}.
\end{enumerate}

Although the number of examples we have examined is relatively small, we find it telling that the structure of even just the quadratic holoraumy tensors is in fact sufficient to distinguish every one of them. Having introduced this new concept and having described its features in considerable computational detail in the case of a handful of well-known supermultiplets, we conclude that holoraumy ought to be further explored, and from both aspects:
\begin{enumerate}\itemsep=-3pt\vspace*{-3mm}
 \item both computationally, by dimensionally reducing all other known supermultiplets to the world-line from the various spacetimes wherein they are defined and assessing the utility of holoraumy in telling inequivalent representations apart,
 \item as well as by more general means of rigorous mathematics.
\end{enumerate}\vspace*{-2mm}
We are looking forward to both of these future efforts.

\bigskip\bigskip
\paragraph{\bfseries Acknowledgments:}
We would like to thank Alison Fink for detailed calculations of a matrix algebra closely related to\eq{e:Dirac},
 and Kevin Iga for helpful discussions on several aspects of the work presented here.
 SJG's research was partially supported by the National Science Foundation 
grants PHY-0652983 and PHY-0354401. This research was also 
supported in part the University of Maryland Center for String \& Particle Theory. 
 TH is grateful to Department of Physics, University of Central Florida, Orlando FL, and the Physics Department of the Faculty of Natural Sciences of the University of Novi Sad, Serbia, for recurring hospitality and resources.

\newpage
\appendix
\section{Matrix Conventions}
 \label{a:M}
Throughout, we have used the $\g$-matrices as defined in Ref.\cite{rUMD09-1}, but have clearly indicated the results that are independent of any such choice; these basis-independent results were also confirmed by repeating the calculations using a handful of different $\g$-matrix bases. In particular, the four $\g$-matrices associated with the 3+1-dimensional spacetime are chosen as
\begin{equation}
  (\g^0)_a{}^b=i[\s^3\,{\otimes}\,\s^2]_a{}^b,~~~
  (\g^1)_a{}^b=[\Ione\,{\otimes}\,\s^1]_a{}^b,~~~
  (\g^2)_a{}^b=[\s^2\,{\otimes}\,\s^2]_a{}^b,~~~
  (\g^3)_a{}^b=[\Ione\,{\otimes}\,\s^3]_a{}^b,
\end{equation}
which clearly implies that all elements of the Dirac algebra are real, \ie, we work in a Majorana representation. The fact that $(\g^0)^2=-\Ione$ while $(\g^m)^2=\Ione$ for $m=1,2,3$ implies that we use the $\h_{\m\n}=\textrm{diag}(-1,1,1,1)$ metric. The $\g^5$-matrix
\begin{equation}
  (\g^5)_a{}^b \Defl i\g^{0123} \Defl i\g^0\g^1\g^2\g^3 = -(\s^1\,{\otimes}\,\s^2)_a{}^b
\end{equation}
is then purely imaginary and antisymmetric. As usual, we employ the fact that the vector space of the Clifford-Dirac algebra is isomorphic to the exterior algebra
\begin{equation}
  \Cl(1,3)=\big(\otimes^*\Span(\g^\m)\big)\big/\big(\{\g^\m,\g^\n\}=2\h^{\m\n}\Ione\big)
  ~\approx~ \wedge^*\Span(\g^\m),
\end{equation}
and use the weighted antisymmetrized products
\begin{equation}
  \g^{\m\n}\Defl\inv2[\g^\m,\g^\n],\quad
  \g^{\m\n\r}\Defl\inv3\big(\g^{\m\n}\g^\r+\g^{\n\r}\g^\m+\g^{\r\m}\g^\n\big),\quad
  \etc
\end{equation}
as the induced basis elements. For convenience, these matrices are tabulated in Table~\ref{t:16}.
\begin{table}[htp]
\caption{A basis of sixteen real invertible matrices\protect\cite{rUMD09-1}, spanning $\wedge^*\Span(\g^0,\g^1,\g^2,\g^3)$, shown both in terms of the outer product of the usual $2\,{\times}\,2$ identity and Pauli matrices, and also explicitly}\vspace*{-4mm}
$$
  \begin{array}{@{} c@{~}c|c@{~}c|c@{~}c|c@{~}c @{}}
 \toprule
  \mM{\ttt(\Ione)_a{}^b\\[1mm]\ttt [\Ione\,{\otimes}\,\Ione]}
  & \Bm{1&0&0&0\\ 0&1&0&0\\ 0&0&1&0\\ 0&0&0&1\\ }
& \mM{\ttt(\g^2)_a{}^b\\[1mm]\ttt [\s^2\,{\otimes}\,\s^2]}
  & \Bm{0&0&0&-1\\ 0&0&1&0\\ 0&1&0&0\\ -1&0&0&0\\ }
& \mM{\ttt(\g^3)_a{}^b\\[1mm]\ttt [\Ione\,{\otimes}\,\s^3]}
  & \Bm{1&0&0&0\\ 0&-1&0&0\\ 0&0&1&0\\ 0&0&0&-1\\ }
& \mM{\ttt(\g^{23})_a{}^b\\[1mm]\ttt i[\s^2\,{\otimes}\,\s^1]}
  & \Bm{0&0&0&1\\ 0&0&1&0\\ 0&-1&0&0\\ -1&0&0&0\\ } \\*[3mm]
  \midrule
  \mM{\ttt(\g^0)_a{}^b\\[1mm]\ttt i[\s^3\,{\otimes}\,\s^2]}
  & \Bm{0&1&0&0\\ -1&0&0&0\\ 0&0&0&-1\\ 0&0&1&0\\ }
& \mM{\ttt(\g^{02})_a{}^b\\[1mm]\ttt [\s^1\,{\otimes}\,\Ione]}
  & \Bm{0&0&1&0\\ 0&0&0&1\\ 1&0&0&0\\ 0&1&0&0\\ }
& \mM{\ttt(\g^{03})_a{}^b\\[1mm]\ttt-[\s^3\,{\otimes}\,\s^1]}
  & \Bm{0&-1&0&0\\ -1&0&0&0\\ 0&0&0&1\\ 0&0&1&0\\ }
& \mM{\ttt(\g^{023})_a{}^b\\[1mm]\ttt [\s^1\,{\otimes}\,\s^3]}
  & \Bm{0&0&1&0\\ 0&0&0&-1\\ 1&0&0&0\\ 0&-1&0&0\\ } \\*[3mm]
  \midrule
  \mM{\ttt(\g^1)_a{}^b\\[1mm]\ttt [\Ione\,{\otimes}\,\s^1]}
  & \Bm{0&1&0&0\\ 1&0&0&0\\ 0&0&0&1\\ 0&0&1&0\\ }
& \mM{\ttt(\g^{12})_a{}^b\\[1mm]\ttt i[\s^2\,{\otimes}\,\s^3]}
  & \Bm{0&0&1&0\\ 0&0&0&-1\\ -1&0&0&0\\ 0&1&0&0\\ }
& \mM{\ttt(\g^{13})_a{}^b\\[1mm]\ttt -i[\Ione\,{\otimes}\,\s^2]}
  & \Bm{0&-1&0&0\\ 1&0&0&0\\ 0&0&0&-1\\ 0&0&1&0\\ }
& \mM{\ttt(\g^{123})_a{}^b\\[1mm]\ttt i[\s^2\,{\otimes}\,\Ione]}
  & \Bm{0&0&1&0\\ 0&0&0&1\\ -1&0&0&0\\ 0&-1&0&0\\ } \\*[3mm]
  \midrule
  \mM{\ttt(\g^{01})_a{}^b\\[1mm]\ttt [\s^3\,{\otimes}\,\s^3]}
  & \Bm{1&0&0&0\\ 0&-1&0&0\\ 0&0&-1&0\\ 0&0&0&1\\ }
& \mM{\ttt(\g^{012})_a{}^b\\[1mm]\ttt-[\s^1\,{\otimes}\,\s^1]}
  & \Bm{0&0&0&-1\\ 0&0&-1&0\\ 0&-1&0&0\\ -1&0&0&0\\ }
& \mM{\ttt(\g^{013})_a{}^b\\[1mm]\ttt [\s^3\,{\otimes}\,\Ione]}
  & \Bm{1&0&0&0\\ 0&1&0&0\\ 0&0&-1&0\\ 0&0&0&-1\\ }
& \mM{\ttt(\g^{0123})_a{}^b\\[1mm]\ttt i[\s^1\,{\otimes}\,\s^2]}
  & \Bm{0&0&0&1\\ 0&0&-1&0\\ 0&1&0&0\\ -1&0&0&0\\ } \\*[3mm]
 \bottomrule
  \end{array}
$$
  \label{t:16}
\end{table}

On the space of Majorana, real 4-component spinors, we choose the metric
\begin{equation}
  C_{ab}\Defl-i[\s^3\,{\otimes}\,\s^2]_{ab},\qquad
  C_{ab}=-C_{ba},
\end{equation}
and which numerically equals $-(\g^0)=(\g^0)^{-1}$. Therefore, $(\g^0)_{ab}=(\g^0)_a{}^c\,C_{cb}=\d_{ab}$. In turn, the inverse spinorial metric is defined by the condition $C_{ac}C^{ab}=\d_c{}^b$.
\begin{table}[htp]
\caption{The ``lowered'' basis for $(\g_{ac}\protect\Defl(\g^*)_a{}^b\,C_{bc}$, with $[C]=[-\g^0]=[\g^0]^{-1}=-i[\s^3\,{\otimes}\,\s^2]$}\vspace*{-4mm}
$$
  \begin{array}{@{} c@{~}c|c@{~}c|c@{~}c|c@{~}c @{}}
 \toprule
  \mM{\ttt C_{ab}\\[1mm]\ttt -i[\s^3\,{\otimes}\,\s^2]}
  & \Bm{0&-1&0&0\\ 1&0&0&0\\ 0&0&0&1\\ 0&0&-1&0\\ }
& \mM{\ttt(\g^2)_{ab}\\[1mm]\ttt [\s^1\,{\otimes}\,\Ione]}
  & \Bm{0&0&1&0\\ 0&0&0&1\\ 1&0&0&0\\ 0&1&0&0\\ }
& \mM{\ttt(\g^3)_{ab}\\[1mm]\ttt-[\s^3\,{\otimes}\,\s^1]}
  & \Bm{0&-1&0&0\\ -1&0&0&0\\ 0&0&0&1\\ 0&0&1&0\\ }
& \mM{\ttt(\g^{23})_{ab}\\[1mm]\ttt-[\s^1\,{\otimes}\,\s^3]}
  & \Bm{0&0&-1&0\\ 0&0&0&1\\ -1&0&0&0\\ 0&1&0&0\\ } \\*[3mm]
  \midrule
  \mM{\ttt (\g^0)_{ab}\\[1mm]\ttt [\Ione\,{\otimes}\,\Ione]}
  & \Bm{1&0&0&0\\ 0&1&0&0\\ 0&0&1&0\\ 0&0&0&1\\ }
& \mM{\ttt(\g^{02})_{ab}\\[1mm]\ttt-[\s^2\,{\otimes}\,\s^2]}
  & \Bm{0&0&0&1\\ 0&0&-1&0\\ 0&-1&0&0\\ 1&0&0&0\\ }
& \mM{\ttt(\g^{03})_{ab}\\[1mm]\ttt-[\Ione\,{\otimes}\,\s^3]}
  & \Bm{-1&0&0&0\\ 0&1&0&0\\ 0&0&-1&0\\ 0&0&0&1\\ }
& \mM{\ttt(\g^{023})_{ab}\\[1mm]\ttt i[\s^2\,{\otimes}\,\s^1]}
  & \Bm{0&0&0&1\\ 0&0&1&0\\ 0&-1&0&0\\ -1&0&0&0\\ } \\*[3mm]
  \midrule
  \mM{\ttt(\g^1)_{ab}\\[1mm]\ttt [\s^3\,{\otimes}\,\s^3]}
  & \Bm{1&0&0&0\\ 0&-1&0&0\\ 0&0&-1&0\\ 0&0&0&1\\ }
& \mM{\ttt(\g^{12})_{ab}\\[1mm]\ttt [\s^1\,{\otimes}\,\s^1]}
  & \Bm{0&0&0&1\\ 0&0&1&0\\ 0&1&0&0\\ 1&0&0&0\\ }
& \mM{\ttt(\g^{13})_{ab}\\[1mm]\ttt-[\s^3\,{\otimes}\,\Ione]}
  & \Bm{-1&0&0&0\\ 0&-1&0&0\\ 0&0&1&0\\ 0&0&0&1\\ }
& \mM{\ttt(\g^{123})_{ab}\\[1mm]\ttt i[\s^1\,{\otimes}\,\s^2]}
  & \Bm{0&0&0&1\\ 0&0&-1&0\\ 0&1&0&0\\ -1&0&0&0\\ } \\*[3mm]
 \midrule
  \mM{\ttt(\g^{01})_{ab}\\[1mm]\ttt-[\Ione\,{\otimes}\,\s^1]}
  & \Bm{0&-1&0&0\\ -1&0&0&0\\ 0&0&0&-1\\ 0&0&-1&0\\ }
& \mM{\ttt(\g^{012})_{ab}\\[1mm]\ttt i[\s^2\,{\otimes}\,\s^3]}
  & \Bm{0&0&1&0\\ 0&0&0&-1\\ -1&0&0&0\\ 0&1&0&0\\ }
& \mM{\ttt(\g^{013})_{ab}\\[1mm]\ttt -i[\Ione\,{\otimes}\,\s^2]}
  & \Bm{0&-1&0&0\\ 1&0&0&0\\ 0&0&0&-1\\ 0&0&1&0\\ }
& \mM{\ttt(\g^{0123})_{ab}\\[1mm]\ttt-i[\s^2\,{\otimes}\,\Ione]}
  & \Bm{0&0&-1&0\\ 0&0&0&-1\\ 1&0&0&0\\ 0&1&0&0\\ } \\*[3mm]
  \bottomrule
  \end{array}
$$
  \label{t:160}
\end{table}
Upon lowering the second index using $C_{ab}$, the matrices in Table~\ref{t:16} turn into those presented in Table~\ref{t:160}. Note that:
\begin{equation}
  (\g^{[p]}\Defl\g^{\m_1\cdots\m_p})_{ab}=(-1)^{\binom{p-1}2}(\g^{\m_1\cdots\m_p})_{ba},\qquad
  (\g^{[0]})_{ab}\Defl C_{ab},
\end{equation}
so that, in particular:
\begin{gather}
  C_{ab}=-C_{ba},\quad
  (\g^{\m\n\r})_{ab}=-(\g^{\m\n\r})_{ba},\quad
  (\g^{\m\n\r\s})_{ab}=-(\g^{\m\n\r\s})_{ba},\\
  (\g^{\m})_{ab}=+(\g^{\m})_{ba},\quad
  (\g^{\m\n})_{ab}=+(\g^{\m\n})_{ba}.
\end{gather}
Also, we have that:
\begin{alignat}9
  \g^5 &\Defl \frc{i}{4!}\ve_{\m\n\r\s}\,\g^{\m\n\r\s}=i\g^{0123},&\quad&\To&\quad
  (\g^5)_{ab} &= -(\g^5)_{ba},\\
  \g^5\g^\l &=\frc{i}{3!}\ve_{\m\n\r}{}^\s\,\g^{\m\n\r},&\quad&\To&\quad
  (\g^5\g^\l)_{ab} &=-(\g^5\g^\l)_{ba}.
\end{alignat}

\section{Twisted-Chiral Supermultiplet}
\label{a:tCS}
This is an abridged derivation of the world-line dimensional reduction of the ``original'' twisted-chiral supermultiplet\cite{rTwSJG0}, since we start from the world-line dimensional reduction of the vector supermultiplet in the Wess-Zumino gauge\eq{e:VS1D}.
 For reasons that will soon become clear, we define $\Tw\j_a\Defl(\g^2\l)_a$, so that:
\begin{equation}
 \begin{aligned}
  &(\Tw\j_1,\Tw\j_2,\Tw\j_3,\Tw\j_4)
     \Defl ({-}\l_4,\l_3,\l_2,{-}\l_1),\\*
  &\det\big[\,(\l_1,\l_2,\l_3,\l_4)\to({-}\l_4,\l_3,\l_2,{-}\l_1)\,\big]=+1.
 \end{aligned}
 \label{e:l=j}
\end{equation}
which turns\eq{e:1DVSt} into:
\begin{subequations}
\begin{equation}\mkern-20mu
  \begin{array}{@{} c|cccc|cccc @{}}
  &\bm{A_1} &\bm{A_2} &\bm{A_3} &\pmb{\sD}
  &\bm{\Tw\j_1} &\bm{\Tw\j_2} &\bm{\Tw\j_3} &\bm{\Tw\j_4} \\ 
    \toprule
\bm{\C1{\rD_1}} & \Tw\j_3 & \Tw\j_1 & -\Tw\j_4 &-\vdt\Tw\j_2
    &i\vdt A_2 &-i\sD & i\vdt A_1 &-i\vdt A_3 \\ 
\bm{\C2{\rD_2}} & -\Tw\j_4 & \Tw\j_2 &-\Tw\j_3 & \vdt\Tw\j_1
    &i\sD & i\vdt A_2 &-i\vdt A_3 &-i\vdt A_1 \\ 
\bm{\C3{\rD_3}} & -\Tw\j_1 & \Tw\j_3 & \Tw\j_2 &-\vdt\Tw\j_4
    &-i\vdt A_1 & i\vdt A_3 & i\vdt A_2 &-i\sD \\ 
\bm{\C4{\rD_4}} & \Tw\j_2 &\Tw\j_4 &\Tw\j_1 & \vdt\Tw\j_3
    &i\vdt A_3 & i\vdt A_1 & i\sD & i\vdt A_2 \\
    \bottomrule
  \end{array}
 \qquad
 \vC{\begin{picture}(35,35)(0,-2)
   \put(0,0){\includegraphics[height=31mm]{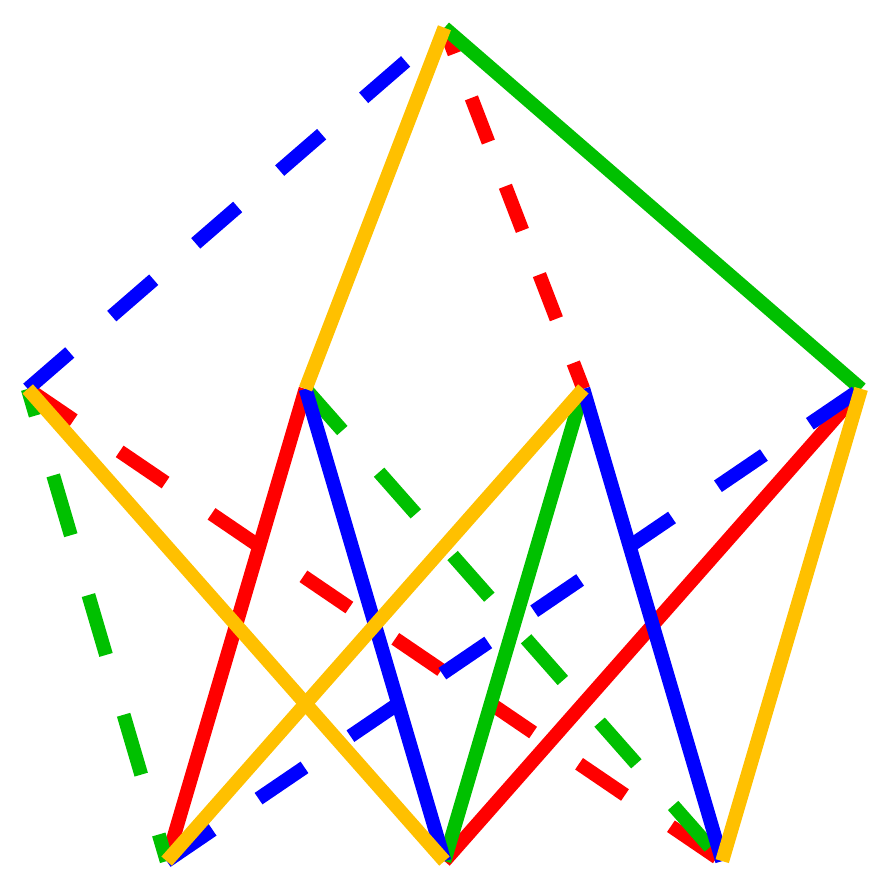}}
    \put(5,0){\cB{$A_1$}}
    \put(15,0){\cB{$A_2$}}
    \put(25,0){\cB{$A_3$}}
    \put(0,16){\bB{$\Tw\j_4$}}
    \put(10,16){\bB{$\Tw\j_3$}}
    \put(20,16){\bB{$\Tw\j_2$}}
    \put(30,16){\bB{$\Tw\j_1$}}
    \put(15,29){\cB{$\sD$}}
 \end{picture}}
 \label{e:1DVSt2}
\end{equation}
or, in tensorial notation:
\begin{align}
  \rD_a\,A_1&=(\g^{12})_a{}^b\,\Tw\j_b,\quad
  \rD_a\,A_2=\Tw\j_a,\quad
  \rD_a\,A_3=-(\g^{23})_a{}^b\,\Tw\j_b,\quad
  \rD_a\,\sD=(\g^{13})_a{}^b\,\vdt\Tw\j_b,\\
  \rD_a\,\Tw\j_b
            &=i(\g^{012})_{ab}\,(\vdt\,A_1) +i(\g^0)_{ab}\,(\vdt\,A_2) 
              -i(\g^{023})_{ab}\,(\vdt\,A_3) +i(\g^{013})_{ab}\,\sD.
\end{align}
\end{subequations}
Aiming to match the unsigned pattern of resulting fields in the transformation of the (now fixed) fermions in\eq{e:1DVSt2} to that found in\eq{e:vCS1Dt}, we define:
\begin{equation}
 \begin{aligned}
  &(\Tw{A},\Tw{B},\Tw{F},\Tw{G})
     \Defl(A_2,A_3,{-}(\ttt\int\!\rd\t\sD),A_1),\\
  &\det\big[\,(\l_1,\l_2,\l_3,\l_4)\to({-}\l_4,\l_3,\l_2,{-}\l_1)\,\big]=+1.
 \end{aligned}
 \label{e:ABFG}
\end{equation}
This turns\eq{e:1DVSt2} into the results\eq{e:vtCS1D} used above.

Analogously to\eq{e:CS1D}, we revert to the original component superfield $\Tw\cF=\vdt\Tw{F}=-\sD$, define $\Tw\cG=\vdt\Tw{G}=\vdt A_1$, and draw their nodes above the fermionic ones to indicate their relative engineering dimensions, $[\Tw\cF]=[\Tw\cG]=[\Tw\j_a]{+}\inv2=[\Tw{A}]{+}1=[\Tw{B}]{+}1$, the system\eq{e:vtCSt1D} becomes:
\begin{equation}\mkern-20mu
  \begin{array}{@{} c|cccc|cccc @{}}
\ha{\text{\bsf tCS}} &\bm{\Tw A} &\bm{\Tw B} &\bm{\Tw\cF} &\bm{\Tw\cG}
                &\bm{\Tw\j_1} &\bm{\Tw\j_2} &\bm{\ha\j_3} &\bm{\Tw\j_4} \\ 
    \toprule
\C1{\bm{\rD_1}} & \Tw\j_1 &-\Tw\j_4 & \vdt\Tw\j_2 &-\vdt\ha\j_3
                & i\vdt\Tw{A} & i\Tw\cF &-i\Tw\cG &-i\vdt\Tw{B} \\ 
\C2{\bm{\rD_2}} & \Tw\j_2 & \ha\j_3 &-\vdt\Tw\j_1 &-\vdt\Tw\j_4
                &-i\Tw\cF & i\vdt\Tw{A} & i\vdt\Tw{B} &-i\Tw\cG \\ 
\C3{\bm{\rD_3}} &-\ha\j_3 & \Tw\j_2 & \vdt\Tw\j_4 &-\vdt\Tw\j_1
                &-i\Tw\cG & i\vdt\Tw{B} &-i\vdt\Tw{A} & i\Tw\cF \\ 
\C4{\bm{\rD_4}} & \Tw\j_4 & \Tw\j_1 & \vdt\ha\j_3 & \vdt\Tw\j_2
                & i\vdt\Tw{B} & i\Tw\cG & i\Tw\cF & i\vdt\Tw{A} \\ 
    \bottomrule
  \end{array}
 \qquad
 \vC{\begin{picture}(40,35)(0,-2)
   \put(0,0){\includegraphics[height=31mm]{tCScNt.pdf}}
    \put(16,0){\cB{$\Tw A$}}
    \put(25,0){\cB{$\Tw B$}}
    \put(1,14){\bB{$\Tw\j_1$}}
    \put(11,14){\bB{$-\Tw\j_4$}}
    \put(30,14){\bB{$\Tw\j_2$}}
    \put(39,14){\bB{$\ha\j_3$}}
    \put(16,29){\cB{$\Tw\cF$}}
    \put(25,29){\cB{$-\Tw\cG$}}
 \end{picture}}
 \label{e:tCS1Dt}
\end{equation}
In our Majorana basis, the twisted-complex chiral supermultiplet is
\begin{equation}
  \big(\,(\Tw{A}{+}i\Tw{B}) ~|~ (\Tw\j_1{-}i\Tw\j_4),~ (\Tw\j_2{+}i\ha\j_3)
        ~|~ (\Tw\cF{-}i\Tw\cG) 
       \,:\, [\C1{\rD_1}{-}i\C4{\rD_4}],[\C2{\rD_2}{-}i\C3{\rD_3}]\,\big),
 \label{e:cpxtCS}
\end{equation}
which indeed differs from\eq{e:cpxCS} only in the sign of $\C3{\rD_3}$.
Besides clustering the edges depicting the actions of the complex super-derivatives $[\C1{\rD_1}-i\C4{\rD_4}]$ and $[\C2{\rD_2}-i\C3{\rD_3}]$, the right-hand side Adinkra\eq{e:tCS1Dt} is also drawn in a way that immediately permits dimensionally extending this world-line supermultiplet to the world-sheet, by defining:
\begin{equation}
  \bD_-\Defl[\C1{\rD_1}-i\C4{\rD_4}],\quad
  \bDb_+\Defl[\C2{\rD_2}-i\C3{\rD_3}],\quad
  \bm{\Tw\j}_-\Defl(\Tw\j_1{-}i\Tw\j_4),\quad
  \bm{\Tw\j}_+\Defl(\Tw\j_2{+}i\ha\j_3),
 \label{e:cpxtCS1}
\end{equation}
where the $\pm$ subscripts indicate spin, in units of $\inv2\hbar$; $(\Tw{A}{+}i\Tw{B})$ and $(\Tw\cF{-}i\Tw\cG)$ have spin 0. Note that the definitions\eq{e:cpxtCS1} are in perfect agreement with\eq{e:cpxCS} and so provide for a direct comparison between\eqs{e:tCS1Dt}{e:cpxtCS} and\eqs{e:CS1D}{e:cpxCS}.  In particular, the annihilation condition specifying the twisted chiral supermultiplet, the analogue of\eq{e:DCS=0}, becomes the well known defining equation\cite{rGHR}:
\begin{equation}
  \bDb_-(\Tw{A}+i\Tw{B}) = 0 = \bD_+(\Tw{A}+i\Tw{B}).
 \label{e:cpxtCSd}
\end{equation}

In fact, the same holds also for\eq{e:vtCS1D} and\eq{e:cpxv2tCSa}, the Adinkra of which differs from\eq{e:tCS1Dt} only in the change $\ha\j_3\to\Tw\j_3=-\ha\j_3$, so that the edges adjacent to the (now) $\Tw\j_3$-node have flipped dashedness. This then ruins the mirror symmetry evident in\eq{e:tCS1Dt} and so obscures the presence of a supersymmetric complex structure\cite{rH-WWS}. Indeed, the fermionic identifications\eq{e:cpxtCS1} would become
\begin{equation}
  \bD_-\Defl[\C1{\rD_1}-i\C4{\rD_4}],\quad
  \bDb_+\Defl[\C2{\rD_2}-i\C3{\rD_3}],\quad
  \bm{\Tw\j}_-\Defl(\Tw\j_1{-}i\Tw\j_4),\quad
  \bm{\Tw\j}^*_+\Defl(\Tw\j_2{-}i\Tw\j_3),
 \label{e:cpxtCS2}
\end{equation}
where the conjugation in $\bm{\Tw\j}^*_+$ is necessary in direct comparison with\eq{e:cpxCS}.

\section{The Flat Metric of Linear Supersymmetry Representations}
\label{s:metric}
Upon dimensional reduction to the world-line, all supersymmetric models result in models of supersymmetric quantum mechanics. After a judicious field redefinition and renaming, the free-field kinetic term for standard off-shell propagating (physical) real bosons $\f_a(\t)$ and real fermions $\j_\a(\t)$ is of course:
\begin{equation}
  \text{KE}[(\f|\j)] = \frac{\k}2\int\!\rd\t~
                      \Big[\d^{ab}\,\Dt\f_a\,\Dt\f_b
                          +\frc{i}2\,\d^{\a\b}\,\big(\j_\a\,\Dt\j_\b
                                                -\Dt\j_\a\,\j_\b\big)\Big]
 \label{e:KE}
\end{equation}
with a suitable parameter $\k$. It is of course supersymmetric with respect to the world-line dimensional reduction of the original, higher-dimensional supersymmetry. Conversely, we have:
\begin{thrm}\label{T:KESZ}
The free-field action\eq{e:KE} is supersymmetric with respect to the maximally $N$-extended world-line supersymmetry transformations
\begin{equation}
  Q_I\f_a =  (\IL_I)_a{}^\a\,\j_\a,\qquad
  Q_I\j_\a= i(\IL^{\!\sss-1}_I)_\a{}^a\,\Dt\f_a,
 \label{e:Qfj}
\end{equation}
without central charges, where
\begin{subequations}
 \label{e:SuSy}
\begin{gather}
  \big\{\,Q_I\,,\,Q_J\,\big\}=2i\d_{IJ}\,\vdt,\qquad
  \big[\,\vdt\,,\,Q_I\,\big]=0,\qquad
  I,J=1,\cdots,N,
 \label{e:SuSyN}\\
  (Q_I)^\dagger=Q_I,\quad H^\dagger=i\vdt,
\end{gather}
\end{subequations}
provided there are as many bosons as fermions, $d_{\sss B}=d_{\sss F}=d$, and they can be partitioned into collections, each satisfying\eq{e:mBF}, given below. Moreover, the field-space flat metric $\d^{ab}\oplus\d^{\a\b}$ is induced {\em\/canonically\/} from the $\d_{IJ}$ specified by the algebra\eq{e:SuSy}.
\end{thrm}
\paragraph{Comments:}
All curvature tensors of the free-field metric of course vanish. An even number of boson and fermion fields can always be arranged into complex pairs, and a complex-paired set of supersymmetries\eq{e:Qfj} that leave the action\eq{e:KE} invariant can also always be found. However, this supersymmetry need not be the result of the dimensional reduction that produced\eq{e:KE}, and the dimensional reduction of that higher-dimensional supersymmetry may not be compatible with any complex pairings of the component fields; see the vector and tensor supermultiplets described in sections~\ref{s:VS} and~\ref{s:TS}, respectively.

The same canonically induced field-space metric, $\d^{ab}\oplus\d^{\a\b}$, on the component field space $(\f|\j)$ also occurs in the supersymmetric super-Zeeman term\cite{r6-7a}:
\begin{equation}
  \text{SZ}[(\f|\j),(\vf|\c)] = \w\int\!\rd\t~
                      \Big[\inv2\,\d^{ab}\,\big(\f_a\,\dt\vf_b-\dt\f_a\,\vf_b\big)
                          -i\,\d^{\a\b}\,\j_\a\,\c_\b\Big]
 \label{e:SZ}
\end{equation}
with a suitable Larmor-like frequency parameter $\w$, as well as various modifications of\eq{e:KE} and\eq{e:SZ} as discussed in Ref.\cite{r6-7a} and below. The combination\eq{e:KE}+(\ref{e:SZ}), with various additionally imposed boundary conditions that also restrict the supersymmetry action\eq{e:SuSy}, is the core world-line framework for all higher-dimensional models, with all additional terms added to the action representing deformations of this core.

\subsection{Proof}
We now derive the $\d^{ab}\oplus\d^{\a\b}$ metric from the metric $\d_{IJ}$ that occurs in the supersymmetry defining relations\eq{e:SuSy}. The proof consists of three stages, relating to:
 ({\bf1})~intact supermultiplets,
 ({\bf2})~projected supermultiplets,
 ({\bf3})~all engineerable supermultiplets,
and we proceed in turn.

\paragraph{Intact supermultiplets:}
We begin with defining the {\em\/intact\/} representation of supersymmetry\eq{e:SuSy}, by starting with, say, a real boson $\f_0$, and defining
\begin{subequations}
 \label{e:wQ}
\begin{alignat}9
 \f_0&,&\qquad
 \j_I&\Defl Q_I\f_0,\\
  F_{[IJ]}&\Defl Q_{[I}Q_{J]}\f_0,&\qquad
 \J_{[IJK]}&\Defl Q_{[I}Q_JQ_{K]}\f_0,\\[-1mm]
  \vdots~~&\qquad\quad\vdots&\qquad\vdots&\qquad\quad\vdots
\end{alignat}
\end{subequations}
which clearly terminates after defining $2^{N-1}$ bosonic and $2^{N-1}$ fermionic component fields. Owing to the relations\eq{e:SuSy}, the result of an infinitesimal supersymmetry transformation, $\d_Q(\e)\Defl\e^IQ_I$, acting on any of these fields is a linear combination of these fields and their $\vdt$-derivatives. Therefore, the collection of fields
\begin{equation}
   \cM_\diamondsuit\Defl\{\f_0|\j_I|F_{[IJ]}|\J_{[IJK]}|\cdots\}
 \label{e:X}
\end{equation}
represents a supermultiplet; it is not hard to see that these component fields are in 1--1 correspondence with the component fields obtained from a Salam-Strathdee superfield\cite{rSSSS4}; see also\cite{r1001,rPW,rWB,rBK}. From the definitions\eq{e:wQ} and the algebra\eq{e:SuSy} alone, it follows that:
\begin{subequations}
 \label{e:dQ}
\begin{alignat}9
 Q_I\f_0&=\j_I,&\qquad
 Q_I\j_J&=F_{[IJ]}+i\,\d_{IJ}\,\dt\f_0,\\
 Q_IF_{[JK]}&=\J_{[IJK]}+i\,\d_{I[J}\,\dt\j_{K]},&\qquad
 Q_I\J_{[JKL]}&={\cal F}_{[IJKL]}+i\,\d_{I[J}\,\dt{F}_{KL]},
 \label{e:X2=}
\end{alignat}
\end{subequations}
and so on. A closer examination shows that this multiplet is {\em\/adinkraic\/}\cite{r6-1}, \ie, {\em\/monomial\/}: the action of each one supercharge $Q_I$ upon each one of the component fields produces a single other of the component fields\eq{e:X} or a $\vdt$-derivative thereof.

From\eq{e:X}, we easily construct the {\em\/valise\/} supermultiplet\Ft{This has been variously also called an ``isoscalar'' and a ``base'' supermultiplet\cite{rGLP}.}:
\begin{subequations}
 \label{e:=}
\begin{alignat}9
 &&\makebox[0pt][c]{$\ddd \cM_=\Defl\{\f_0,\f_{[IJ]},\cdots\,|\,\j_I,\j_{[IJK]},\cdots\},$}\qquad\quad\\[1mm]
  \f_{[IJ]}&\Defl-i\vd_t^{-1}F_{[IJ]},&\qquad
 \j_{[IJK]}&\Defl-i\vd_t^{-1}\J_{[IJK]},\\[-1mm]
  \vdots~~&\qquad\quad\vdots&\qquad\vdots&\qquad\quad\vdots
\end{alignat}
\end{subequations}
where now all the bosons, $\f_0,\f_{[IJ]},\cdots$, have the same engineering dimension, as do all the fermions, $\j_I,\j_{[IJK]},\cdots$ By construction, within this supermultiplet, the supersymmetry acts straightforwardly:
\begin{equation}
  \e^IQ_I~:\quad
   \left\{\begin{array}{rcl}
           \big(\sum_k \a^{[I_1\cdots I_{2k}]}\,\f_{[I_1\cdots I_{2k}]}\big) &\to&
            \big(\sum_k \b^{[I_1\cdots I_{2k+1}]}\,\j_{[I_1\cdots I_{2k+1}]}\big),\\[2mm]
           \big(\sum_k \b^{[I_1\cdots I_{2k+1}]}\,\j_{[I_1\cdots I_{2k+1}]}\big) &\to&
            \big(\sum_k \g^{[I_1\cdots I_{2k}]}\,\dt\f_{[I_1\cdots I_{2k}]}\big),
           \end{array}\right.
\end{equation}
where the $\a,\b,\g$'s are general real coefficients, such that\eq{e:SuSy} is satisfied; indeed, the relations\eq{e:dQ} need only be corrected to accommodate the redefinitions\eq{e:X2=}. In particular, $\cM_=$ is also adinkraic, \ie, monomial.

Finally, we note that the metric $\d_{IJ}$, which is canonical as it is given by the very definition of the supersymmetry algebra\eq{e:SuSy}, induces a metric on the component field space:
\begin{alignat}9
 (f_{[I_1\cdots I_p]},f'_{[J_1\cdots J_p]}) &\Defl
  f_{[I_1\cdots I_p]}\,\d^{I_1J_1}\cdots\d^{I_pJ_p}\,f'_{[J_1\cdots J_p]},
  \qquad\text{for }p=0,\cdots,N,
 \label{e:(p,p)}
\end{alignat}
where $f_{[I_1\cdots I_p]}$ is a bosonic component field for even $p$, and a fermionic one for odd $p$. Since
\begin{equation}
  \sum_{k~\text{even}}\binom{N}{k}=2^{N-1}=\sum_{k~\text{odd}}\binom{N}{k},
\end{equation}
we may order the bosonic (and separately fermionic) component fields lexicographically, and count them using $a,b=1,\cdots,2^{N-1}$ for bosons (and $\a,\b=1,\cdots,2^{N-1}$ for fermions). It is straightforward that the diagonal bilinear form\eq{e:(p,p)}, with this counting, may be chosen to simply give
\begin{equation}
    \Dt\f_a\,\d^{ab}\,\Dt\f_b ~\oplus~ \j_\a\,\d^{\a\b}\,\Dt\j_\b
     \qquad\text{and}\qquad
    \f_a\,\d^{ab}\,\Dt\vf_b ~\oplus~ \j_\a\,\d^{\a\b}\,\c_\b,
\end{equation}
as needed in Eqs.\eq{e:KE} and\eq{e:SZ}, respectively. 
 The bosonic bilinear and the fermionic bilinear of course have different engineering dimensions, to accommodate the differing number of derivatives in either of the two supersymmetric Lagrangians\eq{e:KE} and\eq{e:SZ}.

\paragraph{Projected supermultiplets:}
As originally proven in Refs.\cite{r6-3,r6-3.2}~(see also Ref.\cite{r6-3.1}), all valise supermultiplets may be obtained from $\cM_=$ by means of {\em\/projections\/}. To this end, define the operator
\begin{equation}
  \P^{\bf b}_\pm\Defl\inv2\big[\vdt^{\,|{\bf b}|/2} \pm \bm{Q}^{\bf b}\big],\qquad
  \bm{Q}^{\bf b}\Defl Q_1^{b_1}\cdots Q_N^{b_N},
 \label{e:P}
\end{equation}
where {\bf b} is a length-$N$ doubly even binary number, \ie, the sum of its digits, $|{\bf b}|\Defl\sum_{I=1}^N b_I$, is divisible by 4. Such operators satisfy
\begin{equation}
  \P^{\bf b}_\pm\circ\P^{\bf b}_\pm = \vdt^{\,|{\bf b}|/2}\,\P^{\bf b}_\pm,
   \qquad
  \P^{\bf b}_\pm\circ\P^{\bf b}_\mp=0,
   \quad\text{and}\quad
  \P^{\bf b}_++\P^{\bf b}_-=\vdt^{\,|{\bf b}|/2}.
\end{equation}
They act as quasi-projection operators, in that
\begin{equation}
  \cM_= =
  \cM^{\bf b}_+ + \cM^{\bf b}_-
  \simeq\big(\P^{\bf b}_+\cM_=\big) + \big(\P^{\bf b}_-\cM_=\big)
\end{equation}
is a direct decomposition of the supermultiplet $\cM_=$ into two half-sized supermultiplets, and the relation ``$\simeq$'' here denotes that individual component fields on one and the other side may be equated up to a few initial terms in a Taylor expansion in $\t$. (This reflects the non-locality of the inverses of the field redefinitions\eq{e:X2=} and the ensuing introduction of ``integration constants''.)

To successively apply two such projections, $\P$ and $\P'$, the commutator $[\P,\P']$ must vanish when acting on a supermultiplet annihilated by both $\P$ and $\P'$. Correspondingly,
the bit-wise sum of the corresponding binary exponents must also be a doubly even binary number. Collections of such binary numbers, complete with respect to bit-wise addition, form doubly-even error-detecting and error-correcting binary linear block codes, $\sC$.

 It follows from the classification of these\cite{r6-3,r6-3.2}~(see also Ref.\cite{r6-3.1}) that the number of such simultaneous projections is
\begin{equation}
  k~~\leq~~\vk(N)\Defl \left\{
   \begin{array}{l@{~~\text{for}~~}l}
     0 & N<4;\\[1mm]
     \big\lfloor\frac{(N-4)^2}4\big\rfloor & N=4,5,6,7;\\[1mm]
     \vk(N{-}8)+4 & N>7,~~\text{recursively}.
   \end{array}\right.
\end{equation}
Let $\P^{\sC_k}_\p$ denote the collection of all mutually commuting projections $\P^{\bf b}_\p$ where ${\bf b}\in\sC_k$, where $\sC_k$ is the code formed as a collection of all bit-wise sums of $k$ linearly independent generators of the code, and $\p$ is a choice of signs for each individual projector.
 Since each projection halves the number of component fields, the total number of bosonic+fermionic component fields in a $k$-fold projected (halved) supermultiplet
\begin{equation}
 \begin{aligned}
  (\P^{\sC_k}_{\p}\cM_=)
  =\,&\big(\,\P^{\sC_k}_{\p}\{\f_0,\cdots,\f_{[I_1\cdots I_p]}\}\,\big|
         \,\P^{\sC_k}_{\p}\{\f_I,\cdots,\f_{[I_1\cdots I_q]}\}\,\big)\\
  &~0\leqslant p,q\leqslant N,~~p\cong0\,\text{mod}\,2,~~q\cong1\,\text{mod}\,2,
 \end{aligned}
\end{equation}
is
\begin{equation}
 (d_{\sss B}\,{=}\,2^{N-k-1})+(d_{\sss F}\,{=}\,2^{N-k-1}),
 \label{e:mBF}
\end{equation}
and each such $\sC_k$-projected distinct valise supermultiplet defines a corresponding {\em\/chromotopology\/}; the maximally projected $(\P^{\sC_{\vk(N)}}_{\p}\cM_=)$ are the minimal valise supermultiplets, and are classified by the maximal codes $\sC_{\vk(N)}$\cite{r6-3,r6-3.2,r6-3.1}.

Since each projection defines mutually orthogonal linear combination component fields, the metric $\d^{ab}\oplus\d^{\a\b}$ in $\cM_=$ reduces to a half-size, still diagonal and still positive-definite metric on the halved component field space; the diagonal values can then always be adjusted to unity by properly normalizing the linear combination component fields such as $\f^\pm_0$. For example, for $N=4$, there is a single binary doubly-even linear block code, defining the operators\eq{e:P}
\begin{equation}
  \P^{\sss[1111]}_\pm\Defl\inv2\big[\vdt^2\pm Q_1Q_2Q_3Q_4\big].
 \label{e:Pd4pm}
\end{equation}
Note that, modulo $\q$-dependent terms and up to an overall numerical constant, the operators\eq{e:CSDD}, (\ref{e:VSDD}), (\ref{e:TSDD}), (\ref{e:tCSDD}) and\eq{e:VSDD} all square to\eq{e:Pd4pm}.
 Projecting the $8+8$-component valise supermultiplet\eq{e:=} then produces component fields such as $\f^\pm_0\Defl\fc12(\f_0\pm\f_{[1234]})$, for each of which it is straightforward to prove that
\begin{equation}
   \Dt\f{}_0^{~2} + \Dt\f{}_{\sss[1234]}^{~2} = 2(\Dt\f{}_0^{\,-})^2 + 2(\Dt\f{}_0^{\,+})^2
\end{equation}

Therefore, each of worldline supermultiplets, with any of the $10^{12}$ chromotopologies of Refs.\cite{r6-3,r6-3.2,r6-3.1}, has a ``free'' kinetic action term of the form\eq{e:KE}. In addition, for any pair of such supermultiplets there exists a super-Zeeman action term such as\eq{e:SZ}.

\paragraph{Engineerable Supermultiplets:}
All physical fields have a definite engineering dimension, stemming from their physical units expressed in the natural system where the units $\hbar,c$ are not written explicitly. Supermultiplets wherein every component field has a definite engineering dimension were called {\em\/engineerable\/}\cite{r6-1}. It then follows that the world-line reduction of every physically relevant supermultiplet in any higher-dimensional theory must be engineerable.

In turn, Ref.\cite{rDHIL13} proved that every engineerable, finite-dimensional, off-shell, unitary representation of $N$-extended world-line supersymmetry without central charges, upon component field {\em\/raising\/}\ft{Refs.\cite{rGR0,rPT,rGLP} introduced and used extensively the operation, dubbed variously {\em\/automorphic duality\/}\cite{rGR0}, {\em\/dressing transformation\/}\cite{rKRT}, and {\em\/node-raising/lowering\/}\cite{r6-1}; we use the latter terms for their precision.} of a judicious selection of fields
\begin{equation}
  \f_{\2a}\to F_{\2a}\Defl(\vdt\f_{\2a}),\qquad \text{for some $\2a\in\{1,\cdots,d_{\sss B}\}$},
 \label{e:nRL}
\end{equation}
and judicious real linear combination of so-obtained fields, decomposes into a real linear combination of minimal valises for the specified $N$. We have proven above that all valises (including the minimal ones) do have a {\em\/canonical\/} positive quadratic form (metric)\eq{e:KE}, induced from the quadratic form defined by the supersymmetry algebra itself\eq{e:SuSy}.

As real linear combinations and raising/lowering transformations\eq{e:nRL} cannot change the positive-definiteness of a quadratic form (metric) on the fields (and their derivatives) as used in\eq{e:KE}, it follows that
 all engineerable, finite-dimensional, off-shell, unitary representation of $N$-extended world-line supersymmetry without central charges also have a quadratic form (metric) on the fields (and their derivatives) as used in\eq{e:KE}.

Therefore, the world-line reduction of every physically relevant supermultiplet in any higher-dimensional theory has a supersymmetric kinetic Lagrangian of the form\eq{e:KE}, and the positive-definite quadratic form (metric) used in\eq{e:KE} is {\em\/canonical\/}, in that it is induced from the metric introduced in the supersymmetry algebra\eq{e:SuSy} itself.\QED

\subsection{Completeness and Symmetry}
As rigorously proven in Ref.\cite{r6-1}, all supermultiplets with the same chromotopology may be obtained one from another through the component field redefinitions that generalize\eq{e:X2=} by combinatorially varying the choices of which component fields are transformed by node-raising/lowering. Having obtained the kinetic\eq{e:KE} and super-Zeeman\eq{e:SZ} action terms, the supermultiplets may be adapted to each of the node-raised/lowered variant of the supermultiplets involved. For example, component field raising\eq{e:nRL} a selection of fields
produces
\begin{equation}
  \text{KE}[(\f|\j|F)] = \frac{\k}2\int\!\rd\t~
   \Big[\underbrace{\d^{ab}\,\Dt\f_a\,\Dt\f_b}_{a\neq\2a,~ b\neq\2b}
        +\d^{\2a\,\2b}F_{\2a}F_{\2b}
                          +i\,\d^{\a\b}\,\big(\j_\a\,\Dt\j_\b
                                                -\Dt\j_\a\,\j_\b\big)\Big]
 \label{e:KE2}
\end{equation}
as the resulting kinetic action term. This procedure has been systematically explored for the so-called ultramultiplet in Ref.\cite{rFGH}.

This brings us finally to an alternate consideration: as discussed above, the supersymmetry algebra itself\eq{e:SuSy} is invariant with respect to the $\rO(N)$-transformations $Q_I\to Q'_I\Defl(\IO)_I{}^JQ_J$ and $\d_{IJ}$ in\eq{e:SuSyN} defines the $\rO(N)$-invariant metric. It is not difficult to show that $\{\f_0,\f_{[IJ]},\cdots\}$ and $\{\j_I,\j_{[IJK]},\cdots\}$ span  two spinor representations of this group, which is the reason for writing $\Aut(\SSp^{1|N})=\Pin(N)$ rather than $\Aut(\SSp^{1|N})=\rO(N)$; see also footnote~\ref{f:Pin} on p.~\pageref{f:Pin}.

In turn, given a collection of $m=2^{N-k-1}$ real bosonic and real fermionic component fields\eq{e:=}, the maximal group of transformations of these component fields is $\rO(2^{N-k-1})_{\sss B}\times\rO(2^{N-k-1})_{\sss F}$. Of course, not all such transformations will preserve the action of supersymmetry within the supermultiplet. The collection of component field transformations modulo the ones induced by $\Aut(\SSp^{1|N})=\Pin(N)$ then form the product of cosets
\begin{equation}
  \big[\rO(2^{N-k-1})_{\sss B}\big/\Pin(N)\big]~\times\,\,
  \big[\rO(2^{N-k-1})_{\sss F}\big/\Pin(N)\big].
\end{equation}
We note that these cosets are discrete only for $N=1,2,4,8$ and are continuous for all other $N$. As a coarse estimate, this indicates that the level of difficulty in classifying off-shell supermultiplets radically changes outside the $N=1,2,4,8$ cases. Not coincidentally, the minimal supermultiplets in those cases have $1+1$, $2+2$, $4+4$ and $8+8$ component fields an exhibit real, complex, quaternionic and octonionic structures, respectively.

Accordingly,
\begin{enumerate}\itemsep=-3pt\vspace{-2mm}
 \item by choosing a subset of bosons to redefine as in\eq{e:nRL}, the $\Pin(N)$ symmetry is broken to a subgroup, and different choices of subsets of bosons to redefine correspond to the distinct subgroups of $\Pin(N)$;
 \item the projections obtained using the quasi-projection operators such as\eq{e:P} are spinor-halving projections generalizing the familiar Dirac\,$\to$\,Weyl or the Dirac\,$\to$\,Majorana projection of the so-named spinors in 4d physics\Ft{For such projections of Dirac spinors in spacetime physics, one additionally requires Lorentz-covariance, of which there is no analogue in the present context. This then is what ultimately permits the combinatorially vast plethora of projection possibilities\cite{r6-3,r6-3.2}.}; see also Ref.\cite{r6-3.1}.
\end{enumerate}
The metric $\d^{ab}\oplus\d^{\a\b}$ on the so (possibly iteratively) halved bosonic+fermionic component field space is then the standard, maximally symmetric metric. Indeed, the action terms\eq{e:KE} and\eq{e:SZ} exhibit the much larger $\rO(2^{N-k-1})$ (dynamical/effective) symmetry rather than just $\Pin(N)$ or a subgroup thereof.

For example, the unprojected valise supermultiplet of $N=4$ supersymmetry has $8+8$ boso\-nic+fer\-mi\-onic component fields, whereupon the kinetic action term\eq{e:KE} for each such supermultiplet, and also the super-Zeeman action term\eq{e:SZ} for each pair of such supermultiplets exhibits an $\rO(8)$ (dynamical/effective) symmetry, enlarging the $\Pin(4)$ group of automorphisms of the supersymmetry algebra $\SSp^{1|4}$. Note that $\Pin(4)\subset\rO(8)$ is a {\em\/special\/} (irregular) subgroup\cite{rSsky}, wherein the real vector of $\rO(8)$ transforms as the ($\IC^4\simeq\IR^8$-like) {\em\/spinor\/} of $\Pin(4)\subset\rO(8)$.
 This is also the defining property in the general case, $\Pin(N)\subset\rO(2^{N-1})$, and why we systematically refer to $\Pin(N)=\Aut(\SSp^{1|N})$, but to $\rO(2^{N-1})$ or $\SO(2^{N-1})$ as the\eq{e:KE} and\eq{e:SZ}-preserving group of linear, homogeneous transformations of the bosons and separately the fermions.

Within the free-field limit of any model, the bilinear action terms\eq{e:KE} and \eq{e:SZ} provide the ``supersymmetry-canonical'' kinetic and Zeeman terms; any other action term then may be considered as a deformation thereof.

\begingroup
\small
\raggedright

\endgroup


\begin{thebibliography}{10}
\bibitem{SGG} 
  I.~Chappell, S.~J.~Gates, Jr., W.~D.~Linch, III.,  J.~Parker, S.~Randall, A.~Ridgway and K.~Stiffler, ``4D, N=1 Supergravity Genomics,''
 \href{http://link.springer.com/article/10.1007\%2FJHEP10\%282013\%29004}{ \emph{JHEP} {\bf 1310}, 004 (2013)}, 
 \href{http://arxiv.org/abs/1212.3318}{ arXiv:1212.3318.}

\bibitem{rGR-1}
S.~J. Gates, Jr. and L.~Rana, ``On extended supersymmetric quantum mechanics,''
  {\em University of Maryland Report: UMDPP 93-194} (1994) unpublished.

\bibitem{rGR0}
S.~J. Gates, Jr. and L.~Rana, ``Ultramultiplets: A new representation of rigid
  2-d, ${N}$=8 supersymmetry,'' {\em Phys. Lett.} {\bfseries B342} (1995)
  132--137,
\href{http://arxiv.org/abs/hep-th/9410150}{{\ttfamily arXiv:hep-th/9410150}}.

\bibitem{rGLPR}
S.~J. Gates, Jr., W.~D. Linch, III, J.~Phillips, and L.~Rana, ``The fundamental
  supersymmetry challenge remains,'' {\em Gravit. Cosmol.} {\bfseries 8}
  no.~1-2, (2002) 96--100,
  \href{http://arxiv.org/abs/hep-th/0109109}{{\ttfamily arXiv:hep-th/0109109}}.

\bibitem{rGLP}
S.~J. Gates, Jr., W.~D. Linch, III, and J.~Phillips, ``When superspace is not
  enough,'' \href{http://arxiv.org/abs/hep-th/0211034}{{\ttfamily
  arXiv:hep-th/0211034}}.

\bibitem{rSagn87}
A.~Sagnotti in {\em Cargese 1987 Proceedings ``Non-Perturbative Quantum Field
  Theory''}, G.~{Mack et al.}, ed., p.~521.
\newblock Pergamon Press, 1988.

\bibitem{rHor89}
P.~Ho{\v{r}}ava, ``Strings on world sheet orbifolds,''
  \href{http://dx.doi.org/10.1016/0550-3213(89)90279-4}{{\em Nucl.Phys.}
  {\bfseries B327} (1989) 461}.

\bibitem{rDLP89-OrF}
J.~Dai, R.~G. Leigh, and J.~Polchinski, ``New connections between string
  theories,'' \href{http://dx.doi.org/10.1142/S0217732389002331}{{\em Mod.
  Phys. Lett.} {\bfseries A4} (1989) 2073--2083}.

\bibitem{rL89-OrF}
R.~G. Leigh, ``Dirac-born-infeld action from dirichlet sigma model,''
  \href{http://dx.doi.org/10.1142/S0217732389003099}{{\em Mod. Phys. Lett.}
  {\bfseries A4} (1989) 2767}.

\bibitem{rSJG12}
S.~J. Gates, Jr., ``The search for elementarity among off-shell susy
  representations,'' {\em Korean Institute for Advanced Study Newsletter}
  {\bfseries 5} (2012) .

\bibitem{rGHS-HoloSuSy}
S.~J. Gates, Jr., T.~H{\"u}bsch, and K.~Stiffler, ``Adinkras and {SUSY}
  holography,'' \href{http://dx.doi.org/10.1142/S0217751X14500419}{{\em Int. J.
  Mod. Phys.} {\bfseries A29} no.~7, (2014) 1450041},
  \href{http://arxiv.org/abs/1208.5999}{{\ttfamily arXiv:1208.5999}}.

\bibitem{r6-1}
C.~F. Doran, M.~G. Faux, S.~J. Gates, Jr., T.~H{\"u}bsch, K.~M. Iga, and G.~D.
  Landweber, ``On graph-theoretic identifications of {A}dinkras, supersymmetry
  representations and superfields,''
  \href{http://dx.doi.org/10.1142/S0217751X07035112}{{\em Int. J. Mod. Phys.}
  {\bfseries A22} (2007) 869--930},
  \href{http://arxiv.org/abs/math-ph/0512016}{{\ttfamily
  arXiv:math-ph/0512016}}.

\bibitem{rHTSSp08}
T.~H{\"u}bsch, ``Superspace: A comfortably vast algebraic variety,'' in {\em
  Geometry and Analysis}, L.~Ji, ed., vol.~2 of {\em Advanced Lectures in
  Mathematics}, pp.~39--67.
\newblock International Press, 2010.
\newblock \href{http://arxiv.org/abs/0901.2136}{{\ttfamily arXiv:0901.2136}}.
\newblock Closing address at the {C}onference ``{G}eometric {A}nalysis:
  {P}resent and {F}uture'', Harvard University, 2008.

\bibitem{r1001}
S.~J. Gates, Jr., M.~T. Grisaru, M.~Ro{\v{c}}ek, and W.~Siegel, {\em
  Superspace}.
\newblock Benjamin/Cummings Pub. Co., Reading, MA, 1983.

\bibitem{rWB}
J.~Wess and J.~Bagger, {\em Supersymmetry and Supergravity}.
\newblock Princeton Series in Physics. Princeton University Press, Princeton,
  NJ, 2nd~ed., 1992.

\bibitem{rBK}
I.~L. Buchbinder and S.~M. Kuzenko, {\em Ideas and Methods of Supersymmetry and
  Supergravity}.
\newblock Studies in High Energy Physics Cosmology and Gravitation. IOP
  Publishing Ltd., Bristol, 1998.

\bibitem{rSSSS1}
A.~Salam and J.~Strathdee, ``Supergauge transformations,'' {\em Nucl. Phys.}
  {\bfseries B76} (1974) 477--482.

\bibitem{rDHIL13}
C.~F. Doran, T.~H{\"u}bsch, K.~M. Iga, and G.~D. Landweber, ``On general
  off-shell representations of worldline (1{D}) supersymmetry,''
  \href{http://dx.doi.org/10.3390/sym6010067}{{\em Symmetry} {\bfseries 6}
  no.~1, (2014) 67--88}, \href{http://arxiv.org/abs/1310.3258}{{\ttfamily
  arXiv:1310.3258}}.

\bibitem{rKRT}
Z.~Kuznetsova, M.~Rojas, and F.~Toppan, ``Classification of irreps and
  invariants of the ${N}$-extended supersymmetric quantum mechanics,'' {\em
  JHEP} {\bfseries 03} (2006) 098,
\href{http://arxiv.org/abs/hep-th/0511274}{{\ttfamily arXiv:hep-th/0511274}}.

\bibitem{r6-3.2}
C.~F. Doran, M.~G. Faux, S.~J. Gates, Jr., T.~H{\"u}bsch, K.~M. Iga, G.~D.
  Landweber, and R.~L. Miller, ``Adinkras for clifford algebras, and worldline
  supermultiplets,'' \href{http://arxiv.org/abs/0811.3410}{{\ttfamily
  arXiv:0811.3410}}.

\bibitem{rGR2}
S.~J. Gates, Jr. and L.~Rana, ``A theory of spinning particles for large
  {$N$}-extended supersymmetry. {II},'' {\em Phys. Lett.} {\bfseries B369}
  no.~3-4, (1996) 262--268,
  \href{http://arxiv.org/abs/hep-th/9510151}{{\ttfamily arXiv:hep-th/9510151}}.

\bibitem{r6-3}
C.~F. Doran, M.~G. Faux, S.~J. Gates, Jr., T.~H{\"u}bsch, K.~M. Iga, G.~D.
  Landweber, and R.~L. Miller, ``Topology types of {A}dinkras and the
  corresponding representations of ${N}$-extended supersymmetry,''
  \href{http://arxiv.org/abs/0806.0050}{{\ttfamily arXiv:0806.0050}}.

\bibitem{r6-3.1}
C.~F. Doran, M.~G. Faux, S.~J. Gates, Jr., T.~H{\"u}bsch, K.~M. Iga, G.~D.
  Landweber, and R.~L. Miller, ``Codes and supersymmetry in one dimension,''
  \href{http://dx.doi.org/10.4310/ATMP.2011.v15.n6.a7}{{\em Adv. Theor. Math.
  Phys.} {\bfseries 15} (2011) 1909--1970},
  \href{http://arxiv.org/abs/1108.4124}{{\ttfamily arXiv:1108.4124}}.

\bibitem{rLM}
H.~B. Lawson, Jr. and M.-L. Michelsohn, {\em Spin geometry}, vol.~38 of {\em
  Princeton Mathematical Series}.
\newblock Princeton University Press, Princeton, NJ, 1989.

\bibitem{rFRH}
F.~R. Harvey, {\em Spinors and Calibrations}, vol.~9 of {\em Perspectives in
  Mathematics}.
\newblock Academic Press Inc., Boston, MA, 1990.

\bibitem{rHMP-Inv}
T.~H{\"u}bsch, S.~Meljanac, and S.~Pallua, ``Invariants of self-adjoint
  rank-four ${SU}(n)$ tensors,'' {\em Phys. Rev.} {\bfseries D32} (1985) 1021.

\bibitem{rHE}
S.~W. Hawking and G.~F.~R. Ellis, {\em The Large Scale Structure of
  Space-Time}.
\newblock Cambridge University Press, 1973.

\bibitem{rWyb}
B.~G. Wybourne, {\em Classical Groups for Physicists}.
\newblock John Wiley \&\ Sons Inc., 1974.

\bibitem{rUMD09-1}
S.~J. Gates, Jr., J.~Gonzales, B.~MacGregor, J.~Parker, R.~Polo-Sherk, V.~G.~J.
  Rodgers, and L.~Wassink, ``4{D}, $\mathcal{N}=1$ supersymmetry genomics
  ({I}),'' {\em JHEP} {\bfseries 12} (2009) 009,
  \href{http://arxiv.org/abs/0902.3830}{{\ttfamily arXiv:0902.3830}}.

\bibitem{rHall}
B.~C. Hall, {\em Lie Groups, Lie Algebras, and Representations}.
\newblock Springer-Verlag, 2003.

\bibitem{rPR-GTh}
P.~Ramond, {\em Group Theory: A Physicist's Survey}.
\newblock Cambridge University Press, 2010.

\bibitem{rGH-obs}
S.~J. Gates, Jr. and T.~H{\"u}bsch, ``On dimensional extension of
  supersymmetry: From worldlines to worldsheets,''
  \href{http://dx.doi.org/10.4310/ATMP.2012.v16.n6.a2}{{\em Adv. Theor. Math.
  Phys.} {\bfseries 16} no.~6, (2012) 1619--1667},
  \href{http://arxiv.org/abs/1104.0722}{{\ttfamily arXiv:1104.0722}}.

\bibitem{rFIL}
M.~G. Faux, K.~M. Iga, and G.~D. Landweber, ``Dimensional enhancement via
  supersymmetry,'' {\em Adv. Math. Phys.} {\bfseries 2011} (2011) 259089,
  \href{http://arxiv.org/abs/0907.3605}{{\ttfamily arXiv:0907.3605}}.

\bibitem{rFL}
M.~G. Faux and G.~D. Landweber, ``Spin holography via dimensional
  enhancement,'' {\em Phys. Lett.} {\bfseries B681} (2009) 161--165,
  \href{http://arxiv.org/abs/0907.4543}{{\ttfamily arXiv:0907.4543}}.

\bibitem{r6-1.2}
C.~F. Doran, M.~G. Faux, S.~J. Gates, Jr., T.~H{\"u}bsch, K.~M. Iga, and G.~D.
  Landweber, ``A superfield for every dash-chromotopology,''
  \href{http://dx.doi.org/10.1142/S0217751X09047685}{{\em Int. J. Mod. Phys.}
  {\bfseries A24} (2009) 5681--5695},
  \href{http://arxiv.org/abs/0901.4970}{{\ttfamily arXiv:0901.4970}}.

\bibitem{rTwSJG0}
S.~J. Gates, Jr., ``Superspace formulation of new nonlinear sigma models,''
{\em Nucl. Phys.} {\bfseries B238} (1984) 349--366.

\bibitem{rGHR}
S.~J. Gates, Jr., C.~M. Hull, and M.~Ro{\v{c}}ek, ``Twisted multiplets and new
  supersymmetric nonlinear sigma models,''
{\em Nucl. Phys.} {\bfseries B248} (1984) 157.

\bibitem{rGRS13-RADIO}
S.~J. Gates, Jr., S.~Randall, and K.~Stiffler, ``Reduction redux of
  {A}dinkras,'' \href{http://arxiv.org/abs/1312.2000}{{\ttfamily
  arXiv:1312.2000 [hep-th]}}.

\bibitem{rH-TSS}
T.~H{\"u}bsch, ``On supermultiplet twisting and spin-statistics,''
  \href{http://dx.doi.org/10.1142/S0217732313501472}{{\em Mod. Phys. Lett.}
  {\bfseries 28} no.~33, (2013) 1350147},
  \href{http://arxiv.org/abs/1203.4189}{{\ttfamily arXiv:1203.4189}}.

\bibitem{rPT}
A.~Pashnev and F.~Toppan, ``On the classification of ${N}$-extended
  supersymmetric quantum mechanical systems,'' {\em J. Math. Phys.} {\bfseries
  42} (2001) 5257--5271,
\href{http://arxiv.org/abs/hep-th/0010135}{{\ttfamily arXiv:hep-th/0010135}}.

\bibitem{rT01}
F.~Toppan, ``Classifying ${N}$-extended 1-dimensional supersymmetric systems,''
  in {\em Noncommutative Structures in Mathematics and Physics}, S.~Duplij and
  J.~Wess, eds., p.~195.
\newblock Kluwer Ac. Pub, 2001.
\newblock
\href{http://arxiv.org/abs/hep-th/0109047}{{\ttfamily arXiv:hep-th/0109047}}.
\newblock

\bibitem{rT01a}
F.~Toppan, ``Division algebras, extended supersymmetries and applications,''
  {\em Nucl. Phys. Proc. Suppl.} {\bfseries 102} (2001) 270--277,
\href{http://arxiv.org/abs/hep-th/0109073}{{\ttfamily arXiv:hep-th/0109073}}.

\bibitem{rCRT}
H.~L. Carrion, M.~Rojas, and F.~Toppan, ``Octonionic realizations of
  1-dimensional extended supersymmetries. {A} classification,'' {\em Mod. Phys.
  Lett.} {\bfseries A18} (2003) 787--798,
\href{http://arxiv.org/abs/hep-th/0212030}{{\ttfamily arXiv:hep-th/0212030}}.

\bibitem{rKT07}
Z.~Kuznetsova and F.~Toppan, ``Refining the classification of the irreps of the
  1{D} ${N}$-extended supersymmetry,'' {\em Mod. Phys. Lett.} {\bfseries A23}
  (2008) 37--51, \href{http://arxiv.org/abs/hep-th/0701225}{{\ttfamily
  arXiv:hep-th/0701225}}.

\bibitem{rGKT10}
M.~Gonzales, S.~Khodaee, and F.~Toppan, ``On non-minimal ${N}=4$
  supermultiplets in {1D} and their associated sigma-models,'' {\em J. Math.
  Phys.} {\bfseries 52} (2011) 013514,
  \href{http://arxiv.org/abs/1006.4678}{{\ttfamily arXiv:1006.4678}}.

\bibitem{rFGH}
M.~G. Faux, S.~J. Gates, Jr., and T.~H{\"u}bsch, ``Effective symmetries of the
  minimal supermultiplet of ${N} = 8$ extended worldline supersymmetry,''
  \href{http://dx.doi.org/10.1088/1751-8113/42/41/415206}{{\em J. Phys.}
  {\bfseries A42} (2009) 415206},
  \href{http://arxiv.org/abs/0904.4719}{{\ttfamily arXiv:0904.4719}}.

\bibitem{rH-WWS}
T.~H{\"u}bsch, ``Weaving worldsheet supermultiplets from the worldlines
  within,'' {\em Adv. Theor. Math. Phys.} {\bfseries 17} (2013) 1--72,
  \href{http://arxiv.org/abs/1104.3135}{{\ttfamily arXiv:1104.3135}}.

\bibitem{r6-7a}
C.~F. Doran, M.~G. Faux, S.~J. Gates, Jr., T.~H{\"u}bsch, K.~M. Iga, and G.~D.
  Landweber, ``Super-{Z}eeman embedding models on ${N}$-supersymmetric
  world-lines,'' {\em J. Phys. A} {\bfseries 42} (2008) 065402,
  \href{http://arxiv.org/abs/0803.3434}{{\ttfamily arXiv:0803.3434}}.

\bibitem{rSSSS4}
A.~Salam and J.~Strathdee, ``On superfields and {F}ermi-{B}ose symmetry,'' {\em
  Phys. Rev. D} {\bfseries 11} (1975) 1521--1535.

\bibitem{rPW}
P.~West, {\em Introduction to Supersymmetry and Supergravity}.
\newblock World Scientific Publishing Co. Inc., Teaneck, NJ, 1990.

\bibitem{rSsky}
R.~Slansky, ``Group theory for unified model building,'' {\em Phys. Rep.}
  {\bfseries 79} no.~1, (1981) 1--128.

\end{thebibliography}
\end{document}